\documentclass[screen,format=sigconf,table,xcdraw,9pt,authorversion]{acmart}

\usepackage{graphicx}
\usepackage{subfigure}
\usepackage[justification=RaggedRight,labelfont=bf,font=small,textfont=normalfont,tableposition=below]{caption}
\usepackage{makecell}
\usepackage{amsmath}
\usepackage{url}
\usepackage{multirow}
\usepackage[ruled,vlined]{algorithm2e}
\usepackage{enumitem}
\usepackage{listings}
\usepackage{diagbox}
\usepackage{colortbl}
\usepackage{hyperref}
\usepackage{comment}
\usepackage{balance}
\usepackage{dsfont}
\usepackage{changepage}

\setitemize{leftmargin=4mm}
\setlist{noitemsep,topsep=0pt,parsep=0pt,partopsep=0pt}

\newcommand{\paraskip}{\vspace{4pt plus 2pt minus 2pt}}
\newcommand{\name}{\textsf{Metis}}

\newcommand{\parahead}[1]{\paraskip\noindent\textbf{#1}}

\newcommand{\revise}[1]{#1}

\begin{document}

\title{Interpreting Deep Learning-Based Networking Systems}


\author{Zili Meng$^{1}$, Minhu Wang$^{1}$, Jiasong Bai$^{1, 2}$, Mingwei Xu$^{1, 2}$, Hongzi Mao$^3$, Hongxin Hu$^4$}
\affiliation{$^1$Institute for Network Science and Cyberspace \& BNRist, Tsinghua University\\
	$^2$Department of Computer Science and Technology, Tsinghua University\\
	$^3$Massachusetts Institute of Technology,\qquad$^4$Clemson University\\
	\textsf{zilim@ieee.org, \{\{wangmh19, bjs17\}@mails., xumw@\}tsinghua.edu.cn, hongzi@mit.edu, hongxih@clemson.edu}}



\renewcommand{\authors}{Zili Meng, Minhu Wang, Jiasong Bai, Mingwei Xu, Hongzi Mao, Hongxin Hu.}
\renewcommand{\shortauthors}{Zili Meng, Minhu Wang, Jiasong Bai, Mingwei Xu, Hongzi Mao, Hongxin Hu.}

\begin{abstract}
While many deep learning (DL)-based networking systems have demonstrated superior performance, the underlying Deep Neural Networks (DNNs) remain blackboxes and stay uninterpretable for network operators. The lack of interpretability makes DL-based networking systems prohibitive to deploy in practice. In this paper, we propose \name, a framework that provides interpretability for two general categories of networking problems spanning local and global control. Accordingly, \name\ introduces two different interpretation methods based on decision tree and hypergraph, where it converts DNN policies to interpretable rule-based controllers and highlight critical components based on analysis over hypergraph. We evaluate \name\ over two categories of state-of-the-art DL-based networking systems and show that \name\ provides human-readable interpretations while preserving nearly no degradation in performance. We further present four concrete use cases of \name, showcasing how \name\ helps network operators to design, debug, deploy, and ad-hoc adjust DL-based networking systems.
\end{abstract}

\acmYear{2020}\copyrightyear{2020}
\setcopyright{acmlicensed}
\acmConference[SIGCOMM '20]{Annual conference of the ACM Special Interest Group on Data Communication on the applications, technologies, architectures, and protocols for computer communication}{August 10--14, 2020}{Virtual Event, NY, USA}
\acmBooktitle{Annual conference of the ACM Special Interest Group on Data Communication on the applications, technologies, architectures, and protocols for computer communication (SIGCOMM '20), August 10--14, 2020, Virtual Event, NY, USA}
\acmPrice{15.00}
\acmDOI{10.1145/3387514.3405859}
\acmISBN{978-1-4503-7955-7/20/08}



\maketitle

\section{Introduction}
\label{sec:intro}

Recent years have witnessed a steady trend of applying deep learning (DL) to a diverse set of network optimization problems, including video streaming~\cite{sigcomm2017pensieve, osdi2018nas, abrl}, local traffic control~\cite{sigcomm2018auto, icml2019aurora}, parallel job scheduling~\cite{arxiv201808dq, sigcomm2019decima}, and network resource management~\cite{sosr2019routenet, iwqos2019nfvdeep, icnp2019lego}. 
The key enabler for this trend is the use of Deep Neural Networks (DNNs), thanks to their strong ability to fit complex functions for prediction~\cite{dl_nature, nn_universal}.
Moreover, DNNs are easy to marry with standard optimization techniques such as reinforcement learning (RL)~\cite{sutton2018reinforcement} to allow data-driven and automatic performance improvement.
Consequently, prior work has demonstrated significant improvement with DNNs over hand-crafted heuristics in multiple network applications~\cite{sigcomm2018auto, sigcomm2017pensieve, sigcomm2019decima}.

However, the superior performance of DNNs comes at the cost of using millions or even billions of parameters~\cite{dl_nature, gpt3billion}.
This cost is fundamentally rooted in the design of DNNs, as they typically require numerous parameters to achieve universal function approximation~\cite{nn_universal}.
Therefore, network operators have to consider DNNs as large blackboxes~\cite{netai2019cracking, apnet2018demystifying}, which makes DL-based networking systems incomprehensible to debug, heavyweight to deploy, and extremely difficult to ad-hoc adjust~(\S\ref{sec:motiv}).
As a result, network operators firmly hold a general fear against using DL-based networking systems for critical deployment in practice.

Over the years, the machine learning community has developed several techniques for understanding the behaviors of DNNs in the scope of image recognition~\cite{eccv2014visualizing, cvpr2017netdissect} and language translation~\cite{neurips2019interpretnlp, acl2018semantically}.
These techniques focus on surgically monitoring the activation of neurons to determine the set of features that the neurons are sensitive to~\cite{cvpr2017netdissect}.
However, directly applying these techniques to DL-based networking systems is not suitable\,---\,network operators typically seek simple, deterministic control rules mapped from the input (e.g., scheduling packets with certain headers to a port), as opposed to nitpicking the operational details of DNNs.
Besides, networking systems are diverse in terms of their application settings (e.g., distributed control v.s. centralized decision making) and their input data structure (e.g., time-series of throughput and routing paths in a topology).
The current DNN interpretation tools, designed primarily for well-structured vector inputs (e.g., images, sentences), are not sufficient across diverse networking systems.
Therefore, an interpretable DL framework specifically tailored for the networking domain is much needed.


\begin{figure}
    \vspace{1\baselineskip}
    \centering
	\includegraphics[width=.9\linewidth]{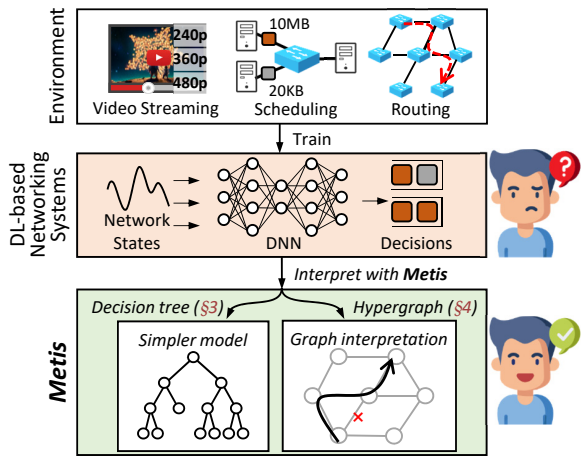}
	\caption{High-level workflow of \name.}
	\label{fig:intro}
\end{figure}

\begin{figure}
	\centering
	\includegraphics[width=\linewidth]{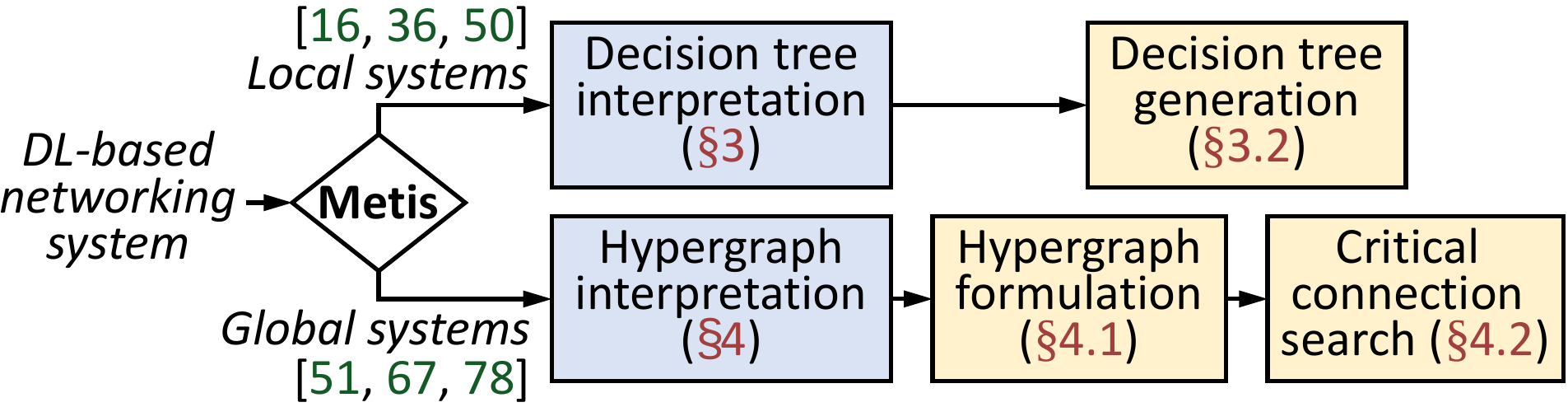}
	\caption{\name\ introduces different interpretation methods for local and global DL-based networking systems.}
	\label{fig:overview}
\end{figure}

In this paper, our high-level design goal is to interpret DL-based networking systems with human-readable control policies so that network operators can easily debug, deploy, and ad-hoc adjust DL-based networking systems.
We develop \name\footnote{Metis is a Greek deity that offers wisdom and consultation.}, a general framework that contains two techniques to provide interpretability.
%
To support a wide range of networking systems, \name\ leverages an abstraction that separates current networking systems into \textit{local systems} and \textit{global systems}~(Figure~\ref{fig:intro}).
In this separation, local systems collect information locally and make decisions for one instance only, such as congestion control agents on end-devices and flow schedulers on switches.
By contrast, global systems aggregate information across the network and make global planning for multiple instances, such as the controller in a software-defined network (SDN).
Table~\ref{tab:control} presents typical examples that fall into these two categories.
For each category, \name\ uses different techniques to achieve interpretability, as depicted in Figure~\ref{fig:overview}.



\begin{table}[t]
	\centering
	\small
	\begin{tabular}{ccc}
		\hline
		\textbf{Category} & \textbf{Scenario} & \textbf{Examples} \\
		\hline
		\multirow{3}{*}{Local} & End-based congestion control & Aurora~\cite{icml2019aurora}\\
		& Client-based video streaming & Pensieve~\cite{sigcomm2017pensieve}\\
		& On-switch flow scheduling & AuTO~\cite{sigcomm2018auto}\\
		\hline
		\multirow{3}{*}{Global} & Cluster job scheduling & Decima~\cite{sigcomm2019decima}\\
		& SDN routing optimization & RouteNet~\cite{sosr2019routenet}\\
		& Network function (NF) placement & NFVdeep~\cite{iwqos2019nfvdeep}\\
		\hline
	\end{tabular}
	\caption{Local systems collect information and make decisions locally (e.g., from end-devices or switches only). Global systems aggregate information and make decisions across the network.}
	\label{tab:control}
\end{table}

\revise{
Specifically, we adopt a decision tree conversion method~\cite{nips2018viper, aistats2011dagger} for local systems.
The main observation behind the design choice is that existing heuristic local systems are usually \textit{rule-based} decision-making systems~(\S\ref{sec:xai}) with a rather simple decision logic (e.g., buffer-based bitrate adaption (ABR)~\cite{sigcomm2014buffer}.)
}
The conversion is built atop a teacher-student training process, where the DNN policy acts as the teacher and generates input-output samples to construct the student decision tree~\cite{aistats2011dagger}.
However, to match the performance with DNNs, traditional decision tree algorithms~\cite{friedman1984cart} usually output an exceedingly large number of branches, which are effectively uninterpretable.
We leverage two important observations to prune the branches down to a tractable number for network operators.
First, sensible policies in local systems often unanimously output the same control action for a large part of the observed states. 
For example, any performant ABR policies~\cite{sigcomm2017pensieve} would keep a low bitrate when both of the bandwidth and the playback buffer are low.
By relying on the data generated by the teacher DNN, the decision tree can easily cut down the decision space.
Second, different input-output pairs have different contributions to the performance of a policy. 
We adopt a special resampling method~\cite{nips2018viper} that allows the teacher DNN to guide the decision tree to prioritize the actions leading to the best outcome.
Empirically, our decision tree can generate human-readable interpretations~(\S\ref{sec:eva-interpret}), and the performance degradation is within 2\% of the original DNNs~(\S\ref{sec:case-lightweight}).

For global systems, our observation is that we can formulate many of them with hypergraphs. 
\revise{
The reason behind the observation is that most global networking systems either have graph-structured inputs or construct a mapping between two variables, both of which could be formulated with hypergraphs~(\S\ref{sec:app}). For example, given routing results of a DL-based routing optimizer~\cite{sosr2019routenet}, we can formulate the interaction between routing paths and links as the relationship between hyperedges\footnote{Similar to an edge connecting two vertices in a graph, a hyperedge covers multiple vertices in the hypergraph (\S\ref{sec:app}).} and vertices. 
The placement of network functions (NFs)~\cite{iwqos2019nfvdeep} could also be formulated as a hypergraph, where NFs and physical servers are hyperedges and vertices, and the placement algorithm constructs a mapping between them~(Appendix~\ref{sec:nfv}).
}
With hypergraph formulations, \name\ computes the importance of each part of the hypergraph by constructing an optimization problem (e.g., finding critical routing decisions to the overall performance)~(\S\ref{sec:action}). With the importance of each decision, network operators can interpret the behaviors of DL-based networking systems~(\S\ref{sec:eva-interpret}). 
%

For concrete evaluation, we generate interpretable policies for two types of DL-based networking systems with \name~(\S\ref{sec:eva-interpret}). For example, we interpret the bitrate adaptation policy of Pensieve~\cite{sigcomm2017pensieve} and recommend a new decision variable. We also present four use cases of \name\ in the design, debugging, deployment, and ad-hoc adjustment of DL-based networking systems. (i)~\name\ helps network operators to redesign the DNN structure of Pensieve with a quality of experience (QoE) improvement by 5.1\%\footnote{Even a 1\% improvement in QoE is significant to current Internet video providers (e.g., YouTube) considering the volume of videos~\cite{abrl}. } on average~(\S\ref{sec:case-modified}). (ii)~\name\ debugs the DNN in Pensieve and improves the average QoE by up to 4\% with only decision trees~(\S\ref{sec:case-troubleshoot}). (iii) \name\ enables a lightweight DL-based flow scheduler (AuTO~\cite{sigcomm2018auto}) and a lightweight Pensieve with shorter decision latency by 27$\times$ and lower resource consumption by up to 156$\times$~(\S\ref{sec:case-lightweight}). (iv) \name\ helps network operators to adjust the routing paths of a DL-based routing optimizer (RouteNet~\cite{sosr2019routenet}) when ad-hoc adjustments are needed~(\S\ref{sec:case-routing}).

We make the following contributions in this paper:
\begin{itemize}
    \item \name, a framework to provide interpretation for two general categories of DL-based networking systems, where it interprets local systems with decision trees~(\S\ref{sec:local}) and global systems with hypergraphs~(\S\ref{sec:global}).
    \item Prototype implementations of \name\ over three DL-based networking systems (Pensieve~\cite{sigcomm2017pensieve}, AuTO~\cite{sigcomm2018auto}, and RouteNet~\cite{sosr2019routenet})~(\S\ref{sec:impl}), and their interpretations with capturing well-known heuristics and discovering new knowledge~(\S\ref{sec:eva-interpret}).
    \item Four use cases on how \name\ can help network operators to design~(\S\ref{sec:case-modified}), debug~(\S\ref{sec:case-troubleshoot}), deploy~(\S\ref{sec:case-lightweight}), and ad-hoc adjust~(\S\ref{sec:case-routing}) DL-based networking systems.
\end{itemize}
To the best of our knowledge, \name\ is the first general framework to interpret diverse DL-based networking systems at deployment. \revise{The source code of \name\ is available at \url{https://github.com/transys-project/metis/}.} We believe that \name\ will accelerate the deployment of DL-based networking systems in practice. 

\section{Motivation}

We motivate the design of \name\ by analyzing (i) the drawbacks of current DL-based networking systems~(\S\ref{sec:motiv}), and (ii) why existing interpretation methods are insufficient for DL-based networking systems~(\S\ref{sec:existing}).

\subsection{Drawbacks of Current Systems}
\label{sec:motiv}

\begin{figure}
	\centering
	\includegraphics[width=.95\linewidth]{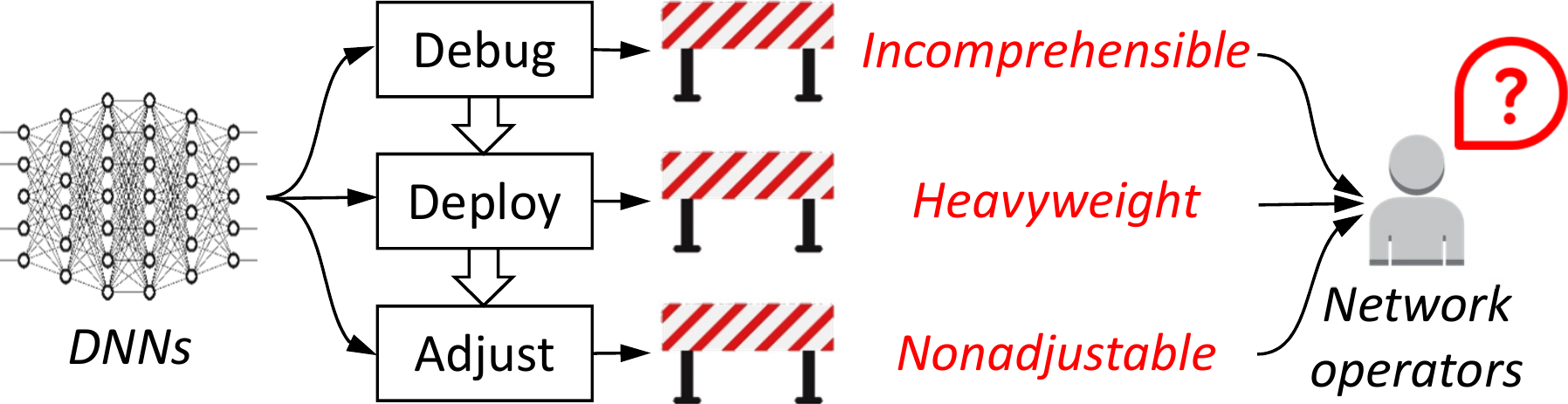}
	\caption{DNNs create barriers for network operators in many stages of the development flow of networking systems.}
	\label{fig:motiv}
\end{figure}

The blackbox property of DNNs lacks interpretability for network operators. Without understanding why DNNs make decisions, network operators might not have enough confidence to adopt them in practice~\cite{apnet2018demystifying}. Moreover, as shown in Figure~\ref{fig:motiv}, the blackbox property brings drawbacks to networking systems in debugging, online deployment, and ad-hoc adjustment due to the following reasons.

\parahead{Incomprehensible structure.} DNNs could contain thousands to billions of neurons~\cite{gpt3billion}, making them incomprehensible for human network operators. Due to the complex structure of DNN, when DL-based networking systems fail to perform as expected, network operators will have difficulty in locating the erroneous component. Even after finding the sub-optimality in the design of DNN structures, network operators are challenged to redesign them for better performance. If network operators could trace the mapping function between inputs and outputs, it would be easier to debug and improve DL-based networking systems. 

\parahead{Heavyweight to deploy.} DNNs are known to be bulky on both resource consumption and decision latency~\cite{mobisys2017deepmon}. Even with advanced hardware (e.g., GPU), DNNs may take tens of milliseconds for decision-making (\S\ref{sec:case-lightweight}). In contrast, networking systems, especially local systems on end devices~(e.g., mobile phones) or in-network devices (e.g., switches), are resource-limited and latency-sensitive~\cite{mobisys2017deepmon}. For example, loading a DNN-based ABR algorithm on mobile clients increases the page load time by around 10 seconds~(\S\ref{sec:case-lightweight}), which will make users leave the page. Existing systems usually provide ``best-effort'' services only and roll back to heuristics when resource and latency constraints can not be met~\cite{sigcomm2018auto}, which degrades the performance of DNNs.

\parahead{Nonadjustable policies.} Practical deployment of networking systems also requires ad-hoc adjustments or adding temporary features. For example, we could adjust the weights for different jobs in fair scheduling to catch up with the fluctuations in workloads~\cite{sigcomm2019decima}. However, the lack of interpretation brings difficulties to network operators when they need to adjust the networking systems. Without understanding why DNNs make such decisions, arbitrary adjustments may lead to severe performance degradation. For example, when network operators want to manually reroute a flow away from a link, without interpretations of decisions, network operators might not know how and where to accommodate that flow.

\begin{figure}
	\centering
	\includegraphics[width=.95\linewidth]{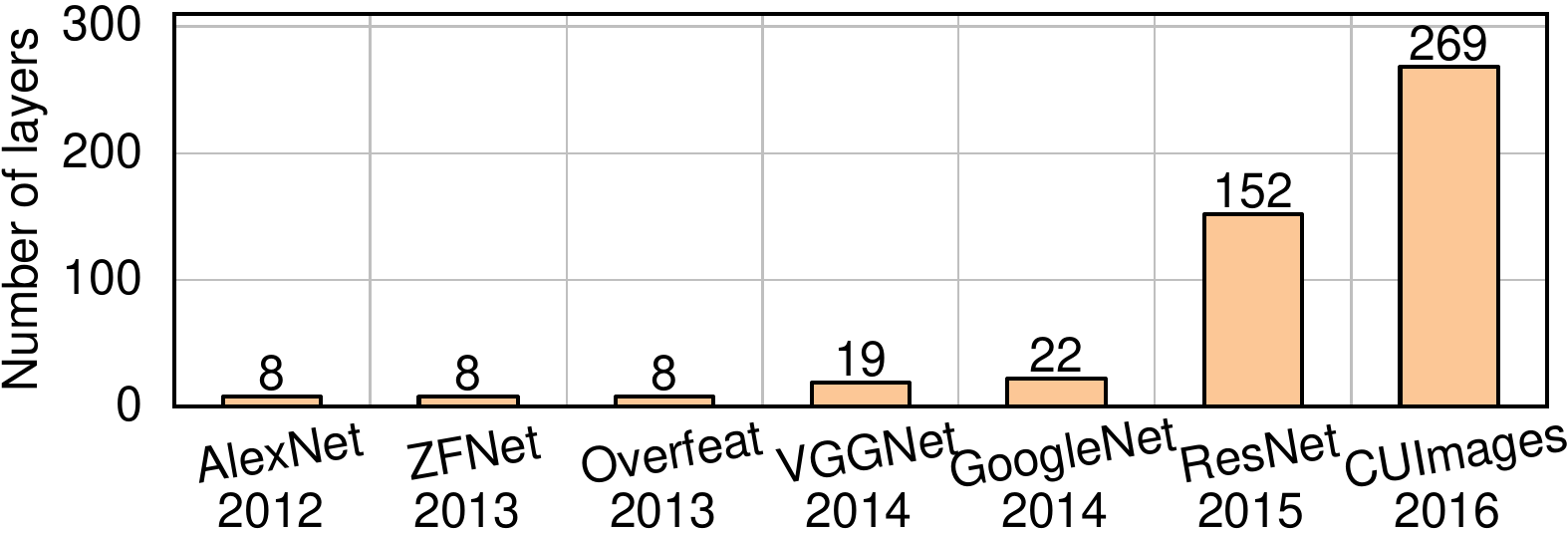}
	\caption{The exponential growth of DNN complexity in ImageNet Challenge winners~\cite{cvpr2009imagenet} (Figure adopted from~\cite{cnndeep}).}
	\label{fig:depth}
\end{figure}

\parahead{Discussions.} The application of DNNs in networking systems is still at a preliminary stage: DNNs in Pensieve~\cite{sigcomm2017pensieve}, AuTO~\cite{sigcomm2018auto}, and RouteNet~\cite{sosr2019routenet} (published in 2017, 2018, and 2019) have less than ten layers. As a comparison, a sharp increase in the number of DNN layers has been observed in other communities (Figure~\ref{fig:depth}). Recent language translation models even contain billions of parameters~\cite{gpt3billion}. Although we are not saying that the larger is the better, it is indisputable that larger DNNs will aggravate the problems and create barriers to deploy DL-based networking systems in practice. 

\subsection{Why Not Existing Interpretations?}
\label{sec:existing}

For DL-based networking systems, existing interpretation methods~\cite{survey2018xai, cacm2020iml} are insufficient in the following aspects:

\parahead{Different interpretation goal.} The question of \textit{why a DNN makes a certain decision} may have answers from two angles. In the machine learning community, the answer could be understanding the \textit{inner mechanism of ML models} (e.g., which neurons are activated for some particular input features)~\cite{cvpr2017netdissect, eccv2014visualizing}. It's like trying to understand how the brain works with surgery. In contrast, the expected answer from network operators is the \textit{relationship between inputs and outputs} (e.g., which input features affect the decision)~\cite{apnet2018demystifying}. What network operators need is a method to interpret the mapping between the input and output for DNNs.

\parahead{Diverse networking systems.} As shown in Table~\ref{tab:control}, DL-based networking systems have different application scenarios and are based on various DL approaches, such as feedforward neural network (FNN)~\cite{sigcomm2017pensieve}, recurrent neural network (RNN)~\cite{access2017rnnids}, and graph neural network (GNN)~\cite{sigcomm2019decima}. Therefore, interpreting diverse DL-based networking systems with one single interpretation method is insufficient. For example, LEMNA~\cite{ccs2018lemna} could only interpret the behaviors of RNN and thus is not suitable for GNN-based networking systems~\cite{sigcomm2019decima}. In \name, we observe that DL-based networking systems can be divided into two categories (local and global) and develop corresponding techniques for each category. 

\parahead{Non-standard state and action spaces.} Existing interpretation methods are usually designed with easy-encoded state and action spaces. For example, methods interpreting image classification tasks are designed for the grid-based RGB encoding~\cite{eccv2014visualizing, cvpr2017netdissect}. The interpretation methods for language translation tasks are also based on vectorized word embeddings~\cite{neurips2019interpretnlp, acl2018semantically}. However, networking systems inherently work with non-standard state and action spaces. For example, RouteNet~\cite{sosr2019routenet} takes the topology as input and generates variable-length routing paths. Specially designed interpretation methods for networking systems are hence needed. 




\paraskip In response, to interpret DL-based networking systems, \name\ introduces a decision tree-based method together with a hypergraph-based method for different systems. Our observation is that although DL-based networking systems are diverse, when divided into two categories (local and global), the commonality inside each category enables us to design specific interpretation methods.

\section{Decision Tree Interpretations}
\label{sec:local}

In this section, we first describe the design choice for choosing decision trees for local systems in \name~(\S\ref{sec:xai}), and then explain the detailed methodology to convert the DNNs to decision trees~(\S\ref{sec:dt}).

\subsection{Design Choice: Decision Tree}
\label{sec:xai}

As introduced in \S\ref{sec:intro}, \name\ converts DNNs into simpler models based on interpretation methods. There are many candidate models, such as (super)linear regression~\cite{ccs2018lemna, kdd2016lime}, decision trees~\cite{nips2018viper, aistats2011dagger}, etc. We refer the readers to~\cite{survey2018xai, cacm2020iml} for a comprehensive review.

In this paper, we decide to convert DNNs to \textit{decision trees} due to three reasons. First, the logic structure of decision trees resembles the policies made by networking systems, which are rule-based policies. For example, flow scheduling algorithms on switches usually depend on a set of forwarding rules, such as shortest-job-first~\cite{nsdi2015pias}. ABR algorithms depend on precomputed rules over buffer occupancy and predicted throughput~\cite{infocom2016bola, sigcomm2015robustmpc}. Second, decision trees have rich expressiveness and high faithfulness because they are non-parametric and can represent very complex policies~\cite{ai1998dt}. We demonstrate the performance of decision trees during conversion compared to other methods~\cite{ccs2018lemna, kdd2016lime} in Appendix~\ref{sec:app-faith}. Third, decision trees are lightweight for networking systems, which will bring further benefits to resource consumption and decision latency~(\S\ref{sec:case-lightweight}). There are also research efforts that interpret DNNs with programming language~\cite{icml2018pirl, pldi2019verifiablerl}. However, designing different primitives for each networking system is time-consuming and inefficient. 

With interpretations of local systems in the form of decision trees, we can interpret the results since the decision-making process is transparent~(\S\ref{sec:eva-interpret}). Also, we can debug the DNN models when they generate sub-optimal decisions~(\S\ref{sec:case-troubleshoot}). Furthermore, since decision trees are much smaller in size, less expensive on computation, we could also deploy the decision trees online instead of deploying heavyweight DNN models. This will result in low decision-making latency and resource consumption~(\S\ref{sec:case-lightweight}). 

\subsection{Conversion Methodology}
\label{sec:dt}


To extract the decision tree from a finetuned DNN, we adopt a teacher-student training methodology proposed in~\cite{nips2018viper}. 
%
We reproduce key conversion steps for networking systems as follows:

\parahead{Step 1: Traces collection.} When training decision trees, it is important to obtain an appropriate dataset from DNNs. Simply covering all possible (state, action) pairs is too costly and does not faithfully reflect the state distribution from the target policy. Thus, \name\ follows the trajectories generated by the teacher DNNs. Moreover, networking systems are sequential decision processes, where each action has long-lasting effects on future states. Therefore, the decision tree can deviate significantly from the trajectories of DNNs due to imperfect conversion~\cite{nips2018viper}. To make the converted policy more robust, we let the DNN policy take over the control on the deviated trajectory and re-collect (state, action) pair to refine the conversion training. We iterate the process until the deviation is confined (i.e., the converted policy closely tracks the DNN trajectory).

\begin{table*}
	\small
	\centering
	\begin{tabular}{cccccc}
		\hline
		& Scenario & Vertex & Hyperedge & Meaning of $I_{ev}=1$ & Details\\
		\hline
		\#1 & SDN routing optimization & Physical link & Path (src-dst pairs) & Path $e$ contains link $v$. & \S~\ref{sec:app}\\
		\#2 & Network function placement & Physical server & Network function & One instance of NF $e$ is on server $v$. & Appendix~\ref{sec:nfv}\\
		\#3 & Ultra-dense cellular network & Mobile user & Base station coverage & Base station $e$ covers user $v$. & Appendix~\ref{sec:coverage}\\
		\#4 & Cluster job scheduling & Job node & Dependency & Dependency $e$ is related to node $v$. & Appendix~\ref{sec:scheduling}\\
		\hline
	\end{tabular}
	\caption{Several hypergraph-based models in different scenarios.}
	\label{tab:hypergraph}
	\vspace{-1\baselineskip}
\end{table*}

\parahead{Step 2: Resampling.} Local systems usually optimize \textit{policies} instead of independent actions~\cite{sigcomm2018auto, icml2019aurora, sigcomm2017pensieve}. In this case, different actions of networking systems may have different importance to the optimization goal. For example, an ABR algorithm downloading a huge chunk at extremely low buffer will lead to a long stall, resulting in severe performance degradation. Meanwhile, downloading a little larger chunk when network condition and buffer are moderate will not have drastic effects. However, decision tree algorithms are designed to optimize the accuracy of a single action and treat all actions the same. Therefore, their optimization goals do not match. Existing DL-based local systems adopt reinforcement learning (RL) to optimize the policy instead of single actions, where the \textit{advantage} of each (state, action) represents the importance to the optimization goal. Therefore, we follow recent advances in converting DNNs in RL policies into decision trees~\cite{nips2018viper} and resample $\mathcal{D}$ according to the advantage function. For each pair $(s, a)$, the sampling probability $p(s,a)$ could be expressed as:
\begin{equation}
\label{eq:prob}
p(s,a) \propto \left(V^{\left(\pi^*\right)}(s) - \min_{a'\in A} Q^{\left(\pi^*\right)}(s,a')\right)\cdot \mathds{1}\left((s,a)\in\mathcal{D}\right)
\end{equation}
where $V(s)$ and $Q(s,a)$ are the value function and $Q$-function of RL~\cite{sutton2018reinforcement}. Value function represents the expected total reward starting at state $s$ and following the policy $\pi$. $Q$-function further specifies the next step action $a$. $\pi^*$ is the DNN policy, and $A$ is the action space. $\mathds{1}(x)$ is the indicator function, which equals to 1 if and only if $x$ is true. We analyze Equation~\ref{eq:prob} with more details in Appendix~\ref{sec:rl}. We then retrain the decision tree on the resampled dataset. Our empirical results demonstrate that the resampling step can improve the QoE over 73\% of the traces (Appendix~\ref{sec:rl}).

\parahead{Step 3: Pruning.} As the size of the decision tree sometimes becomes much larger than network operators can understand, we adopt cost complexity pruning (CCP)~\cite{friedman1984cart} to reduce the number of branches according to the requirements from network operators. Compared with other pruning methods, CCP empirically achieves a smaller decision tree with a similar error rate~\cite{mingers1989empirical}. At its core, CCP creates a cost function of the complexity of the pruned decision tree to balance between accuracy and complexity. Moreover, for the continuous outputs in networking systems (e.g., queue thresholds~\cite{sigcomm2018auto}), we employ the design of the \textit{regression tree} to generate real value outputs~\cite{venables2002tree}. In our experiments, for Pensieve, the size of leaf nodes may be up to 1000 without pruning (Appendix~\ref{sec:app-sensitivity}). With CCP, pruning the decision tree down to 200 leaf nodes only results in a performance degradation of less than 0.6\% (\S\ref{sec:case-lightweight}).

\parahead{Step 4: Deployment.} Finally, network operators could deploy the converted model online and enjoy both the performance improvement brought by deep learning and the interpretability provided by the converted model. Our evaluation shows that the performance degradation of decision trees is less than 2\% for two DL-based networking systems~(\S\ref{sec:case-lightweight}). We also present further benefits of converting DNNs of networking systems into decision trees (easy debugging and lightweight deployment) in \S\ref{sec:case-troubleshoot} and \S\ref{sec:case-lightweight}.


\section{Hypergraph Interpretations}
\label{sec:global}

We first briefly introduce hypergraph and present several applications on how to formulate networking systems with hypergraphs~(\S\ref{sec:app}), and then introduce our interpretation methods to find critical components in hypergraphs~(\S\ref{sec:action}).

\subsection{Hypergraph Formulation}
\label{sec:app}

\begin{figure}
	\centering
	\subfigure{\includegraphics[height=2cm]{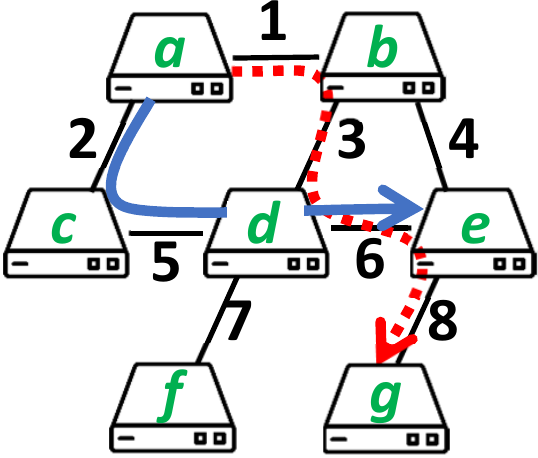}}\hfill
	\subfigure{\includegraphics[height=2cm]{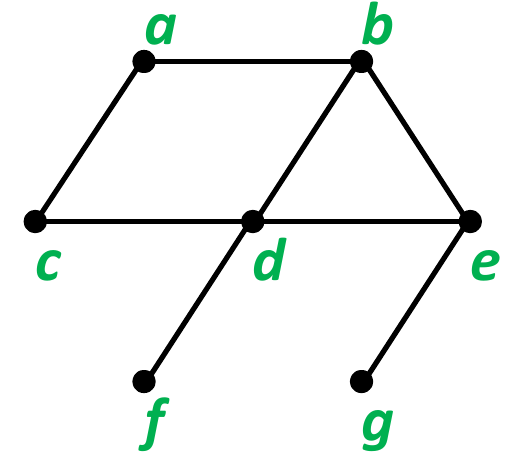}}\hfill
	\subfigure{\includegraphics[height=2cm]{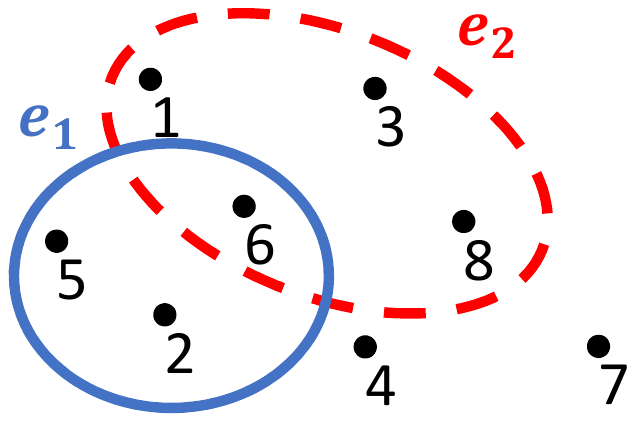}}
	\\
	\addtocounter{subfigure}{-3}
	\subfiguretopcaptrue
	\vspace{-.5\baselineskip}
	\subfigure[Physical topology.]{
	    \hbox to 2.34cm{\hfil\null}
		\label{fig:routing-original}
	}\hfill
	\subfigure[Graph representation of Figure~\ref{fig:routing-original}.]{
	    \hbox to 2.24cm{\hfil\null}
		\label{fig:routing-graph}
	}\hfill
	\subfigure[Paths and links form a hypergraph.]{
	    \hbox to 2.77cm{\hfil\null}
		\label{fig:routing-hypergraph}
	}
	\subfiguretopcapfalse
	\vspace{-1\baselineskip}
	\caption{The hypergraph representation of the SDN routing model. Hypergraph could efficiently represent path information.}
	\label{fig:routing}
\end{figure}

A hypergraph is composed of \textit{vertices} and \textit{hyperedges}. The main difference between the edge in a graph and the hyperedge in a hypergraph is that a hyperedge can cover \textit{multiple} vertices, as shown in Figures~\ref{fig:routing-graph} and~\ref{fig:routing-hypergraph}. We denote the set of all vertices and all hyperedges as $\mathcal{V}$ and $\mathcal{E}$. Each vertex $v$ and hyperedge $e$ may also attach their features, denoted as $f_v$ and $f_e$. We denote the matrix of features of all vertices and hyperedges as $F_V$ and $F_E$, respectively. 

With hypergraph, we can formulate many global systems uniformly, as shown in Table~\ref{tab:hypergraph}. In the following, we will introduce the formulation of SDN routing optimization (scenario \#1) in detail and leave other formulations in Appendix~\ref{sec:models}.

\parahead{Case study: SDN routing optimization.} We first present a case study of formulating SDN routing with hypergraph. The SDN controller collects the information from all data plane switches. In this case, an SDN routing optimizer analyzes the traffic demands for each src-dst pair and generates the \textit{routing paths} for all src-dst traffic demands based on the topology structure and link capacity. However, composed of variant-length switches and links, routing paths are high-order information and are difficult to be efficiently expressed. Previous research efforts try to represent the paths with integer programming~\cite{infocom2005optimal}, which is hard to be efficiently optimized within a limited time. RouteNet~\cite{sosr2019routenet} designs a DNN-based optimization algorithm to continuously select the best routing paths for each src-dst traffic demand pair. 

To formulate the system with hypergraph, we consider the paths as hyperedges and physical links as vertices. A hyperedge covering a vertex indicates the path of that pair of demand contains a link. An illustration of hypergraph mapping results is shown in Figure~\ref{fig:routing}. Links (1, 2,$\cdots$, 8) are modeled as vertices. Two pairs of transmission demand ($a$$\Rightarrow$$e$ and $a$$\Rightarrow$$g$) are modeled as hyperedges (denoted as $e_1$ and $e_2$). Vertex features $F_V$ are the link capacity. Hyperedge features $F_E$ are the traffic demand volume between each pair of switches. If a hyperedge $e$ covers vertex $v$, the respective flow of $e$ should go through the respective link of $v$. 

RouteNet generates the overall routing results, i.e., the path of all traffic demands. For example, assume that RouteNet decides the demand from $a$ to $e$ going through link 2, 5, 6 (path in blue), and the demand from $a$ to $g$ going through link 1, 3, 6, 8, the respective hypergraph should be Figure~\ref{fig:routing-hypergraph}. Hyperedge $e_1$ covers vertices 2, 5, 6, and hyperedge $e_2$ covers 1, 3, 6, 8. All vertex-hyperedge connections $\{(v, e)\}$ are:
\begin{equation}
\label{eq:connection}
\{(2, e_1), (5, e_1), (6, e_1), (1, e_2), (3, e_2), (6, e_2), (8, e_2)\}
\end{equation}

Later in \S\ref{sec:action}, we are going to find out which connections are critical to the overall routing decisions of the topology.

\parahead{Capability of hypergraph representation.} We empirically summarize two key features that enable global systems to be formulated with hypergraph:
\begin{itemize}
	\item \textit{Graph-structured inputs or outputs.} Since a graph is a simple form of a hypergraph, if the inputs or outputs of a global system are graph-structured (e.g., network topology~\cite{sosr2019routenet}, dataflow computation graph~\cite{sigcomm2019decima}), this system can be naturally formulated with hypergraph. 
	\item \textit{Bivariate mapping.} If a global system constructs a mapping between two variables, those two variables could be formulated with vertices and hyperedges. The mapping could be formulated the connection relationship in the hypergraph. Many resource allocation systems construct the mapping between resources (e.g., physical servers) and requests (e.g., network functions)~\cite{iwqos2019nfvdeep}.
\end{itemize}
As long as a global system has one of the features above, we can formulate it with hypergraphs and interpret it with \name. We find that many global systems have at least one feature. For example, in Table~\ref{tab:hypergraph}, scenario \#1 processes network topology and scenario \#4 processes dataflow graph, both of which are graph-structured. Scenario \#2 maps the NF instances to servers and scenario \#3 maps each mobile user to a base station, both following bivariate mappings.

\subsection{Critical Connections Search}
\label{sec:action}

Next, we are going to find out which vertex-hyperedge connections are critical to the optimization result of the original system. We first introduce the \textit{incidence matrix} representation of a hypergraph. Incidence matrix $I$ (with the size of ${|\mathcal{E}| \times |\mathcal{V}|}$) is a 0-1 matrix to represent the connection relationship between vertices and hyperedges. $I_{ev}=1$ indicates hyperedge $e$ contains vertex $v$. For example, the incidence matrix of the hypergraph in Figure~\ref{fig:routing-hypergraph} is:
\begin{equation}
\label{eq:incidence}
I = \left(
\begin{array}{cccccccc}
0 & 1 & 0 & 0 & 1 & 1 & 0 & 0 \\
1 & 0 & 1 & 0 & 0 & 1 & 0 & 1 \\
\end{array}
\right)
\end{equation}

Our design goal is to evaluate how each connection is critical to the optimization results of the original system. Taking the case of SDN routing as an example, \name\ is going to evaluate how each (link, path) connection in Equation~\ref{eq:connection} is critical to the overall routing result. We allow a fractional incidence matrix $W\in [0,1]^{|\mathcal{E}|\times|\mathcal{V}|}$ to represent the significance of each hyperedge-vertex connection. $W_{ev}=0$ if there is no connection between $v$ and $e$. We first present the overview of the critical connection searching algorithm in Figure~\ref{fig:opt}. The optimization objective in Equation~\ref{eq:objective} consists of the following three parts:

\begin{figure}
	\small
    \begin{equation}
    \label{eq:objective}
    \min \ell(W) \quad\text{s.t. }~~0 \leqslant W_{ev} \leqslant I_{ev},~\forall v\in\mathcal{V},e\in\mathcal{E}
    \end{equation}
    where\hfill\null
    \begin{align}
    \label{eq:loss} \ell(W) &= D(Y_W, Y_I) + \lambda_1  ||W|| + \lambda_2 H(W)\\
    \label{eq:faith} D(Y_W, Y_I) &= \left\{
        \begin{array}{ll}
            \sum Y_W \log\frac{Y_W}{Y_I} & (\text{discrete}) \\
            \sum ||Y_W - Y_I ||^2 & (\text{continuous})
        \end{array}
    \right.\\
    \label{eq:concise} ||W|| &= \sum_{v,e} |W_{ev}|\\
    \label{eq:confident}H(W) &= -\sum_{v,e} \left(W_{ev} \log W_{ev} + (1-W_{ev}) \log(1 - W_{ev})\right)
    \end{align}
    \vspace{-1\baselineskip}
    \caption{Formulation of critical connection search optimization.}
    \label{fig:opt}
\end{figure}


\parahead{Performance degradation ($\mathbf{D(Y_W, Y_I)}$).} The critical connections should be those connections that have a great influence on the output of the networking system, which is task-independent. Therefore, we need to measure the output of the original DL-based networking system when input features of hyperedges and vertices are weighted by the mask $W$. Taking the SDN routing case in \S\ref{sec:app} as an example, routing decisions generated by the masked features (demands, capacities) should be similar to the original ones. We denote the decisions generated by the original inputs and inputs with mask $W$ as $Y_I$ and $Y_W$. Thus, we maximize the similarity between the $Y_W$ and $Y_I$, denoted as $D(Y_W, Y_I)$ in Equation~\ref{eq:loss}. We adopt KL-divergence~\cite{kldivergence1951} to measure discrete outputs~(e.g., sequences of routing decisions) and mean square error for continuous outputs, both of which are common similarity metrics in the DL community~\cite{iclr2014vae}, as shown in Equation~\ref{eq:faith}.

\parahead{Interpretation conciseness ($\mathbf{||W||}$).} Usually, the number of interpretations that humans can understand is budgeted~\cite{kdd2016lime}. Therefore, the number of critical connections should also be concise enough to be understandable for network operators. If the algorithm provides too many ``critical'' connections, network operators will be confused and cannot easily interpret the networking systems. In \name, we measure the conciseness of $W$ as the sum of all elements (the scale of the matrix). We also need to penalize the scale of mask $W$ in the optimization goal, as shown in Equation~\ref{eq:concise}.

\parahead{Determinism ($\textbf{H(W)}$).} Moreover, we also expect the results of $W$ to be \textit{deterministic}, i.e., for each connection $(v, e)$, it is either seriously suppressed ($W_{ev}$ close to 0) or almost unaffected ($W_{ev}$ close to 1). Otherwise, the crafty agent will learn to mask all connections with the same weight and generate meaningless interpretations. In this paper, \name\ optimizes the \textit{entropy} of mask $W$ to encourage the connections in $W$ to be close to 1 or 0, where the entropy is a measure of uncertainty in the information theory~\cite{shannon1948entropy}, as shown in Equation~\ref{eq:confident}.

\paraskip To balance the optimization goals above, we provide two customizable hyperparameters ($\lambda_1$ and $\lambda_2$) for network opreators due to the differences in operators' understandability and application scenarios of systems. For example, an online monitor of routing results may only need the most critical information for fast decisions, while an offline analyzer of routing results requires more detailed interpretations for further improvement. In this case, network operators can increase (or decrease) $\lambda_2$ to reduce (or increase) the number of undetermined connections with median mask values. \name\ will then expose less (or more) critical connections to network operators. We empirically study the effects of setting $\lambda_1$ and $\lambda_2$ for network operators in Appendix~\ref{sec:app-gcs-sensitivity}. 

In this way, we can quantitatively know how critical the connection contributes to the output. In the SDN routing case, instead of trivially identifying links where many flows run through, \name\ can provide finer-grained interpretations by further identifying \textit{which flow on which link} plays a dominant role in the overall result. We present the interpretations and further improvements in this case in \S\ref{sec:eva-interpret} and \S\ref{sec:case-routing}. 

\section{Implementation}
\label{sec:impl}

We interpret two local systems, Pensieve~\cite{sigcomm2017pensieve} and AuTO~\cite{sigcomm2018auto}, and one global system, RouteNet~\cite{sosr2019routenet}, with \name. Testbed settings are introduced in Appendix~\ref{sec:app-param}. 
\revise{
For other types of DL-based networking systems, please also refer to  Appendix~\ref{sec:app-param} for network operators and our project page\footnote{\url{https://github.com/transys-project/metis/}} for a detailed implementation guideline.
}

\parahead{Pensieve implementation.} In current Internet video transmissions, each video consists of many \textit{chunks} (a few seconds of playtime), and each chunk is encoded at multiple bitrates~\cite{sigcomm2017pensieve}. Pensieve is a deep RL-based ABR system to optimize bitrates with network observations such as past chunk throughput, buffer occupancy.

We use the same video in Pensieve unless other specified. The chunk size, bitrates of the video are respectively set to 4 seconds and \{300, 750, 1200, 1850, 2850, 4300\} kbps. Real-world network traces include 250 HSDPA traces~\cite{mmsys2013norway} and 205 FCC traces~\cite{fcc}. We integrate DNNs into JavaScript with \texttt{tf.js}~\cite{sysml2019tfjs} to run Pensieve in the browser. We set up the same environment and QoE metric with Pensieve. 

We then implement \name+Pensieve. We use the finetuned model provided by~\cite{sigcomm2017pensieve} to generate the decision tree. We use five baseline ABRs (BB~\cite{sigcomm2014buffer}, RB~\cite{sigcomm2017pensieve}, Festive~\cite{conext2012festive}, BOLA~\cite{infocom2016bola}, rMPC~\cite{sigcomm2015robustmpc}) as Pensieve and migrate them into \texttt{dash.js}~\cite{dashjs}. 

\parahead{AuTO implementation.} AuTO is a flow scheduling system to optimize flow completion time (FCT) based on deep RL. Limited by the long decision latency of DNN, AuTO can only optimize long flows individually with a long-flow RL agent (lRLA). For short flows, AuTO makes decisions locally with multi-level feedback queues~\cite{nsdi2015pias} and optimizes the queue thresholds with a short-flow RL agent (sRLA). lRLA takes \{5-tuples, priorities\} of running long flows, and \{5-tuples, FCTs, flow sizes\} of finished long flows as states and decides the \{priority, rate limit, routing path\} for each running long flow. sRLA observes \{5-tuples, FCTs, flow sizes\} of finished short flows and outputs the queue thresholds. 

We use the same 16-server one-switch topology and traces evaluated in AuTO: web search (WS) traces~\cite{sigcomm2009vl2} and data mining (DM) traces~\cite{sigcomm2010dctcp}. We train the DNNs following the instructions in~\cite{sigcomm2018auto}. All other configurations (e.g., link capacity, link load, DNN structure) are set the same as AuTO. We then evaluate the decision tree generated by \name~(\name+AuTO). 


\parahead{RouteNet* implementation.} We train the model on the traffic dataset of the NSFNet topology provided by RouteNet, as presented in Figure~\ref{fig:topo}. We adopt the close-loop routing system in RouteNet, denoted as RouteNet*, which concatenates latency predictions with routing decisions.

As for \name+RouteNet*, to implement the constraint of $W$ in Equation~\ref{eq:objective}, we adopt the \textit{gating} mechanism used in the machine learning community~\cite{nips2014gru}. Specifically, the incidence matrix value $I_{ve}$ acts as a gate to bound the mask value $W_{ve}$. We construct a matrix $W' \in \mathbb{R}^{|E|\times|V|}$ and get mask matrix $W$ by the following equation:
\begin{equation}
W = I \circ \text{sigmoid}(W') 
\end{equation}
$\circ$ means element-wise multiplication, and $\text{sigmoid}$ function is applied to each element separately. Since the output of sigmoid function is limited in $(0,1)$, $W_{ve}$ will always be less than or equal to $I_{ve}$. In this case, the constraint in Equation~\ref{eq:objective} will be followed during the optimization. 

\section{Experiments}
\label{sec:eva}

In this section, we first empirically evaluate the interpretability of  \name\ with two types of DL-based networking systems. Subsequently, we showcase how \name\ addresses the drawbacks of existing DL-based networking systems (\S\ref{sec:motiv}). We finally benchmark the interpretability of \name. Overall, our experiments cover the following aspects:
\begin{itemize}
	\item \textbf{System interpretations.} We demonstrate the effectiveness of \name\ by presenting the interpretations of one local system (Pensieve) and one global system (RouteNet*) with newly discovered knowledge~(\S\ref{sec:eva-interpret}).
	\item \textbf{Guide for model design.} We present a case on how to improve the DNN structure of Pensieve for better performance based on the interpretations of \name~(\S\ref{sec:case-modified}).
	\item \textbf{Enabling debuggability.} With a use case of Pensieve, \name\ debugs a problem and improves its performance by adjusting the structure of decision trees~(\S\ref{sec:case-troubleshoot}).
	\item \textbf{Lightweight deployment.} For local systems (AuTO and Pensieve), network operators could directly deploy the converted decision trees provided by \name\ online and achieve benefits enabled by lightweight deployments~(\S\ref{sec:case-lightweight}).
	\item \textbf{Ad-hoc adjustments.} We provide a case study on how network operators can adjust the decision results of RouteNet* based on the interpretations provided by \name~(\S\ref{sec:case-routing}).
	\item \textbf{\name\ deep dive.} We finally evaluate the interpretation performance, parameter sensitivity, and computation overhead of \name\ under different settings~(\S\ref{sec:eva-microbenchmark}).
\end{itemize}

\subsection{System Interpretations}
\label{sec:eva-interpret}

With \name, we interpret the DNN policy learned by a local system, Pensieve, and a global system, RouteNet*.

\begin{figure}
	\centering
	\includegraphics[width=\linewidth]{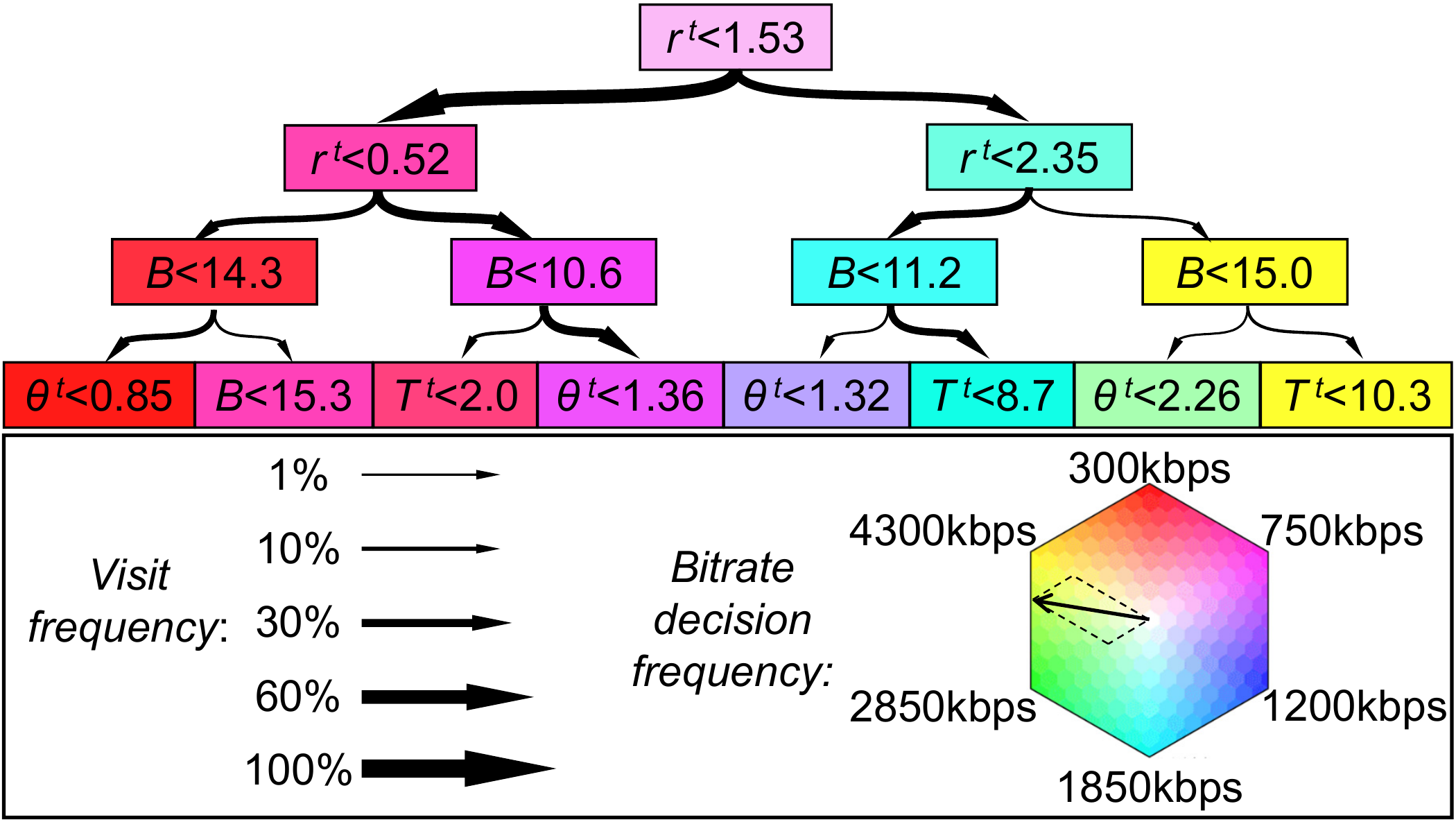}
	\caption{Top 4 layers of the decision tree of \name+Pensieve. The color represents the frequency of bitrate selections at that node. For example, the arrow in the palette represents that 67\% states traversing a node with that color are finally decided as 4300kbps, and 33\% states are 2850kbps. Better viewed with color.} 
	\label{fig:pensieve-dt}
\end{figure}

\parahead{Local system interpretations.} We present the top 4 layers of the decision tree of \name+Pensieve in Figure~\ref{fig:pensieve-dt}. The decision variables of each node include the last chunk bitrate ($r^t$), previous throughput ($\theta^t$), buffer occupancy~($B$), and last chunk download time ($T_t$). Since we only present the top 4 layers of the decision tree, we represent the frequency of final decisions of each node with the color on the palette in Figure~\ref{fig:pensieve-dt}. 

From the interpretations in Figure~\ref{fig:pensieve-dt}, we can know the reasons behind the superior performance of Pensieve in two directions. (i) \textit{Discovering new knowledge.} On the top two layers, \name+Pensieve first classifies inputs into four branches based on the \textit{last chunk bitrate}, which is different from existing methods. The information contained in the last bitrate choice affects the output QoE significantly. Based on this observation, we recommend that network operators could improve ABR algorithms with particular focus on the last chunk bitrate. We present a use case on how to utilize this observation to improve the DNN structure in \S\ref{sec:case-modified}. (ii)~\textit{Capturing existing heuristics.} Similar to existing methods, \name+Pensieve makes decisions based on buffer occupancy~\cite{sigcomm2014buffer, infocom2016bola} and predicted throughput~\cite{sigcomm2015robustmpc, dashjs}. With the interpretations provided by \name, network operators can understand how Pensieve makes decisions. 

\begin{table}
    \centering
    \begin{tabular}{ccccc}
        \hline
        & Routing path & Link & Mask $M_{ve}$ & Interpretation type\\
        \hline
        \#1 & 6$\to$7$\to$10$\to$9 & 6$\to$7 & 0.886 & Shorter\\
        \#2 & 1$\to$7$\to$10$\to$9 & 1$\to$7 & 0.880 & Shorter\\
        \#3 & 7$\to$10$\to$9$\to$12 & 10$\to$9 & 0.878 & Less congested\\
        \#4 & 8$\to$3$\to$0$\to$2 & 8$\to$3 & 0.875 & Shorter\\
        \#5 & 6$\to$4$\to$3$\to$0 & 6$\to$4 & 0.874 & Less congested\\
        \hline
    \end{tabular}
    \vspace{-.5\baselineskip}
    \caption{Top 5 mask value interpretations in Figure~\ref{fig:topo}.}
    \label{tab:mask}
\end{table}

\begin{figure}
	\centering
	\subfigure[Connection \#1.]{
	    \includegraphics[height=2.3cm]{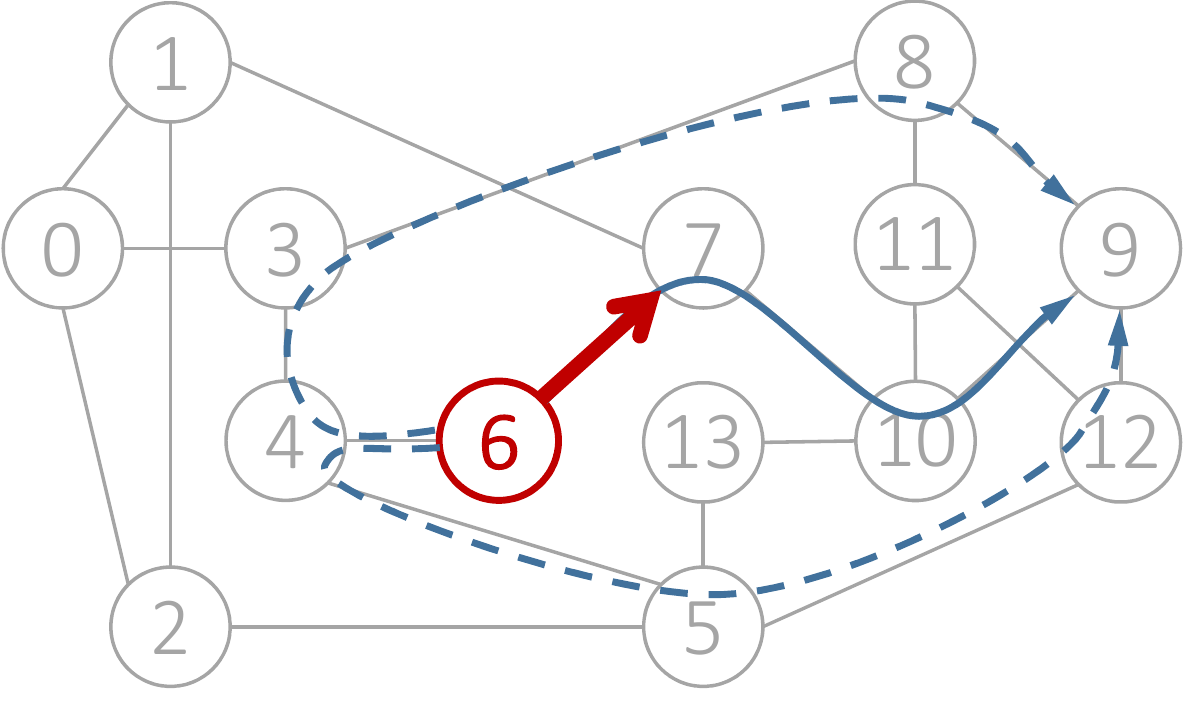}
	    \label{fig:topo-short}
	}
	\subfigure[Connection \#3. Traffic load of two links are in the bottom-right corner.]{
	    \includegraphics[height=2.3cm]{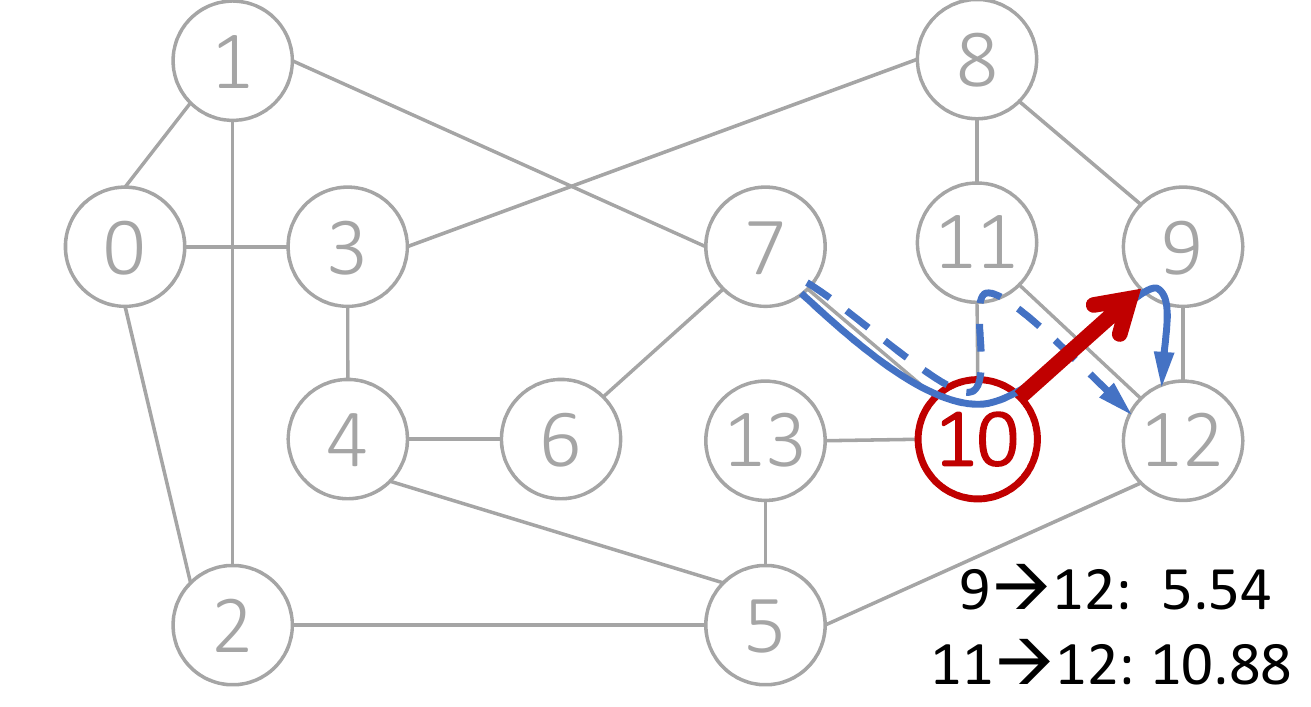}
	    \label{fig:topo-congested}
	}
	\caption{Solid blue paths are the results generated by RouteNet. Dashed paths are candidates serving the same src-dst demand. Critical decisions interpreted by \name\ are colored red.} 
	\label{fig:topo}
\end{figure}

\begin{figure}
	\centering
	\subfigure{
		\includegraphics[height=3cm]{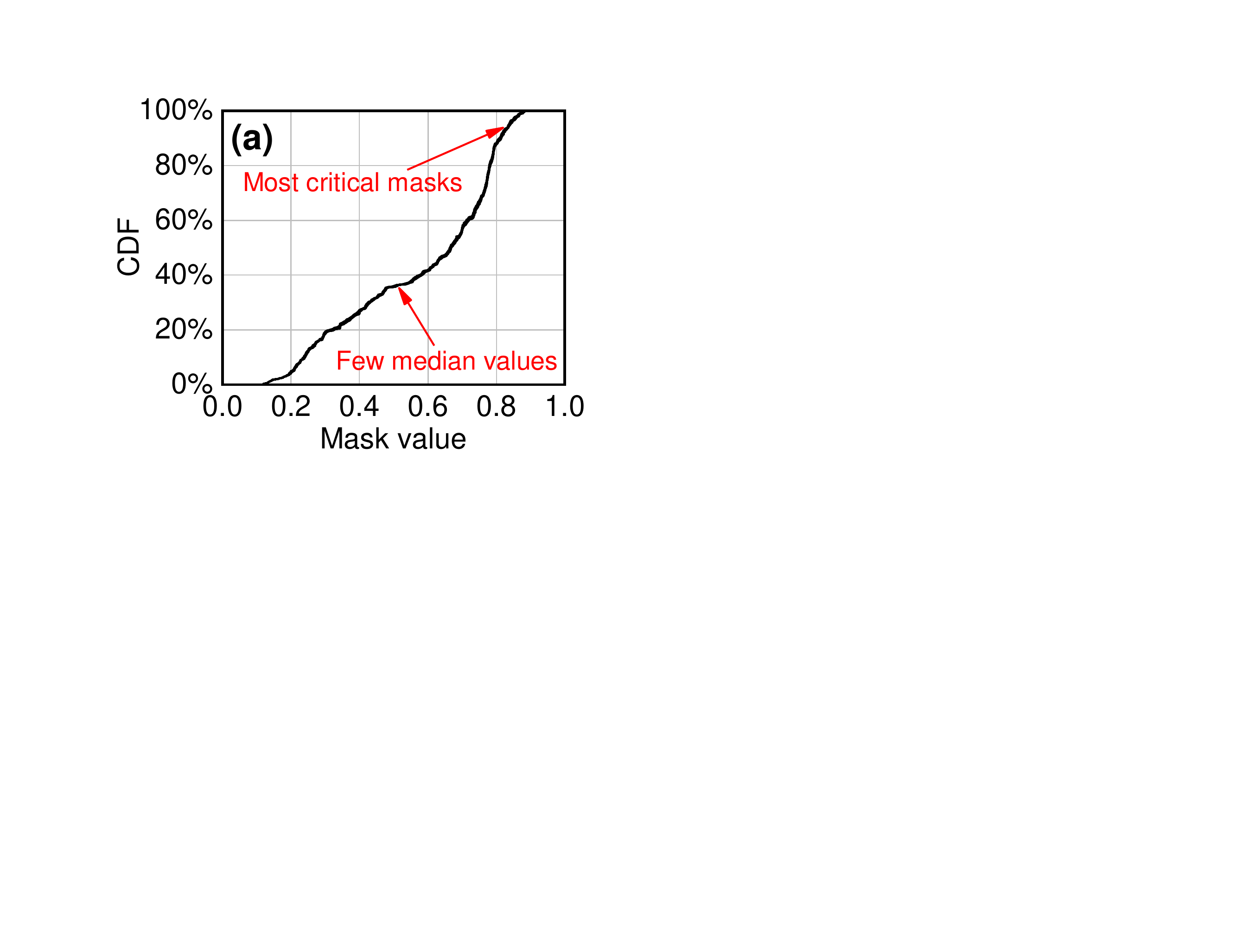}
		\label{fig:mask-cdf}
	}
	\subfigure{
		\includegraphics[height=3cm]{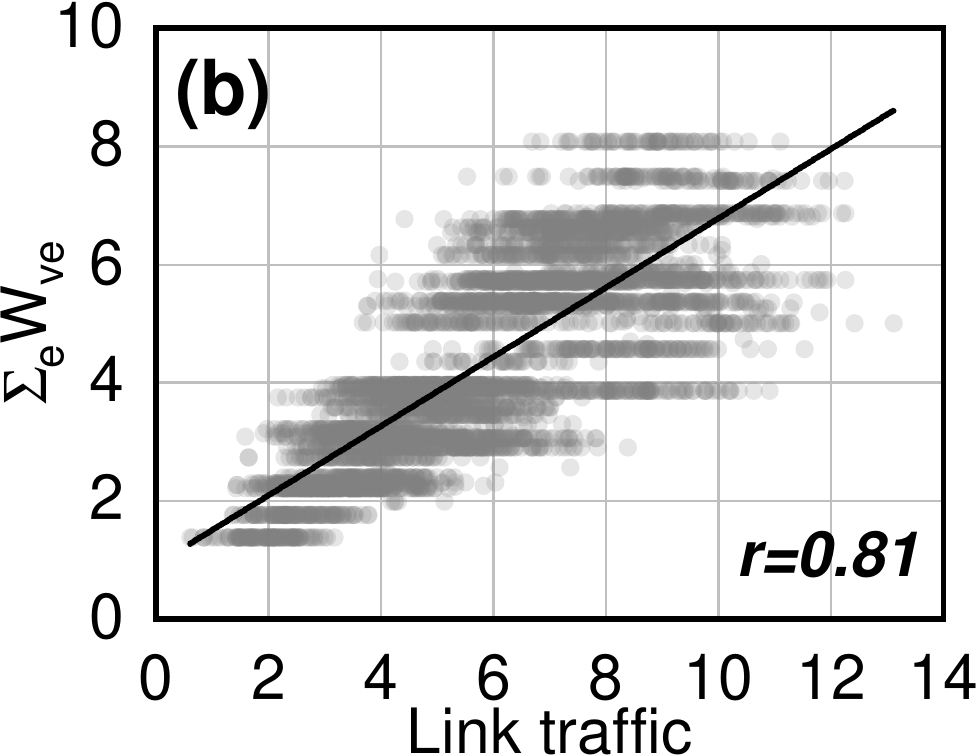}
		\label{fig:mask-traffic}
	}
	\caption{(a) The distribution of mask values in 50 experiments. (b) Sum of mask values ($\sum_e W_{v,e}$) is correlated to the link traffic.}
	\label{fig:interpret-gcs}
\end{figure}

\parahead{Global system interpretations.} We interpret RouteNet* with \name\ and present the top-5 mask values in Table~\ref{tab:mask}. For each path-link connection, there are two common reasons behind selecting path $a$ instead of path $b$. (\textit{i}) \textit{Path $a$ is shorter than path $b$.} For example, connection \#1 in Table~\ref{tab:mask} (path 6$\to$7$\to$10$\to$9 $+$ link 6$\to$7) has a high mask value, indicating selecting 6$\to$7 is a critical decision for the performance. As shown in Figure~\ref{fig:topo-short}, among three candidate paths (colored blue), the shortest path (solid path) has the first hop of 6$\to$7 while the others (dashed path) have 6$\to$4. Thus, \name\ discovers that selecting the first hop is important in deciding the path from 6 to 9, and 6$\to$7 is selected. (\textit{ii}) \textit{Path $a$ is less congested than path $b$.} For example, for connection \#3 in Table~\ref{tab:mask}, there are two paths with the same length, as shown in Figure~\ref{fig:topo-congested}. However, according to the traffic load, link 11$\to$12 is severely congested. Therefore, path 7$\to$10$\to$11$\to$12 should be avoided. \name\ correctly identifies the critical branch and finds that 10$\to$9 is an important decision to avoid the congested path (the red link in Figure~\ref{fig:topo-congested}). 

Besides the individual interpretations over connections, we also analyze the overall behaviors of \name. We present the distribution of the mask values in Figure~\ref{fig:mask-cdf}. Results demonstrate our optimization goal in \S\ref{sec:action} that the number of median mask values is reduced so that network operators can focus on the most critical connections. We also sum up all mask values on each link (vertex in the hypergraph) $\sum_e W_{v,e}$, and measure their relationship with the traffic on each link. As shown in Figure~\ref{fig:mask-traffic}, the sum of mask values and link traffic have a Pearson's correlation coefficient of $r=0.81$. Thus, the sum of mask values and link traffic are statistically correlated, which indicates that the interpretations provided by \name\ are reasonable. Note that \name\ can provide connection-level interpretations as presented above, which is finer-grained than the information from link traffic.

\begin{figure}
	\centering
	\subfigure[Original structure.]{
		\includegraphics[width=4cm]{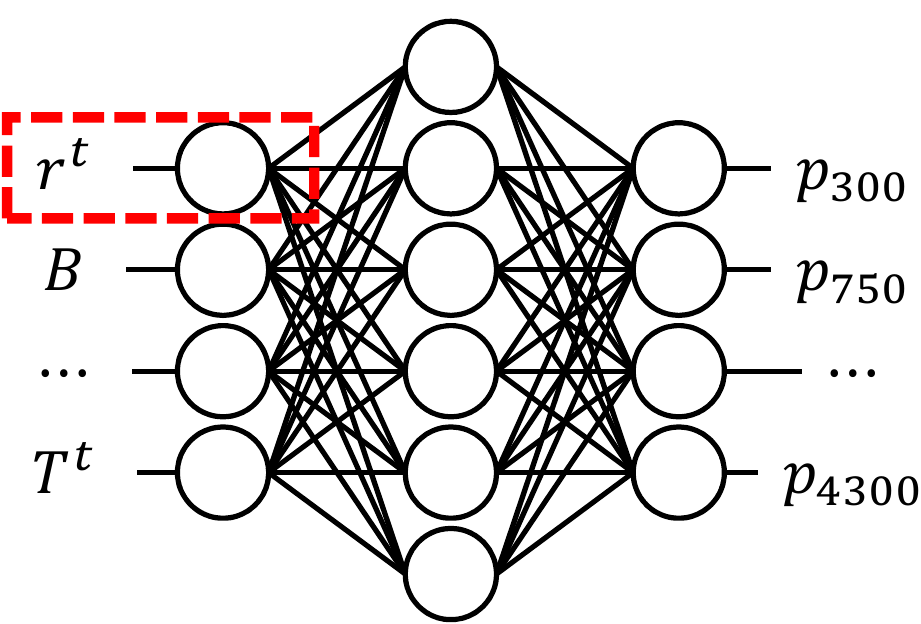}
		\label{fig:nn-original}
	}
	\subfigure[Modified structure.]{
		\includegraphics[width=4cm]{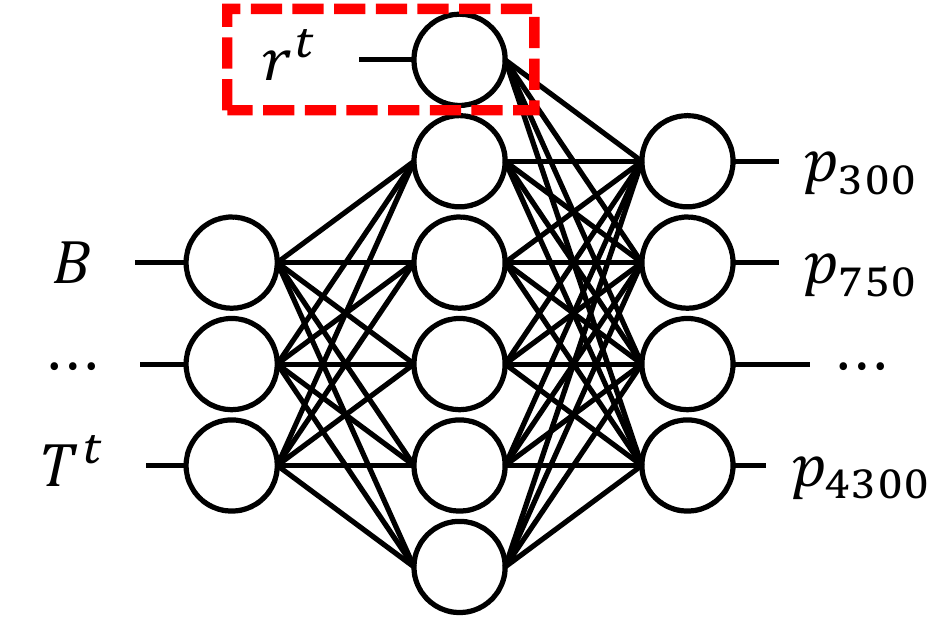}
		\label{fig:nn-modified}
	}
	\caption{We modify the DNN structure of Pensieve based on the interpretations in \S\ref{sec:eva-interpret}. Although two structures are equivalent for the expressive ability, putting significant inputs near to the output will make the DNN optimize easier and better.}
	\label{fig:modified}
\end{figure}

\begin{figure}
	\centering
	\subfigure[Training set.]{
		\includegraphics[width=4cm]{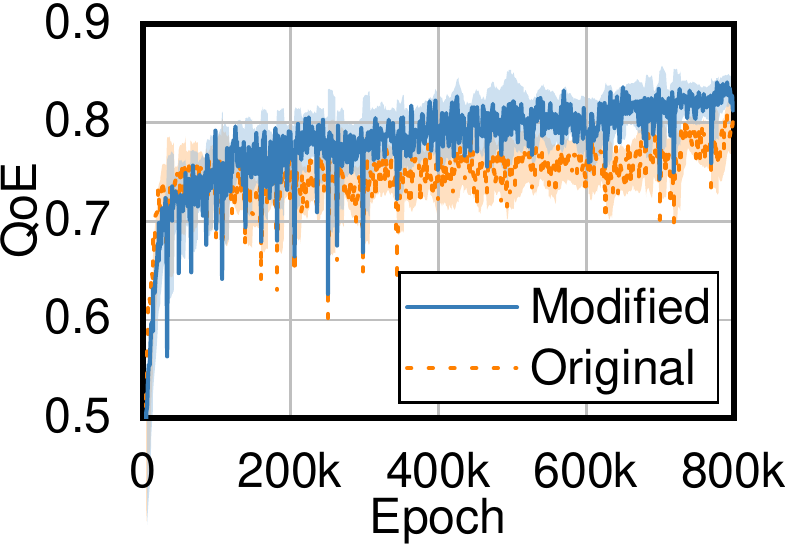}
		\label{fig:modified-train}
	}
	\subfigure[Test set.]{
		\includegraphics[width=4cm]{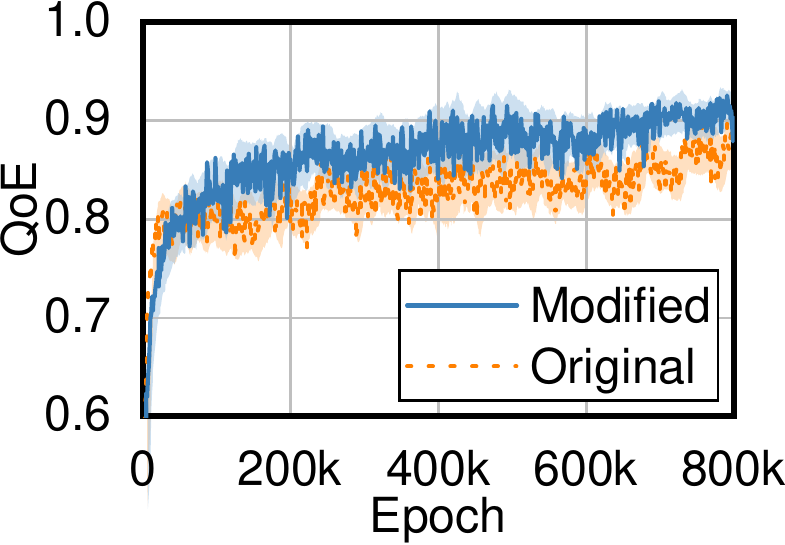}
		\label{fig:modified-test}
	}
	\caption{The modification in Figure~\ref{fig:modified} could improve both the QoE and the training efficiency. Shaded area spans $\pm$ std.}
	\label{fig:eva-modified}
\end{figure}

\begin{figure*}
	\centering
	\subfigure[HSDPA traces (12250 actions).]{
		\includegraphics[width=5.6cm]{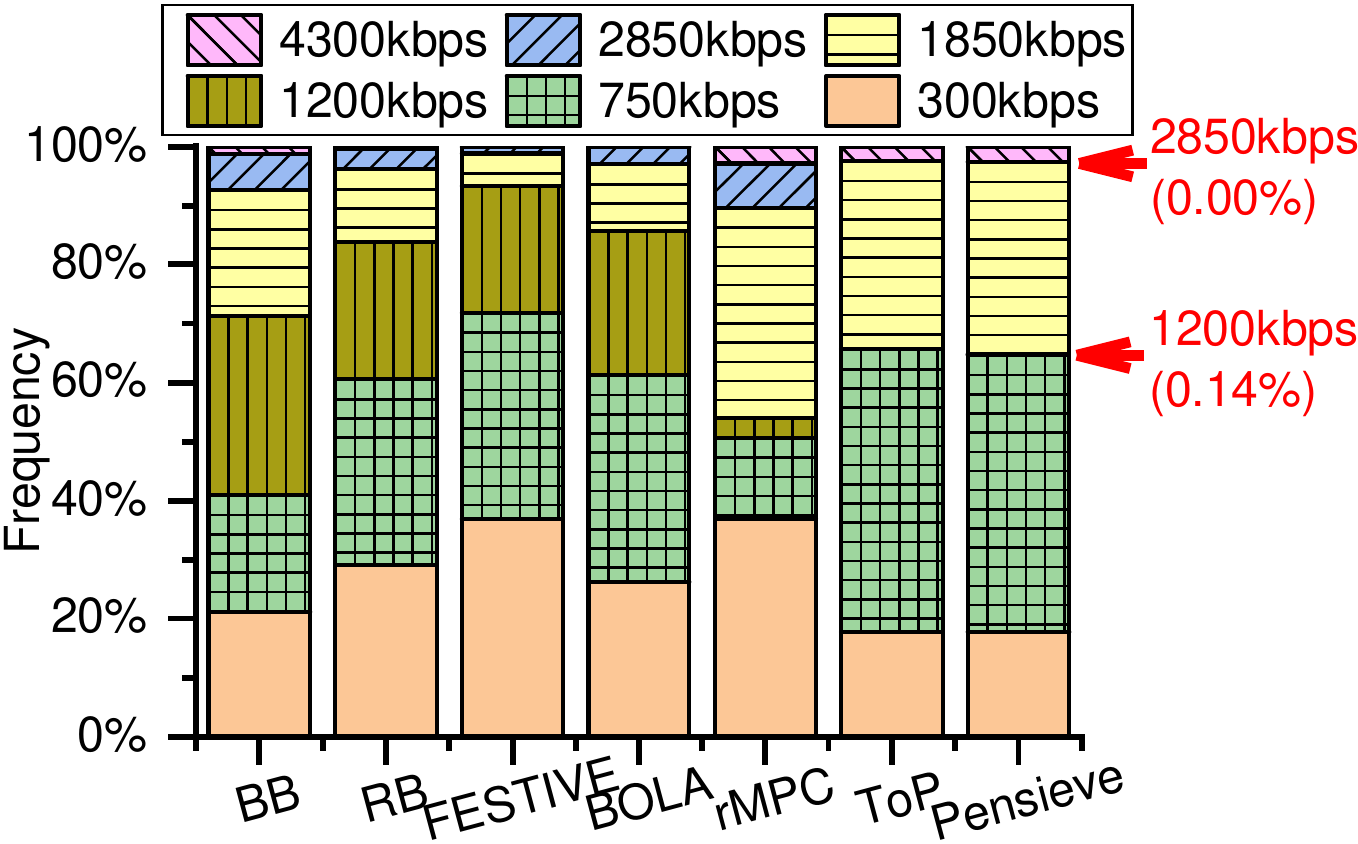}
		\label{fig:frequency-norway}
	}
	\subfigure[FCC traces (10045 actions).]{
		\includegraphics[width=5.6cm]{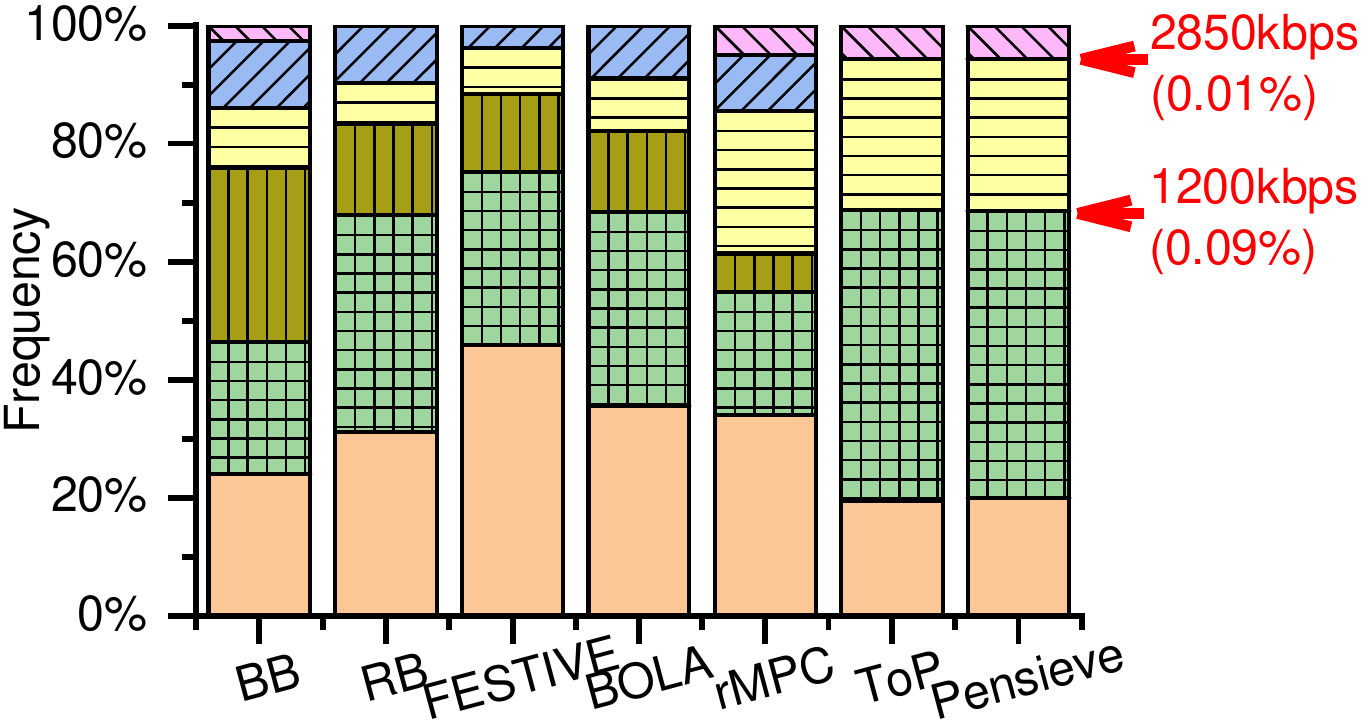}
		\label{fig:frequency-fcc}
	}
	\subfigure[Fixed bandwidth with Pensieve.]{
		\includegraphics[width=5.4cm]{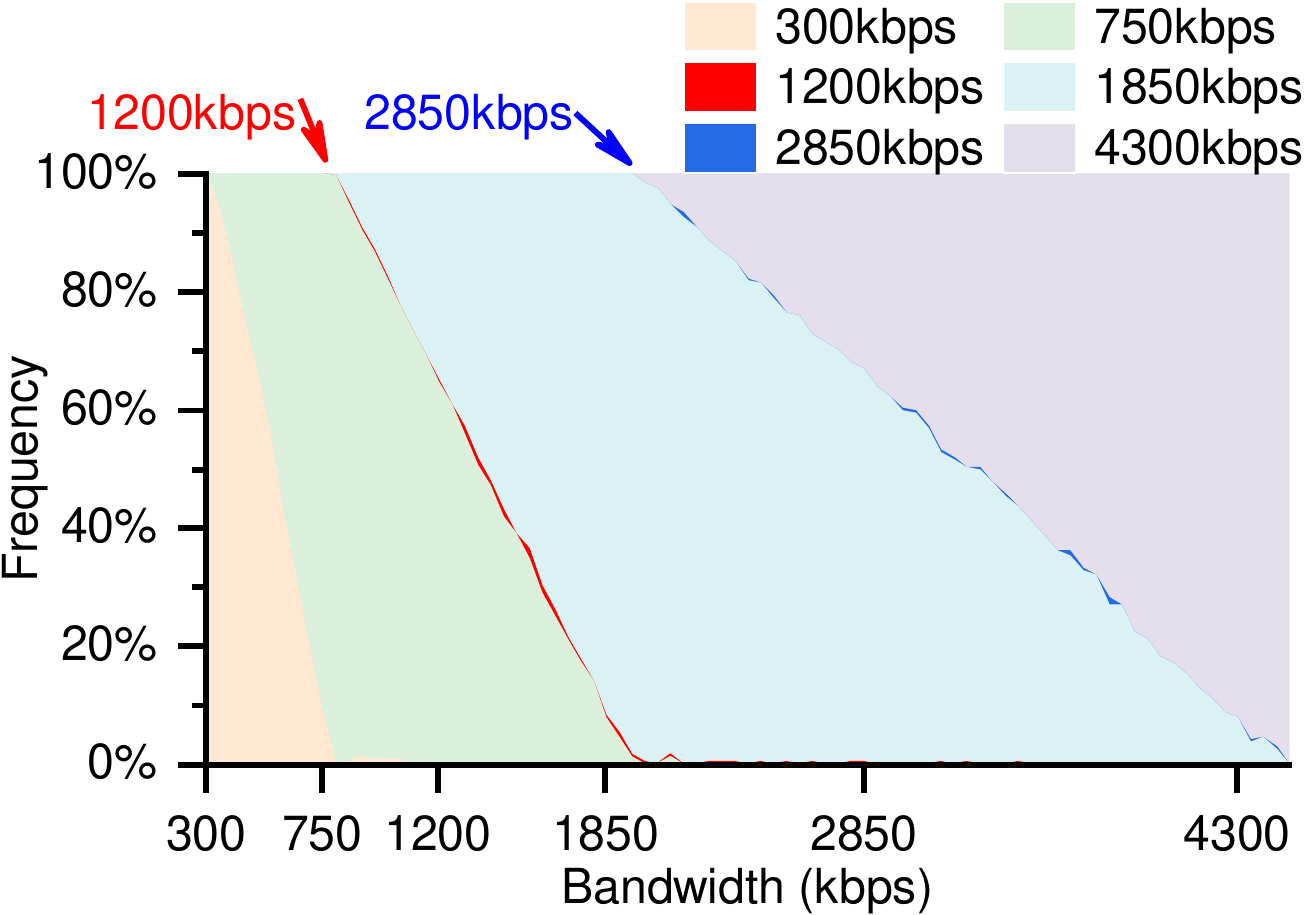}
		\label{fig:freq-bw}
	}
	\caption{For (a) and (b), \name+Pensieve generates almost the same results with Pensieve, where 1200kbps and 2850kbps are rarely selected. (c) On a set of fixed-bandwidth links, 1200kbps and 2850kbps are still not preferred. Better viewed in color.}
	\label{fig:frequency}
	\vspace{-1\baselineskip}
\end{figure*}

\subsection{Guide for Model Design}
\label{sec:case-modified}

We present a use case to demonstrate that the interpretations of \name\ can help the design of the DNN structure of Pensieve. As interpreted in \S\ref{sec:eva-interpret}, \name\ finds that Pensieve significantly relies on the last chunk bitrate ($r^t$) when making decisions. This indicates that $r^t$ may contain important information to the optimization. 

To utilize this observation, we modify the DNN structure of Pensieve to enlarge the influence of $r^t$ on the output result. As shown in Figure~\ref{fig:nn-modified}, we directly concatenate the $r^t$ to the output layer so that it can affect the prediction result more directly. Although the two DNN structures are mathematically equivalent, they will lead to different optimization performance and training efficiency due to the huge search space of DNNs~\cite{jmlr2019nas}. After putting the significant feature nearer to the output layer (thus simplifying the relationship between the significant feature and results), the modified DNN will focus more on that significant feature.

We retrain the two DNN models on the same training and test sets and present the results in Figure~\ref{fig:eva-modified}. From the curves of the original model and the modified model, we can see that the modification in Figure~\ref{fig:modified} improves both the training speed and the final QoE. For example, on the test set, the modified DNN achieves 5.1\% higher QoE on average than the original DNN\footnote{\revise{The offline optimality gap of Pensieve reported in~\cite{sigcomm2017pensieve} is 9.6\%-14.3\%.}}. Considering the scale of views (millions of hours of video watched per day~\cite{facebook_video}) for video providers, even a small improvement in QoE is significant~\cite{abrl}. Moreover, the modified DNN can save 550k epochs on average to achieve the same QoE, which saves 23 hours on our testbed.

\subsection{Enabling Debuggability}
\label{sec:case-troubleshoot}

\revise{
When interpreting Pensieve, as also reported in~\cite{netai2019cracking}, we observe that some bitrates are rarely selected by Pensieve. The frequencies of selected bitrates of the experiments in \S\ref{sec:eva-interpret} are presented in Figures~\ref{fig:frequency-norway} and \ref{fig:frequency-fcc}. Among six bitrates from 300kbps to 4300kbps, two bitrates (1200kbps and 2850kbps) are rarely selected by Pensieve. The imbalance raises our interests since missing bitrates are \textit{median} bitrates: the highest or lowest bitrates may not be selected due to network conditions, but not median ones.
}

\begin{figure}
	\centering
	\includegraphics[width=\linewidth]{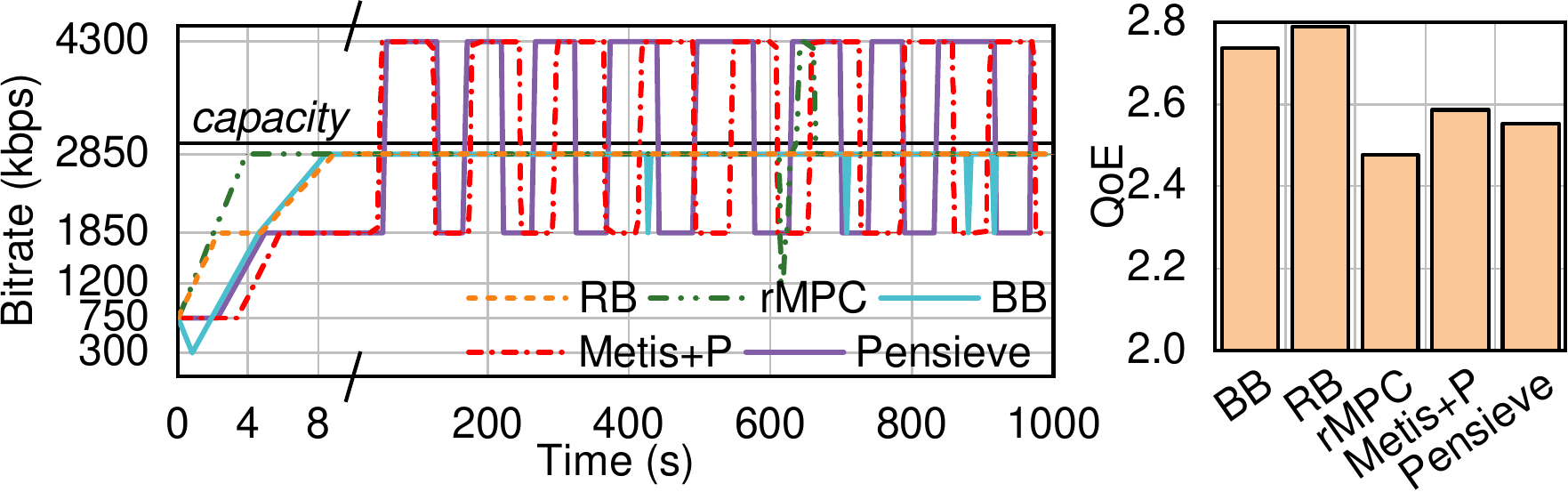}
	\caption{On a 3000kbps link, BB, RB, and rMPC learn the optimal policy and converge to 2850kbps. \name+Pensieve (\name+P) and Pensieve oscillate between 1850kbps and 4300kbps, degrading the QoE. Better viewed in color.}
	\label{fig:bw-variation}
\end{figure}

To further explore the reasons, we emulate Pensieve on a set of links with fixed bandwidth ranging from 300kbps to 4500kbps. As the sample video used by~\cite{sigcomm2017pensieve} is too short for illustration, we replace the test video with a video of 1000 seconds and keep all other configurations the same with the original experiment. As shown in Figure~\ref{fig:freq-bw}, 1200kbps and 2850kbps are still not preferred by Pensieve. For example, on a fixed 3000kbps\footnote{The goodput (bitrate) in this case is roughly 2850kbps.} link, the optimal decision of which should always select 2850kbps. However, in this case, only 0.4\% of selections made by Pensieve are 2850kbps, while the remaining decisions are divided between 1850kbps and 4300kbps. As shown in Figure~\ref{fig:bw-variation}, Pensieve oscillates between 1850kbps and 4300kbps, which is also mimicked by \name+Pensieve. However, such a policy is sub-optimal. In contrast, other baselines learn the optimal selection policy and fix their decisions to 2850kbps, achieving a higher QoE. Similar observations can also be observed on a 1200kbps link~(Appendix~\ref{sec:missing}). 

Studying the raw outputs of Pensieve, we find that Pensieve does not have enough confidence in either choice and therefore oscillates between them. The probability of selecting the optimal bitrate is at a surprisingly low level (Figure~\ref{fig:confidence} in Appendix~\ref{sec:missing}). The training mechanism of Pensieve may cause this problem. At each step, the agent tries to \textit{reinforce} particular actions that lead to larger rewards. In this case, when the agent discovers that four out of six actions can achieve a relatively good reward, it will keep reinforcing this discovery by continuously selecting those actions and finally abandon the others. Making decisions with fewer actions brings higher confidence to the agent, but also makes the agent converge to a local optimum in this case. 

\begin{figure}
	\centering
	\includegraphics[width=7cm]{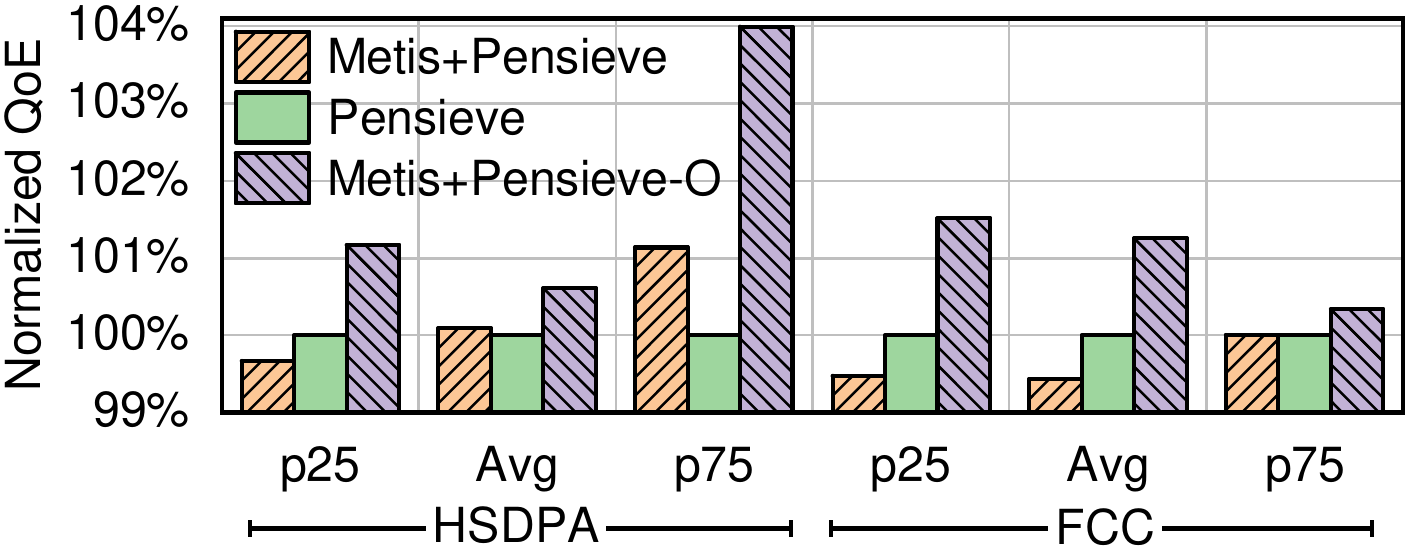}
	\caption{When converting DNNs to decision trees in \name, oversampling the missing bitrates (\name+Pensieve-O) improves the QoE by around 1\% on average compared to the original DNN in Pensieve. QoE is normalized by Pensieve.}
	\label{fig:troubleshooted}
\end{figure}

\revise{
Beyond discovering the problem as~\cite{netai2019cracking}, \name\ can also help fix the problem. Without \name, since Pensieve is designed based on RL, network operators do not have an explicit dataset of bitrates. Network operators may have to penalize the imbalance of bitrate in the reward and retrain the DNN model for hours to days, without knowing whether the RL agent can learn to escape the local optimum itself.
} 
With \name, the conversion from DNN to decision tree exposes an interface for network operators to debug the model. Since the dataset $\mathcal{D}$ to train the decision tree is highly \textit{imbalanced}, as a straightforward solution, we oversample the missing bitrates to make sure their frequencies after sampling are around 1\%. As shown in Figure~\ref{fig:troubleshooted}, the oversampled decision tree (\name+Pensieve-O) outperforms DNNs by about 1\% on average and 4\% at the 75$^{th}$ percentile on HSDPA traces. 

\subsection{Lightweight Deployment}
\label{sec:case-lightweight}

For local systems, decision trees provided by \name\ are also lightweight to deploy. We first demonstrate that the performance degradation between the decision tree and the original DNN is negligible (less than 2\%). Therefore, directly deploying decision trees of Pensieve and AuTO online will (i) shorten the decision latency, (ii) reduce the resource consumption and bring further performance benefits, and (iii) enable implementations onto advanced devices. 

\begin{figure}
	\centering
	\subfigure[\name\ over Pensieve. Error bars represent 25\%ile and 75\%ile.]{
		\includegraphics[height=2.4cm]{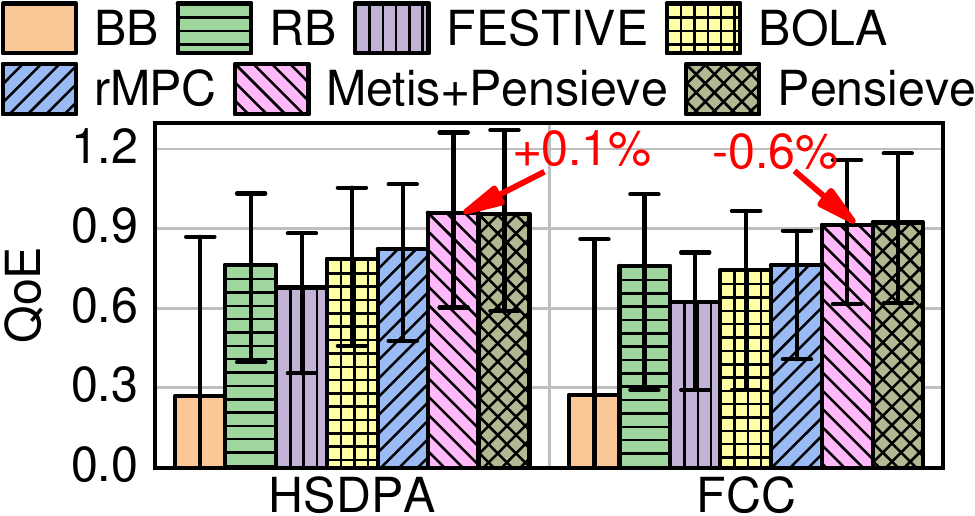}
		\label{fig:qoe}
	}
	\subfigure[\name\ over AuTO.]{
		\includegraphics[height=2.4cm]{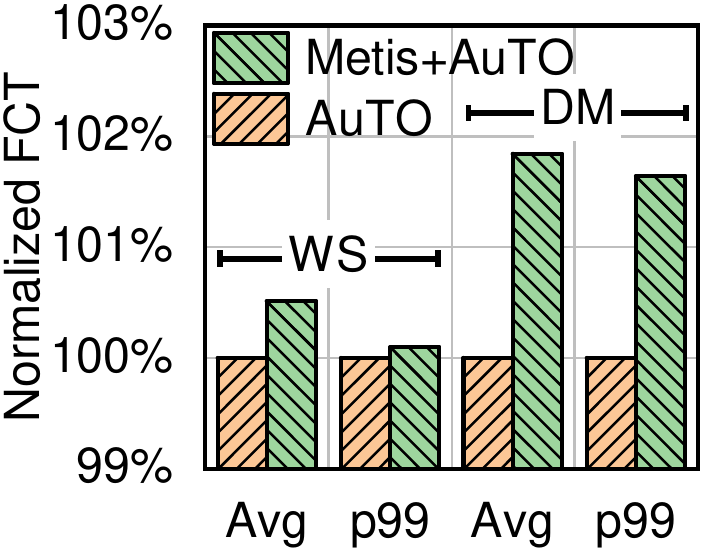}
		\label{fig:fct}
	}
	\caption{The performance degradation between the original DNN and the decision tree with \name\ is less than 2\% for Pensieve and AuTO.}
	\label{fig:eva-faith}
\end{figure}

\parahead{Performance maintenance.} The performance of \name-based systems is comparable to the original systems for both Pensieve and AuTO. As shown in Figure~\ref{fig:qoe}, the differences in average QoE between the decision tree interpreted by \name\ and the original DNN of Pensieve are less than 0.6\% on both traces. Similarly, as shown in Figure~\ref{fig:fct}, the decision tree interpreted from AuTO (\name+AuTO) degrades the performance within 2\% compared to the original DNN. The performance loss is much less than the gain of introducing DNN (Pensieve by 14\%, AuTO by up to 48\%). Therefore, \name\ could maintain the performance of the original DNNs with negligible degradation. 

\begin{figure}
	\centering
	\subfigure[Decision latency.]{
		\includegraphics[width=2.9cm]{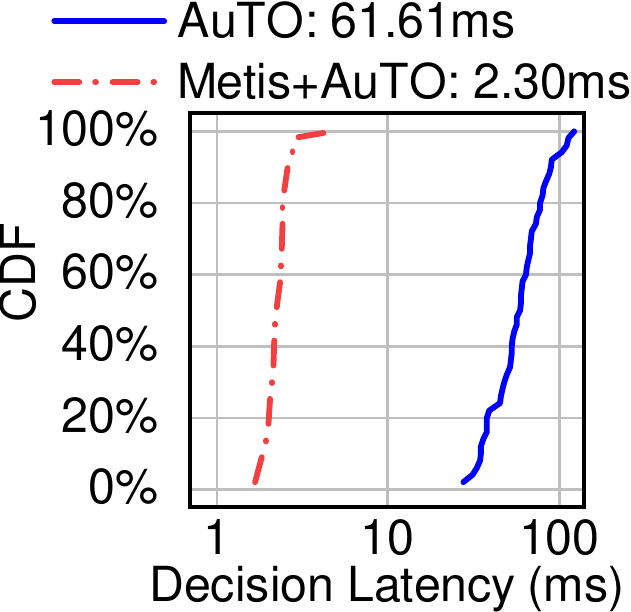}
		\label{fig:http-latency}
	}
	\subfigure[Coverage of the per-flow decision.]{
		\includegraphics[width=4.6cm]{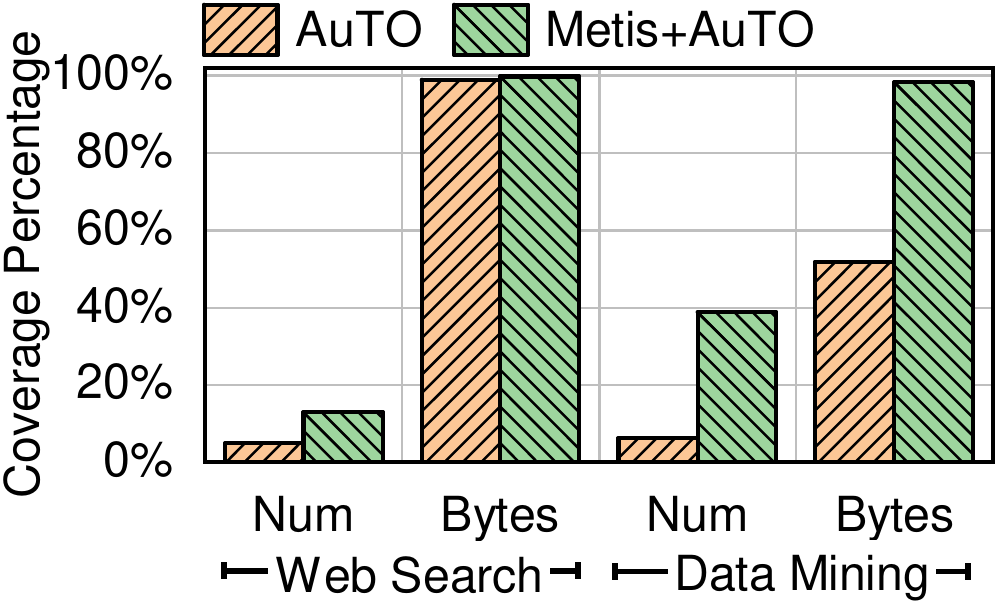}
		\label{fig:coverage}
	}
	\caption{By converting DNNs to decision trees, \name\ could (a) shorten the decision latency by 26.8$\times$, and therefore (b) enlarge the coverage of the per-flow decision.}
	\label{fig:auto-latency}
\end{figure}

\parahead{Decision latency.} We showcase how \name\ helps improve the decision latency of AuTO. The per-flow decision latency of AuTO is 62ms on average, during which short flows in data centers will run out. Converting DNNs into decision trees enables us to make per-flow decisions for more flows since the decision latency is shortened. As shown in Figure~\ref{fig:http-latency}, when replacing DNNs with decision trees, the decision latency of per-flow scheduling could be reduced by 26.8$\times$. In this case, compared to AuTO, \name+AuTO will cover more flows by 33\% and more bytes by 46\% for DM traces~\cite{sigcomm2018auto}, as shown in Figure~\ref{fig:coverage}. 

\begin{figure}
	\centering
	\subfigure{
		\includegraphics[height=2.35cm]{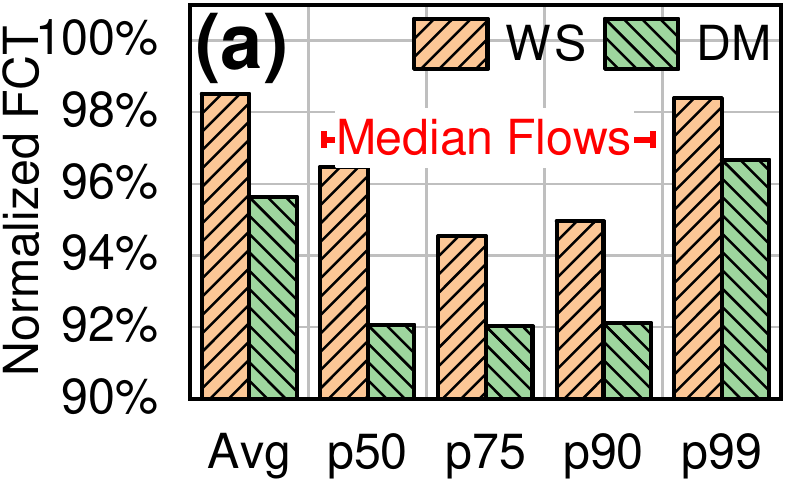}
		\label{fig:all-long}
	}
	\subfigure{
		\includegraphics[height=2.4cm]{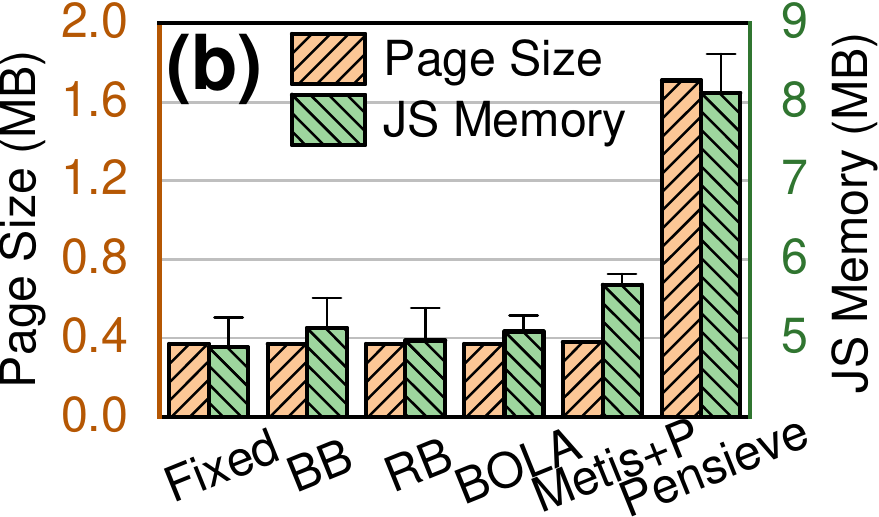}
		\label{fig:memory-pagesize}
	}
	\caption{(a) With precise per-flow optimizations, \name+AuTO could reduce the FCT for median flows. FCT is normalized by the performance of the unmodified AuTO system. (b) Compared to the original Pensieve model, \name+Pensieve could reduce both page size and JS memory.}
\end{figure}

By covering more flows, \name+AuTO can perform optimized per-flow scheduling for not only long flows but also \textit{median flows}, which will improve the overall performance. We modify our prototype of \name+AuTO to allow the decision tree to schedule median flows and present the FCT results in Figure~\ref{fig:all-long}. Although the decision tree has not experienced the scheduling of median flows during training, it can still improve the average performance by 1.5\% and 4.4\% on two traces. We also observe significant performance improvements for median flows (from the 50$^{th}$ to the 90$^{th}$ percentile) by up to 8.0\%. This indicates that median flows enjoy the benefits of precise per-flow scheduling. Improvements in DM traces are better than WS since the coverage increase of DM is larger than that of WS~(cf. Figure~\ref{fig:coverage}). 

\parahead{Resource consumption.} We evaluate the resource consumption (specifically, \textit{page load time} and \textit{memory consumption}) of \name+Pensieve. To eliminate the influence of other modules in the DASH player, we compare these ABR algorithms with a \texttt{fixed} algorithm, which always selects the lowest bitrate.

For page load time, if the HTML page size is too large, users have to wait for a long time before the video starts to play. As shown in Figure~\ref{fig:memory-pagesize}, Fixed, BB, RB, and BOLA have almost the same page size because of their simple processing logic. Pensieve increases the page size by 1370KB since it needs to download the DNN model first. In contrast, \name+Pensieve has a similar page size with the heuristics. When the goodput is 1200kbps (the average bandwidth of Pensieve's evaluation traces), the \textit{additional} page load time of ABR algorithms compared to \texttt{fixed} is reduced by 156$\times$: Pensieve introduces an additional page load time of 9.36 seconds, while \name+Penseve only adds 60ms. 

We then measure the runtime memory and present the results in Figure~\ref{fig:memory-pagesize}. Due to the complexity of forward propagation in the neural networks, Pensieve consumes much more memory than other ABR algorithms. In contrast, the additional memory introduced by \name+Pensieve is reduced by 4.0$\times$ on average and 6.6$\times$ on the peak, which is at the same level as other heuristics.

\parahead{On-device implementation.} Besides, converting DNNs into decision trees also make the model implementable on data plane devices. For example, DNNs are hardly possible to be implemented even with advanced devices (e.g., SmartNICs~\cite{nfp4000-too} and programmable switches~\cite{sigcomm2014p4}) since there are a lot of complicated operations (e.g., floating numbers)~\cite{arxiv2019pforest}. In contrast, decision trees could be implemented with branching clauses only. This enables the offloading of decision trees onto data planes devices. We preliminarily demonstrate the potential by implementing the decision tree onto a Netronome NFP-4000 SmartNIC~\cite{nfp4000-too}. The decision tree interpretations enable us to deploy the \name+AuTO-lRLA with 1,000 LoCs. Evaluation results also show that the decision latency of \name+AuTO on SmartNICs is only 9.37$\mu$s on average. The latency might be further reduced with programmable switches. We leave the deployment of decision trees on programmable switches~\cite{sigcomm2014p4} and the comparison with other baselines for future work.

\subsection{Ad-Hoc Adjustments}
\label{sec:case-routing}

\begin{figure}
	\centering
	\subfigure[Candidate paths.]{
		\includegraphics[height=3.5cm]{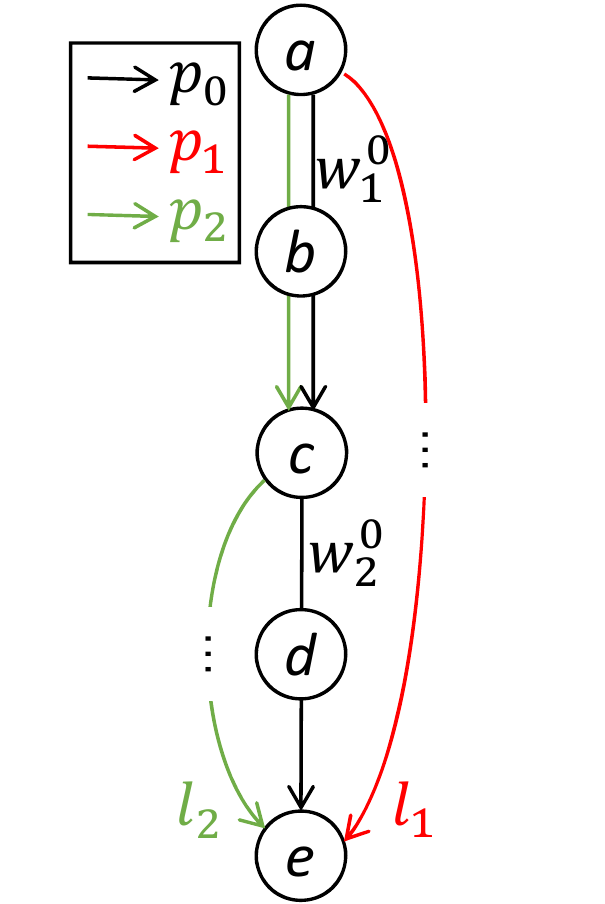}
		\label{fig:reroute}
	}
	\subfigure[Results of $(w_1^0-w_2^0,l_1-l_2)$.]{
		\includegraphics[height=3.5cm]{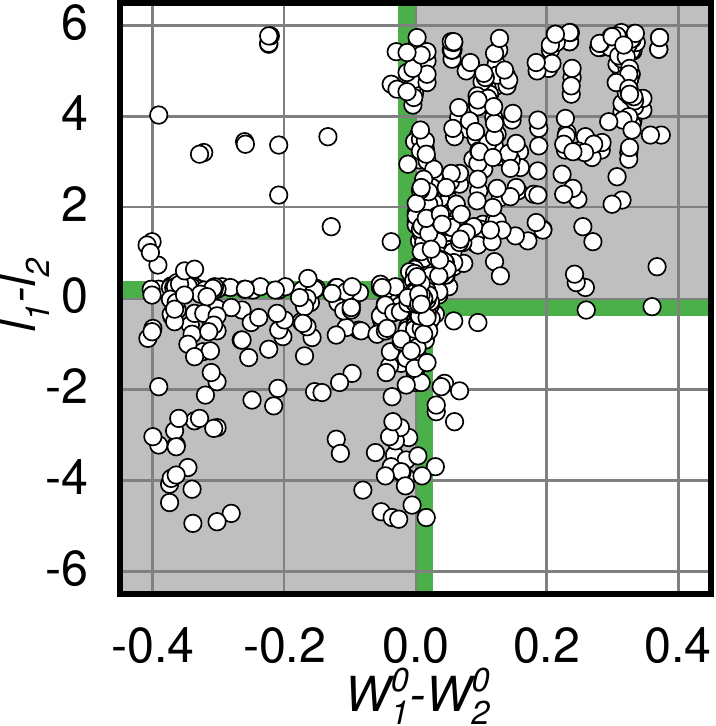}
		\label{fig:mask-case}
	}
	\caption{For RouteNet, the mask value provided by \name\ could help network operators select the better path ($l_2$) during ad-hoc adjustments.}
\end{figure}

We present a use case of \name\ on how network operators can execute ad-hoc adjustments onto RouteNet* based on the interpretations of \name. In the routing case, network operators might need to reroute a flow to another path due to external reasons (e.g., pricing). As shown in Figure~\ref{fig:reroute}, when the demand from node $a$ to node $e$ needs rerouting away from the original path $p_0$, there are several candidates paths ($p_1$ and $p_2$). Since the actual performance of each path is unknown until rerouting rules are installed, deciding which path to reroute is challenging. Such a scenario with multiple similar paths is common in topologies such as fat-trees. 

Our observation is that since the candidate paths divert at different nodes from the original path, we could estimate their performance by the mask value of the connection between the diverting node and its next-hop link. For example, in Figure~\ref{fig:reroute}, $p_1$ diverts from $p_0$ at node $a$ and $p_2$ diverts from $p_0$ at node $c$. $w^0_1$ is the mask value of the connection between $p_0$ and link $a$$\to$$b$. Since $w^0_1$ represents the significance of selecting $a$$\to$$b$ rather than other links, it is correlated to the possibility that there is also a relatively good path if $a$$\to$$b$ is not selected. Recalling the optimization in Equation~\ref{eq:objective}, a lower mask value of a connection means that the selection is not critical in deciding the routing path. Thus we have:

\begin{adjustwidth}{6mm}{6mm}
	\smallskip
	\textsc{Observation.} \textit{If $w^0_1 > w^0_2$, the latency of $p_1$ (denoted as $l_1$) is likely to larger than the latency of $p_2$ (denoted as $l_2$).}
\end{adjustwidth}

\smallskip We verify the observation above with the NSFNet topology in Figure~\ref{fig:topo}. Since the optimal path may not be the shortest path, we consider all paths that are $\leqslant$1 hop longer than the shortest path as candidates. For example, for path 0$\to$2$\to$5$\to$12, path 0$\to$3$\to$4$\to$5$\to$12 is considered as a candidate, but path 0$\to$1$\to$ 7$\to$10$\to$11$\to$12 is not. We go through all pairs of demand in the NSFNet topology in Figure~\ref{fig:topo} and measure all the path latency for such candidate scenarios and the mask values at the diverting node. We repeat the experiments with all the 50 traffic samples provided by \cite{sosr2019routenet}. 

For each routing path $p_0$ generated by RouteNet*, we collect all $(p_0, p_1, p_2)$ that satisfy the conditions in Figure~\ref{fig:reroute}, and measure their end-to-end latency $(l_0, l_1, l_2)$. We also measure the mask values at the diverting nodes ($w_1^0$ and $w_2^0$) and plot the $(w_1^0-w_2^0, l_1-l_2)$. For simplicity, we present the results of all paths originating from nodes 0,1,2,3 in Figure~\ref{fig:mask-case}~(750 points in total). Most points (72\%) fall into quadrants I and III (shaded gray) with another 19\% points very close to quadrants I and III (shaded green), which verifies our observation above. Thus, we provide an indicator for network operators to decide which path to reroute without estimating end-to-end path latency. 

\subsection{\name\ Deep Dive}
\label{sec:eva-microbenchmark}

Finally, we overview the experiments that benchmark the interpretability of \name. The detailed experimentation setup and more empirical results are deferred to the appendix.

\parahead{Interpretation baselines comparison.} We compare the performance of the decision tree in \name\ against two baselines in the DL community. We implement LIME~\cite{kdd2016lime}, one of the most typical blackbox interpretation methods in the DL community, and LEMNA~\cite{ccs2018lemna}, an interpretation method specifically designed for time sequences in RNN. We measure the misprediction rate and errors of three interpretation methods. The misprediction rates on two systems with \name-based methods are reduced by 1.2$\times$-1.7$\times$ compared to two baselines. The root-mean-square errors (RMSEs) are reduced by 1.2$\times$-3.2$\times$. Experiments are presented in Appendix~\ref{sec:app-faith} in detail. The decision tree outperforms the other two interpretation methods, which confirms our design choice in \S\ref{sec:xai}.

\parahead{Sensitivity analysis.} We test the robustness of hyperparameters of \name\ in Appendix~\ref{sec:app-sensitivity}. For decision tree interpretations, we test the robustness of the number of leaf nodes. Results show that a wide range of settings (from 10 to 5000) perform well for Pensieve and AuTO (accuracy variations within 10\%). For hypergraph interpretations, we vary the two hyperparameters $\lambda_1$ and $\lambda_2$ in Equation~\ref{eq:objective}. We then measure how the interpreted mask values respond to the variations of hyperparameters. Results show that network operators could effectively adjust respective hyperparameters according to their needs. For example, when network operators want to inspect less critical connections, they can increase the value of $\lambda_1$ to penalize the scale of mask values.

\parahead{Computation overhead.} In Appendix~\ref{sec:app-overhead}, our evaluation shows that converting finetuned DNNs into decision trees for Pensieve and AuTO takes less than 40 seconds under different settings. For hypergraph interpretations, the computation time of generating the mask values for RouteNet* is 80 seconds on average. This offline computation time is negligible compared to the training time of DNN models, which may take several hours to days. 

\section{Discussion}
\label{sec:discuss}

\revise{
In this section, we discuss some design choices, the generalization ability, limitations, and potential future directions of \name.
}


\parahead{Why not directly train a decision tree?} As shown in~\S\ref{sec:case-lightweight}, converted decision trees exhibit comparable performance to larger models. However, \emph{directly} training the simpler model from scratch is difficult to achieve the same performance. 
\revise{
We hypothesize that the first reason is that decision trees are non-parametric models, which are not designed for continuously parameter updating and structure adjusting. Even with recent advances in decision tree adjusting~\cite{kdd2018efdt}, the efficient adjustment relies on massive amount of training data.
Another possible explanation behind this phenomenon is the \textit{lottery ticket hypothesis}~\cite{iclr2019lottery, iclr2020lotteryrl}: training deep models is analogous to winning the lottery by buying a very large number of tickets (i.e., building a large neural network). However, we cannot know the winning ticket configuration in advance. Therefore, directly training a simpler model is similar to buying one lottery ticket only, which has little chance to achieve satisfying performance.
}

\parahead{Can \name\ interpret \textit{all} types of networking systems?} Admittedly, \name\ cannot interpret all DL-based networking systems. For example, network intrusion detection systems (NIDSes) are used to detect malicious packets with regular expression matching on the packet payload~\cite{jnca2013surveyids}. Prior DL-based methods introduced RNN to improve the performance of NIDSes~\cite{access2017rnnids}. However, since RNN (and other DNNs with recurrent structures) fundamentally contains \textit{implicit} memory units, decision trees cannot faithfully capture the policy with only \textit{explicit} decision variables. In the future, we aim to combine \name\ with recurrent units, e.g., employing recurrent decision trees~\cite{cvpr2016rdt}. We also clarify the capability of hypergraph formulation in \S\ref{sec:action}. 


\revise{
\parahead{How to interpret deeper DNNs?} Although our evaluation shows satisfying performance on three DL-based networking systems, compare to the applications of DNNs in other communities (Figure~\ref{fig:depth}), those in networking systems are still at a preliminary stage: both Pensieve and AuTO have less than 10 hidden layers.  Whether current approaches could scale to network systems with more complicated neural networks remains unknown. Nonetheless, on one hand, \name\ might be scalable to deeper neural networks because deeper neural networks (regardless of training difficulty) sometimes have the same level of expressiveness compared to shallower ones~\cite{nips2014deep, icml2019complexity}. On the other hand, as a preliminary attempt, we adopt the traditional CART algorithm in decision tree training. More optimized decision tree representations~\cite{ijcai2018odt} with tree-based regularization~\cite{aaai2018treereg} during the training process of DNNs might interpret the policies more faithfully. 
	
\parahead{Will the generalization ability of DNNs be impaired?} Although the generalization ability of DNNs is still under exploration, it is indisputable that the generalization ability of DNNs roots in the massive amount of parameters~\cite{iclr2018sensitivity}. Despite that \name\ performs well in our experiment settings as demonstrated in \S\ref{sec:eva}, the generalization ability of interpretations still needs investigation. There are two ways to further address the generalization ability of interpretations on different traces. On one hand, researchers can analyze the theoretic performance bounds of the interpretation~\cite{mm2019pitree}. On the other hand, network operators can deploy the interpretation results into the production environments and evaluate the online performance. We call on the community to devote more research efforts in this direction.
}

\parahead{Will interpretations always be correct?} \name\ is designed to offer a sense of confidence by helping network operators understand (and further troubleshoot) DL-based networking systems. However, the interpretations themselves can also make mistakes. In fact, researchers have recently discovered attacks against the interpreting systems for image classification~\cite{sec2020adv2, neurips2019fooling}. Nonetheless, interpretations from our experiments are empirically sane (\S\ref{sec:eva}). Since the interpretations are concise and well understood, human operators could easily spot the rare case of erroneous interpretation.

\vspace{0.5em}
\section{Related Work}
\label{sec:related}

There is little prior work on interpreting DL-based networking systems. 
\revise{
Some previous workshop papers discuss the problems of interpreting DL-based networking systems~\cite{netai2019cracking, apnet2018demystifying}, which innovatively shed light on the direction of interpretability in the networking community. However, those research efforts still remain preliminary regarding the design of interpretation methods, the application scenario, and the implementation on real systems.
}
In terms of approach, the closest work is NeuroCuts~\cite{sigcomm2019neurocuts}, which optimizes a decision tree for packet classification with deep RL and is therefore self-interpretable. However, NeuroCuts directly trains the decision tree from scratch for packet classification only while \name\ interprets existing diverse DL-based networking systems.
In the following, we survey the practicality of prior work on using DL in networking applications and alternative methods to apply DNNs interpretations in other domains.


\parahead{Practicality of DL-based networking systems.} There are also some other issues of DL-based networking systems that need to be addressed before deployed in practice. Some recent work focuses on the verification of DL-based networking systems~\cite{netai2019verily}, which is orthogonal to our work and could be adopted together for a more practical system. Recent solutions also address the heavyweight issue of specific networking systems~\cite{mm2019pitree, hotnets2018lfo, mobisys2017deepmon}, which do not focus on interpretability and are difficult to support complex DL-based networking systems. \name\ provides a systematic solution to effectively interpret diverse DL-based networking systems with high quality for practical deployment. \name\ could also be integrated with research efforts on the training phase of DL-based networking systems~\cite{neurips2019park} to achieve a practical system at the design phase.

\parahead{Interpretation methods.} 
\revise{
As discussed in \S\ref{sec:existing}, many interpretability approaches focus on understanding the mechanism of DNNs, such as convolutional neural networks (CNN)~\cite{cvpr2017netdissect, cvpr2018interpretablecnn}, RNN~\cite{ccs2018lemna, icml2019interpretlstm}, GNN~\cite{neurips2019gnnexplainer, arxiv2020graphlime}, which is not the goal of \name. 
}
Besides interpretation methods introduced in \S\ref{sec:existing} and \S\ref{sec:xai}, there are also some research efforts to interpret existing applications in many domains. Examples include image analysis~\cite{eccv2014visualizing, cvpr2017netdissect}, neural language translation~\cite{neurips2019interpretnlp, acl2018semantically}, recommendation systems~\cite{www2018narre, aaai2019dynamic}, and security applications~\cite{sec2020adv2, ccs2018lemna}. However, as discussed in \S\ref{sec:existing}, existing methods are insufficient for networking systems. There still lacks an effective interpretation method for the networking community. \name\ sheds light on the interpretability of DL-based networking systems with our specially designed framework.

\parahead{Hypergraph learning.} In the machine learning community, the hypergraph structure has many applications in the modeling of high-order correlation in social network and image recognition. The message passing process on hypergraph structure is first introduced in~\cite{nips2007hypergraph}. The most recent efforts combine the hypergraph structure with convolution~\cite{aaai2019hgnn, neurips2019hypergcn} and attention mechanisms~\cite{iclr2020hypersagnn} to further improve the model performance. The objective there is to \textit{directly optimize} the model performance (e.g., prediction accuracy). 
\revise{
In contrast, \name\ employs the hypergraph structure to \textit{formulate} various outputs of global networking system outputs. \name\ also \textit{interprets} the critical components in hypergraphs with the searching algorithm in \S\ref{sec:action}.
}
A possible future direction is to \textit{design} more DL-based networking systems with our hypergraph formulation~(\S\ref{sec:app}) and hypergraph learning methods above, which is left as our future work.

\vspace{0.5em}
\section{Conclusion}
\label{sec:concl}

In this paper, we propose \name, a new framework to interpret diverse DL-based networking systems. \name\ categorizes DL-based networking systems and provides respective solutions by modeling and analyzing the commonplaces of them. We apply \name\ over several typical DL-based networking systems. Evaluation results show that \name-based systems can interpret the behaviors of DL-based networking systems with high quality. Further use cases demonstrate that \name\ could help network operators design, debug, deploy, and ad-hoc adjust DL-based networking systems.

 This work does not raise any ethical issues. 

\revise{
\section*{Acknowledgement}
We thank our shepherd, Arpit Gupta, and anonymous SIGCOMM reviewers for their valuable comments. We also thank for the suggestions from Mohammad Alizadeh, Chen Sun, Jun Bi, Shuhe Wang, Yangyang Wang, and all members of NetLab in Tsinghua University. The research is supported by the National Natural Science Foundation of China (No. 61625203 and 61832013), and the National Key R\&D Program of China (No. 2017YFB0801701). Mingwei Xu is the corresponding author.
}

\bibliographystyle{plain}
\bibliography{bibfile}

\begin{thebibliography}{10}

\bibitem{fcc}
Raw data - measuring broadband america.
\newblock
  \url{https://www.fcc.gov/reports-research/reports/measuring-broadband-america/raw-data-measuring-broadband-america-2016},
  2016.

\bibitem{dashjs}
Dash.js.
\newblock \url{https://github.com/Dash-Industry-Forum/dash.js}, 2018.

\bibitem{sigcomm2010dctcp}
Mohammad Alizadeh, Albert Greenberg, David~A Maltz, Jitendra Padhye, Parveen
  Patel, Balaji Prabhakar, Sudipta Sengupta, and Murari Sridharan.
\newblock Data center tcp (dctcp).
\newblock In {\em Proc. ACM SIGCOMM}, 2010.

\bibitem{sigcomm2013pfabric}
Mohammad Alizadeh, Shuang Yang, Milad Sharif, Sachin Katti, Nick McKeown,
  Balaji Prabhakar, and Scott Shenker.
\newblock pfabric: Minimal near-optimal datacenter transport.
\newblock In {\em Proc. ACM SIGCOMM}, 2013.

\bibitem{nips2014deep}
Jimmy Ba and Rich Caruana.
\newblock Do deep nets really need to be deep?
\newblock In {\em Proc. NIPS}, 2014.

\bibitem{nsdi2015pias}
Wei Bai, Li~Chen, Kai Chen, Dongsu Han, Chen Tian, and Hao Wang.
\newblock Information-agnostic flow scheduling for commodity data centers.
\newblock In {\em Proc. USENIX NSDI}, 2015.

\bibitem{nips2018viper}
Osbert Bastani, Yewen Pu, and Armando Solar-Lezama.
\newblock Verifiable reinforcement learning via policy extraction.
\newblock In {\em Proc. NeurIPS}, 2018.

\bibitem{cvpr2017netdissect}
David Bau, Bolei Zhou, Aditya Khosla, Aude Oliva, and Antonio Torralba.
\newblock Network dissection: Quantifying interpretability of deep visual
  representations.
\newblock In {\em Proc. IEEE CVPR}, 2017.

\bibitem{hotnets2018lfo}
Daniel~S Berger.
\newblock Towards lightweight and robust machine learning for cdn caching.
\newblock In {\em Proc. ACM HotNets}, 2018.

\bibitem{ai1998dt}
Hendrik Blockeel and Luc De~Raedt.
\newblock Top-down induction of first-order logical decision trees.
\newblock {\em Artificial intelligence}, 101(1-2):285--297, 1998.

\bibitem{sigcomm2014p4}
Pat Bosshart, Dan Daly, Glen Gibb, Martin Izzard, Nick McKeown, Jennifer
  Rexford, Cole Schlesinger, Dan Talayco, Amin Vahdat, George Varghese, and
  David Walker.
\newblock P4: Programming protocol-independent packet processors.
\newblock {\em ACM SIGCOMM Computer Communication Review}, 44(3):87--95, 2014.

\bibitem{gpt3billion}
Tom~B. Brown, Benjamin Mann, Nick Ryder, Melanie Subbiah, Jared Kaplan,
  Prafulla Dhariwal, Arvind Neelakantan, Pranav Shyam, Girish Sastry, Amanda
  Askell, Sandhini Agarwal, Ariel Herbert-Voss, Gretchen Krueger, Tom Henighan,
  Rewon Child, Aditya Ramesh, Daniel~M. Ziegler, Jeffrey Wu, Clemens Winter,
  Christopher Hesse, Mark Chen, Eric Sigler, Mateusz Litwin, Scott Gray,
  Benjamin Chess, Jack Clark, Christopher Berner, Sam McCandlish, Alec Radford,
  Ilya Sutskever, and Dario Amodei.
\newblock Language models are few-shot learners.
\newblock {\em arXiv preprint 2005.14165}, 2020.

\bibitem{arxiv2019pforest}
Coralie Busse-Grawitz, Roland Meier, Alexander Dietm{\"u}ller, Tobias
  B{\"u}hler, and Laurent Vanbever.
\newblock pforest: In-network inference with random forests.
\newblock {\em arXiv preprint 1909.05680}, 2019.

\bibitem{www2018narre}
Chong Chen, Min Zhang, Yiqun Liu, and Shaoping Ma.
\newblock Neural attentional rating regression with review-level explanations.
\newblock In {\em Proc. WWW}, 2018.

\bibitem{cvpr2016rdt}
Jianhui Chen, Hoang~M Le, Peter Carr, Yisong Yue, and James~J Little.
\newblock Learning online smooth predictors for realtime camera planning using
  recurrent decision trees.
\newblock In {\em Proc. IEEE CVPR}, 2016.

\bibitem{sigcomm2018auto}
Li~Chen, Justinas Lingys, Kai Chen, and Feng Liu.
\newblock Auto: Scaling deep reinforcement learning for datacenter-scale
  automatic traffic optimization.
\newblock In {\em Proc. ACM SIGCOMM}, 2018.

\bibitem{aaai2019dynamic}
Xu~Chen, Yongfeng Zhang, and Zheng Qin.
\newblock Dynamic explainable recommendation based on neural attentive models.
\newblock In {\em Proc. AAAI}, 2019.

\bibitem{nips2014gru}
Junyoung Chung, Caglar Gulcehre, Kyunghyun Cho, and Yoshua Bengio.
\newblock Empirical evaluation of gated recurrent neural networks on sequence
  modeling.
\newblock In {\em NIPS Workshop on Deep Learning}, 2014.

\bibitem{cvpr2009imagenet}
Jia Deng, Wei Dong, Richard Socher, Li-Jia Li, Kai Li, and Li~Fei-Fei.
\newblock Imagenet: A large-scale hierarchical image database.
\newblock In {\em Proc. IEEE CVPR}, 2009.

\bibitem{netai2019cracking}
Arnaud Dethise, Marco Canini, and Srikanth Kandula.
\newblock Cracking open the black box: What observations can tell us about
  reinforcement learning agents.
\newblock In {\em Proc. ACM NetAI}, 2019.

\bibitem{cacm2020iml}
Mengnan Du, Ninghao Liu, and Xia Hu.
\newblock Techniques for interpretable machine learning.
\newblock {\em Commun. ACM}, pages 68--77, 2020.

\bibitem{cnndeep}
Pierre Ecarlat.
\newblock Cnn - do we need to go deeper?
\newblock
  \url{https://medium.com/finc-engineering/cnn-do-we-need-to-go-deeper-afe1041e263e},
  2017.

\bibitem{jmlr2019nas}
Thomas Elsken, Jan~Hendrik Metzen, and Frank Hutter.
\newblock Neural architecture search: A survey.
\newblock {\em Journal of Machine Learning Research}, 2019.

\bibitem{aaai2019hgnn}
Yifan Feng, Haoxuan You, Zizhao Zhang, Rongrong Ji, and Yue Gao.
\newblock Hypergraph neural networks.
\newblock In {\em Proc. AAAI}, 2019.

\bibitem{iclr2019lottery}
Jonathan Frankle and Michael Carbin.
\newblock The lottery ticket hypothesis: Finding sparse, trainable neural
  networks.
\newblock In {\em Proc. ICLR}, 2019.

\bibitem{friedman1984cart}
Jerome~H Friedman, Richard~A Olshen, Charles~J Stone, et~al.
\newblock Classification and regression trees.
\newblock {\em Wadsworth \& Brooks}, 1984.

\bibitem{sigcomm2009vl2}
Albert Greenberg, James~R Hamilton, Navendu Jain, Srikanth Kandula, Changhoon
  Kim, Parantap Lahiri, David~A Maltz, Parveen Patel, and Sudipta Sengupta.
\newblock Vl2: a scalable and flexible data center network.
\newblock In {\em Proc. ACM SIGCOMM}, 2009.

\bibitem{survey2018xai}
Riccardo Guidotti, Anna Monreale, Salvatore Ruggieri, Franco Turini, Fosca
  Giannotti, and Dino Pedreschi.
\newblock A survey of methods for explaining black box models.
\newblock {\em ACM Computing Surveys (CSUR)}, 2018.

\bibitem{icml2019interpretlstm}
Tian Guo, Tao Lin, and Nino Antulov-Fantulin.
\newblock Exploring interpretable {LSTM} neural networks over multi-variable
  data.
\newblock In {\em Proc. ICML}, 2019.

\bibitem{ccs2018lemna}
Wenbo Guo, Dongliang Mu, Jun Xu, Purui Su, Gang Wang, and Xinyu Xing.
\newblock Lemna: Explaining deep learning based security applications.
\newblock In {\em Proc. ACM CCS}, 2018.

\bibitem{icml2019complexity}
Boris Hanin and David Rolnick.
\newblock Complexity of linear regions in deep networks.
\newblock In {\em Proc. ICML}, 2019.

\bibitem{neurips2019fooling}
Juyeon Heo, Sunghwan Joo, and Taesup Moon.
\newblock Fooling neural network interpretations via adversarial model
  manipulation.
\newblock In {\em Proc. NeurIPS}, 2019.

\bibitem{arxiv2020graphlime}
Qiang Huang, Makoto Yamada, Yuan Tian, Dinesh Singh, Dawei Yin, and Yi~Chang.
\newblock Graphlime: Local interpretable model explanations for graph neural
  networks.
\newblock {\em arXiv Preprint 2001.06216}, 2020.

\bibitem{sigcomm2014buffer}
Te-Yuan Huang, Ramesh Johari, Nick McKeown, Matthew Trunnell, and Mark Watson.
\newblock A buffer-based approach to rate adaptation: Evidence from a large
  video streaming service.
\newblock In {\em Proc. ACM SIGCOMM}, 2014.

\bibitem{mobisys2017deepmon}
Loc~N Huynh, Youngki Lee, and Rajesh~Krishna Balan.
\newblock Deepmon: Mobile gpu-based deep learning framework for continuous
  vision applications.
\newblock In {\em Proc. ACM MobiSys}, 2017.

\bibitem{icml2019aurora}
Nathan Jay, Noga Rotman, Brighten Godfrey, Michael Schapira, and Aviv Tamar.
\newblock A deep reinforcement learning perspective on internet congestion
  control.
\newblock In {\em Proc. ICML}, 2019.

\bibitem{conext2012festive}
Junchen Jiang, Vyas Sekar, and Hui Zhang.
\newblock Improving fairness, efficiency, and stability in http-based adaptive
  video streaming with festive.
\newblock In {\em Proc. ACM CoNEXT}, 2012.

\bibitem{survey2016udn}
Mahmoud Kamel, Walaa Hamouda, and Amr Youssef.
\newblock Ultra-dense networks: A survey.
\newblock {\em IEEE Communications Surveys \& Tutorials}, 2016.

\bibitem{netai2019verily}
Yafim Kazak, Clark Barrett, Guy Katz, and Michael Schapira.
\newblock Verifying deep-rl-driven systems.
\newblock In {\em Proc. ACM NetAI}, 2019.

\bibitem{iclr2014vae}
Diederik~P Kingma and Max Welling.
\newblock Auto-encoding variational bayes.
\newblock In {\em Proc. ICLR}, 2014.

\bibitem{arxiv201808dq}
Sanjay Krishnan, Zongheng Yang, Ken Goldberg, Joseph Hellerstein, and Ion
  Stoica.
\newblock Learning to optimize join queries with deep reinforcement learning.
\newblock {\em arXiv:1808.03196}, 2018.

\bibitem{kldivergence1951}
S.~Kullback and R.~A. Leibler.
\newblock On information and sufficiency.
\newblock {\em The Annals of Mathematical Statistics}, 22(1):79--86, 03 1951.

\bibitem{dl_nature}
Yann LeCun, Yoshua Bengio, and Geoffrey Hinton.
\newblock Deep learning.
\newblock {\em Nature}, 521(7553):436, 2015.

\bibitem{nn_universal}
Moshe Leshno, Vladimir~Ya Lin, Allan Pinkus, and Shimon Schocken.
\newblock Multilayer feedforward networks with a nonpolynomial activation
  function can approximate any function.
\newblock {\em Neural networks}, 6(6):861--867, 1993.

\bibitem{sigcomm2019neurocuts}
Eric Liang, Hang Zhu, Xin Jin, and Ion Stoica.
\newblock Neural packet classification.
\newblock In {\em Proc. ACM SIGCOMM}, 2019.

\bibitem{macqueen1967kmeans}
James MacQueen et~al.
\newblock Some methods for classification and analysis of multivariate
  observations.
\newblock In {\em Proceedings of Berkeley symposium on mathematical statistics
  and probability}, 1967.

\bibitem{kdd2018efdt}
Chaitanya Manapragada, Geoffrey~I Webb, and Mahsa Salehi.
\newblock Extremely fast decision tree.
\newblock In {\em Proc. ACM KDD}, 2018.

\bibitem{abrl}
Hongzi Mao, Shannon Chen, Drew Dimmery, Shaun Singh, Drew Blaisdell, Yuandong
  Tian, Mohammad Alizadeh, and Eytan Bakshy.
\newblock Real-world video adaptation with reinforcement learning.
\newblock In {\em ICML Reinforcement Learning for Real Life Workshop}, 2019.

\bibitem{neurips2019park}
Hongzi Mao, Parimarjan Negi, Akshay Narayan, Hanrui Wang, Jiacheng Yang, Haonan
  Wang, Ryan Marcus, ravichandra addanki, Mehrdad Khani~Shirkoohi, Songtao He,
  Vikram Nathan, Frank Cangialosi, Shaileshh Venkatakrishnan, Wei-Hung Weng,
  Song Han, Tim Kraska, and Mohammad Alizadeh.
\newblock Park: An open platform for learning augmented computer systems.
\newblock In {\em Proc. NeurIPS}, 2019.

\bibitem{sigcomm2017pensieve}
Hongzi Mao, Ravi Netravali, and Mohammad Alizadeh.
\newblock Neural adaptive video streaming with pensieve.
\newblock In {\em Proc. ACM SIGCOMM}, 2017.

\bibitem{sigcomm2019decima}
Hongzi Mao, Malte Schwarzkopf, Shaileshh~Bojja Venkatakrishnan, Zili Meng, and
  Mohammad Alizadeh.
\newblock Learning scheduling algorithms for data processing clusters.
\newblock In {\em Proc. ACM SIGCOMM}, 2019.

\bibitem{icc2018coco}
Zili Meng, Jun Bi, Haiping Wang, Chen Sun, and Hongxin Hu.
\newblock Coco: Compact and optimized consolidation of modularized service
  function chains in nfv.
\newblock In {\em Proc. IEEE ICC}, 2018.

\bibitem{mm2019pitree}
Zili Meng, Jing Chen, Yaning Guo, Chen Sun, Hongxin Hu, and Mingwei Xu.
\newblock Pitree: Practical implementation of abr algorithms using decision
  trees.
\newblock In {\em Proc. ACM MM}, 2019.

\bibitem{mingers1989empirical}
John Mingers.
\newblock An empirical comparison of pruning methods for decision tree
  induction.
\newblock {\em Machine learning}, 4(2):227--243, 1989.

\bibitem{nips2013atari}
Volodymyr Mnih, Koray Kavukcuoglu, David Silver, Alex Graves, Ioannis
  Antonoglou, Daan Wierstra, and Martin Riedmiller.
\newblock Playing atari with deep reinforcement learning.
\newblock In {\em NIPS Deep Learning Workshop}, 2013.

\bibitem{nature2015drl}
Volodymyr Mnih, Koray Kavukcuoglu, David Silver, Andrei~A Rusu, Joel Veness,
  Marc~G Bellemare, Alex Graves, Martin Riedmiller, Andreas~K. Fidjeland, Georg
  Ostrovski, Stig Petersen, Charles Beattie, Amir Sadik, Ioannis Antonoglou,
  Helen King, Dharshan Kumaran, Daan Wierstra, Shane Legg, and Demis Hassabis.
\newblock Human-level control through deep reinforcement learning.
\newblock {\em Nature}, 518(7540):529, 2015.

\bibitem{jnca2013surveyids}
Chirag Modi, Dhiren Patel, Bhavesh Borisaniya, Hiren Patel, Avi Patel, and
  Muttukrishnan Rajarajan.
\newblock A survey of intrusion detection techniques in cloud.
\newblock {\em Elsevier Journal of network and computer applications}, pages
  42--57, 2013.

\bibitem{ijcai2018odt}
Nina Narodytska, Alexey Ignatiev, Filipe Pereira, Joao Marques-Silva, and
  IS~RAS.
\newblock Learning optimal decision trees with sat.
\newblock In {\em Proc. IJCAI}, 2018.

\bibitem{ton2007fir}
Srihari Nelakuditi, Sanghwan Lee, Yinzhe Yu, Zhi-Li Zhang, and Chen-Nee Chuah.
\newblock Fast local rerouting for handling transient link failures.
\newblock {\em IEEE/ACM Transactions on Networking}, pages 359--372, 2007.

\bibitem{nfp4000-too}
Netronome.
\newblock Wihte paper: Nfp-4000 theory of operation.
\newblock \url{https://www.netronome.com/media/documents/WP\_NFP4000\_TOO.pdf},
  2016.

\bibitem{iclr2018sensitivity}
Roman Novak, Yasaman Bahri, Daniel~A Abolafia, Jeffrey Pennington, and Jascha
  Sohl-Dickstein.
\newblock Sensitivity and generalization in neural networks: an empirical
  study.
\newblock In {\em Proc. ICLR}, 2018.

\bibitem{scikit-learn}
Fabian Pedregosa, Ga\"{e}l Varoquaux, Alexandre Gramfort, Vincent Michel,
  Bertrand Thirion, Olivier Grisel, Mathieu Blondel, Peter Prettenhofer, Ron
  Weiss, Vincent Dubourg, Jake Vanderplas, Alexandre Passos, David Cournapeau,
  Matthieu Brucher, Matthieu Perrot, and \'{E}douard Duchesnay.
\newblock Scikit-learn: Machine learning in {P}ython.
\newblock {\em Journal of Machine Learning Research}, 2011.

\bibitem{kdd2016lime}
Marco~Tulio Ribeiro, Sameer Singh, and Carlos Guestrin.
\newblock Why should i trust you?: Explaining the predictions of any
  classifier.
\newblock In {\em Proc. ACM KDD}, 2016.

\bibitem{acl2018semantically}
Marco~Tulio Ribeiro, Sameer Singh, and Carlos Guestrin.
\newblock Semantically equivalent adversarial rules for debugging nlp models.
\newblock In {\em Proc. ACL}, 2018.

\bibitem{mmsys2013norway}
Haakon Riiser, Paul Vigmostad, Carsten Griwodz, and P{\aa}l Halvorsen.
\newblock Commute path bandwidth traces from 3g networks: Analysis and
  applications.
\newblock In {\em Proc. ACM MMSys}, 2013.

\bibitem{aistats2011dagger}
St{\'e}phane Ross, Geoffrey Gordon, and Drew Bagnell.
\newblock A reduction of imitation learning and structured prediction to
  no-regret online learning.
\newblock In {\em Proc. AISTATS}, 2011.

\bibitem{sosr2019routenet}
Krzysztof Rusek, Jos{\'e} Su{\'a}rez-Varela, Albert Mestres, Pere Barlet-Ros,
  and Albert Cabellos-Aparicio.
\newblock Unveiling the potential of graph neural networks for network modeling
  and optimization in sdn.
\newblock In {\em Proc. ACM SOSR}, 2019.

\bibitem{shannon1948entropy}
Claude~Elwood Shannon.
\newblock A mathematical theory of communication.
\newblock {\em Bell System Technical Journal}, 27(3), 1948.

\bibitem{sysml2019tfjs}
Daniel Smilkov, Nikhil Thorat, Yannick Assogba, Ann Yuan, Nick Kreeger, Ping
  Yu, Kangyi Zhang, Shanqing Cai, Eric Nielsen, David Soergel, Stan Bileschi,
  Michael Terry, Charles Nicholson, Sandeep~N. Gupta, Sarah Sirajuddin,
  D.~Sculley, Rajat Monga, Greg Corrado, Fernanda~B. Viégas, and Martin
  Wattenberg.
\newblock Tensorflow. js: Machine learning for the web and beyond.
\newblock In {\em Proc. SysML}, 2019.

\bibitem{mmsys2018sabre}
Kevin Spiteri, Ramesh Sitaraman, and Daniel Sparacio.
\newblock From theory to practice: improving bitrate adaptation in the dash
  reference player.
\newblock In {\em Proc. ACM MMSys}, 2018.

\bibitem{infocom2016bola}
Kevin Spiteri, Rahul Urgaonkar, and Ramesh~K Sitaraman.
\newblock Bola: Near-optimal bitrate adaptation for online videos.
\newblock In {\em Proc. IEEE INFOCOM}, 2016.

\bibitem{sutton2018reinforcement}
Richard~S Sutton and Andrew~G Barto.
\newblock {\em Reinforcement Learning (Second Edition): An Introduction}.
\newblock MIT press, 2018.

\bibitem{neurips2019interpretnlp}
Mariya Toneva and Leila Wehbe.
\newblock Interpreting and improving natural-language processing (in machines)
  with natural language-processing (in the brain).
\newblock In {\em Proc. NeurIPS}, 2019.

\bibitem{venables2002tree}
William~N Venables and Brian~D Ripley.
\newblock Tree-based methods.
\newblock In {\em Modern Applied Statistics with S}, pages 251--269. Springer,
  2002.

\bibitem{icml2018pirl}
Abhinav Verma, Vijayaraghavan Murali, Rishabh Singh, Pushmeet Kohli, and Swarat
  Chaudhuri.
\newblock Programmatically interpretable reinforcement learning.
\newblock In {\em Proc. ICML}, 2018.

\bibitem{facebook_video}
Kurt Wagner.
\newblock Facebook says video is huge -- 100-million-hours-per-day huge.
\newblock \url{https://www.vox.com/2016/1/27/11589140/}, 2016.

\bibitem{aaai2018treereg}
Mike Wu, Michael~C Hughes, Sonali Parbhoo, Maurizio Zazzi, Volker Roth, and
  Finale Doshi-Velez.
\newblock Beyond sparsity: Tree regularization of deep models for
  interpretability.
\newblock In {\em Proc. AAAI}, 2018.

\bibitem{iwqos2019nfvdeep}
Yikai Xiao, Qixia Zhang, Fangming Liu, Jia Wang, Miao Zhao, Zhongxing Zhang,
  and Jiaxing Zhang.
\newblock Nfvdeep: Adaptive online service function chain deployment with deep
  reinforcement learning.
\newblock In {\em Proc. IEEE/ACM IWQoS}, 2019.

\bibitem{neurips2019hypergcn}
Naganand Yadati, Madhav Nimishakavi, Prateek Yadav, Vikram Nitin, Anand Louis,
  and Partha Talukdar.
\newblock Hypergcn: A new method for training graph convolutional networks on
  hypergraphs.
\newblock In {\em Proc. NeurIPS}, 2019.

\bibitem{osdi2018nas}
Hyunho Yeo, Youngmok Jung, Jaehong Kim, Jinwoo Shin, and Dongsu Han.
\newblock Neural adaptive content-aware internet video delivery.
\newblock In {\em Proc. USENIX OSDI}, 2018.

\bibitem{access2017rnnids}
Chuanlong Yin, Yuefei Zhu, Jinlong Fei, and Xinzheng He.
\newblock A deep learning approach for intrusion detection using recurrent
  neural networks.
\newblock {\em IEEE Access}, pages 21954--21961, 2017.

\bibitem{sigcomm2015robustmpc}
Xiaoqi Yin, Abhishek Jindal, Vyas Sekar, and Bruno Sinopoli.
\newblock A control-theoretic approach for dynamic adaptive video streaming
  over http.
\newblock In {\em Proc. ACM SIGCOMM}, 2015.

\bibitem{neurips2019gnnexplainer}
Zhitao Ying, Dylan Bourgeois, Jiaxuan You, Marinka Zitnik, and Jure Leskovec.
\newblock Gnn explainer: A tool for post-hoc explanation of graph neural
  networks.
\newblock In {\em Proc. NeurIPS}, 2019.

\bibitem{iclr2020lotteryrl}
Haonan Yu, Sergey Edunov, Yuandong Tian, and Ari~S Morcos.
\newblock Playing the lottery with rewards and multiple languages: lottery
  tickets in rl and nlp.
\newblock In {\em Proc. ICLR}, 2020.

\bibitem{eccv2014visualizing}
Matthew~D Zeiler and Rob Fergus.
\newblock Visualizing and understanding convolutional networks.
\newblock In {\em Proc. ECCV}, 2014.

\bibitem{infocom2005optimal}
Chun Zhang, Yong Liu, Weibo Gong, Jim Kurose, Robert Moll, and Don Towsley.
\newblock On optimal routing with multiple traffic matrices.
\newblock In {\em Proc. IEEE INFOCOM}, 2005.

\bibitem{mcom2017hudn}
Hongliang Zhang, Lingyang Song, Yonghui Li, and Geoffrey~Ye Li.
\newblock Hypergraph theory: Applications in 5g heterogeneous ultra-dense
  networks.
\newblock {\em IEEE Communications Magazine}, 55(12):70--76, 2017.

\bibitem{icnp2019lego}
Menghao Zhang, Jiasong Bai, Guanyu Li, Zili Meng, Hongda Li, Hongxin Hu, and
  Mingwei Xu.
\newblock When nfv meets ann: Rethinking elastic scaling for ann-based nfs.
\newblock In {\em Proc. IEEE ICNP}, 2019.

\bibitem{cvpr2018interpretablecnn}
Quanshi Zhang, Ying~Nian Wu, and Song-Chun Zhu.
\newblock Interpretable convolutional neural networks.
\newblock In {\em Proc. IEEE CVPR}, 2018.

\bibitem{iclr2020hypersagnn}
Ruochi Zhang, Yuesong Zou, and Jian Ma.
\newblock Hyper-sagnn: a self-attention based graph neural network for
  hypergraphs.
\newblock In {\em Proc. ICLR}, 2020.

\bibitem{sec2020adv2}
Xinyang Zhang, Ningfei Wang, Shouling Ji, Hua Shen, and Ting Wang.
\newblock Interpretable deep learning under fire.
\newblock In {\em Proc. USENIX Security}, 2020.

\bibitem{apnet2018demystifying}
Ying Zheng, Ziyu Liu, Xinyu You, Yuedong Xu, and Junchen Jiang.
\newblock Demystifying deep learning in networking.
\newblock In {\em Proc. ACM APNet}, 2018.

\bibitem{nips2007hypergraph}
Dengyong Zhou, Jiayuan Huang, and Bernhard Sch{\"o}lkopf.
\newblock Learning with hypergraphs: Clustering, classification, and embedding.
\newblock In {\em Proc. NIPS}, 2007.

\bibitem{pldi2019verifiablerl}
He~Zhu, Zikang Xiong, Stephen Magill, and Suresh Jagannathan.
\newblock An inductive synthesis framework for verifiable reinforcement
  learning.
\newblock In {\em Proc. ACM PLDI}, 2019.

\end{thebibliography}

\appendix

\section*{Appendices}

Appendices are supporting material that has not been peer-reviewed.

\section{Resampling in Decision Tree Training}
\label{sec:rl}

\begin{figure}[h]
    \centering
    \includegraphics[width=7cm]{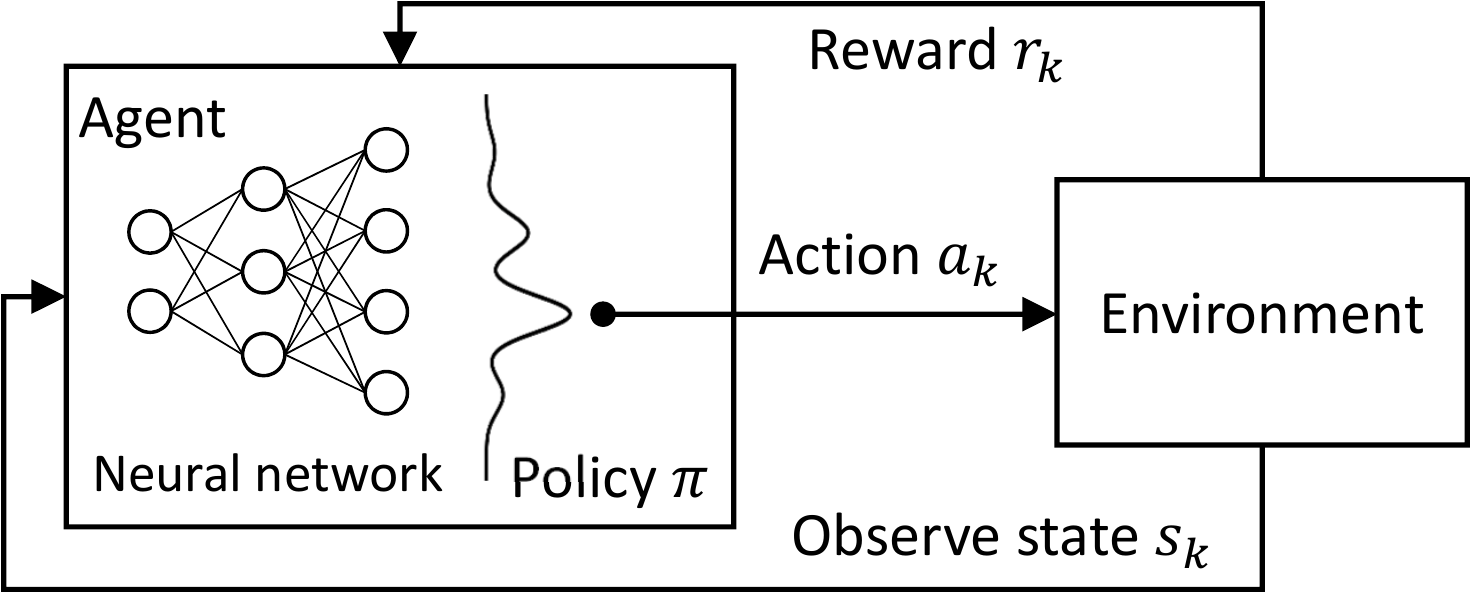}
    \caption{RL with neural networks as policy.}
    \label{fig:rl}
\end{figure}

To explain the resampling equation during decision tree training (Equation~\ref{eq:prob}), we first briefly introduce the basic knowledge about RL used in this paper. We refer the readers to~\cite{sutton2018reinforcement} for a more comprehensive understanding of RL.

In RL, at each iteration $t$, the \textit{agent} (e.g., a flow scheduler~\cite{sigcomm2018auto}) first observers a \textit{state} $s_t\in\mathcal{S}$ (e.g., remaining flow sizes) from the surrounding \textit{environment}. The agent then takes an \textit{action} $a_t\in\mathcal{A}$ (e.g., scheduling a flow to a certain port) according to its \textit{policy} $\pi$ (e.g., shortest flow first). The environment then returns a \textit{reward} $r_t$ (e.g., FCTs of finished flows) and updates its state to $s_{t+1}$. Reward is used to indicate how good is the current decision. The goal is to learn a policy $\pi$ to optimize \textit{accumulated future discounted reward} $\mathbb{E}\left[\sum_t \gamma^t r_t\right]$ with the discounting factor $\gamma\in(0,1]$. $\pi_\theta(s,a)$ is the probability of taking action $a$ at state $s$ with policy $\pi_\theta$ parameterized by $\theta$, which is usually represented with DNNs to solve large-scale practical problems~\cite{nips2013atari, nature2015drl}. An illustration of RL is presented in Figure~\ref{fig:rl}.

However, it is not easy for the agent to find out the actual reward of a state or an action in the training process since the reward is usually \textit{delayed}. For example, the FCT can only be observed after the flow is completed. Therefore, we need to estimate the potential \textit{value} of a state.\textit{Value function} $V_t^{(\pi)}(s)$ is introduced to determine the potential future reward of a state $s$ at the time $t$ with the policy $\pi$:
\begin{equation}
    \label{eq:value}
    V^{(\pi)}(s) = R(s) + \sum_{s'\in \mathcal{S}} p\left(s' | s, \pi(s)\right)V^{(\pi)}(s')
\end{equation}
where $p(s'|s,a)$ is the transition probability onto state $s'$ given state $s$ and subsequent action $a$. Similarly, \textit{$Q$-function} $Q_t^{(\pi)}(s,a)$ is to estimate the value of how a certain action $a$ at state $s$ may contribute to the future reward:
\begin{equation}
    \label{eq:q-func}
    Q^{(\pi)}(s,a) = R(s) + \sum_{s'\in \mathcal{S}} p\left(s' | s, a\right)V^{(\pi)}(s')
\end{equation}
Therefore, a good action $a$ at the state $s$ would maximize the difference between the value function and $Q$-function, i.e., the optimization loss $\ell (s, \pi)$ of RL could be written as:
\begin{equation}
\label{eq:rl-loss}
\ell(s,\pi) = V^{(\pi)}(s) - Q^{(\pi)}(s,a)
\end{equation}

In the teacher-student learning optimization in \S\ref{sec:dt}, to make the loss independent of $\pi$ and therefore easy to optimize, Bastani et al.~\cite{nips2018viper} bounded the loss above with:
\begin{equation}
\label{eq:bounded-loss}
\tilde{\ell}(s) =V^{\left(\pi\right)}(s) - \min_{a'\in A} Q^{\left(\pi\right)}(s,a') \geqslant  V^{(\pi)}(s) - Q^{(\pi)}(s,a)
\end{equation}
Therefore, we can resample the (state, action) pairs with the loss function above, which explains the sampling probability in Equation~\ref{eq:prob}. The sampling probability $p(s,a)$ in Equation~\ref{eq:prob} is proportional to but not equal to the loss function due to the normalization of probability.

\begin{figure}
	\centering
	\includegraphics[width=5.5cm]{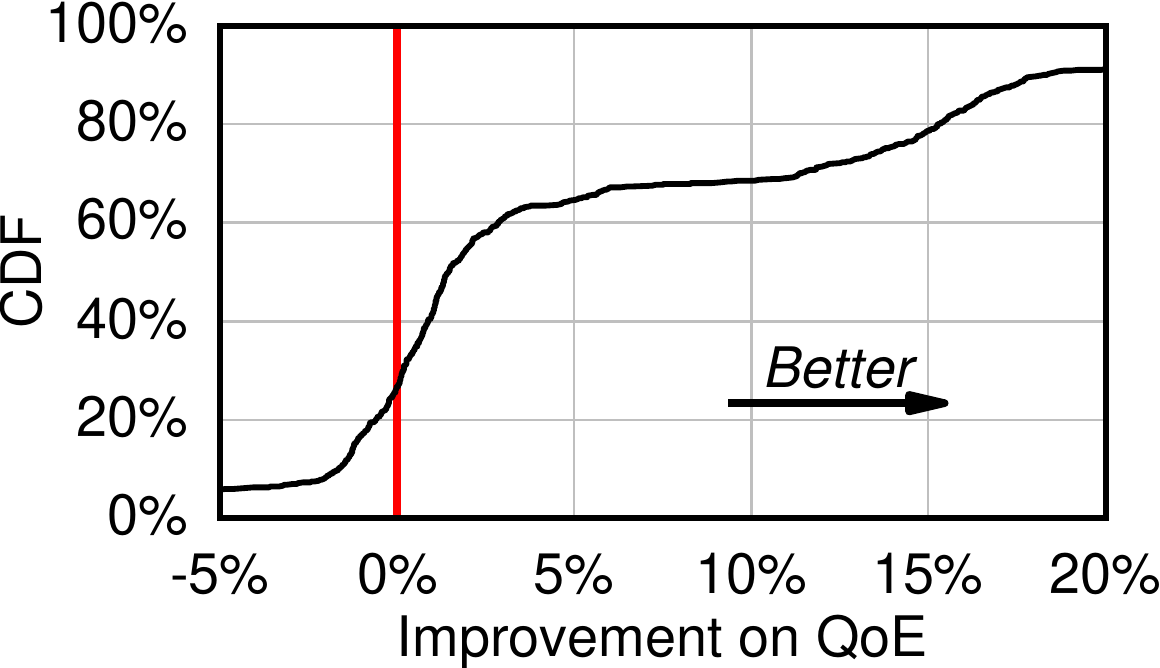}
	\caption{The resampling step could improve the QoE of 73\% of the traces, with the median improvement of 1.5\%.}
	\label{fig:improve-resample}
\end{figure}

We further empirically evaluate the improvement on QoE of the resampling step. We measure the QoE of the decision trees with and without the resampling step. As shown in Figure~\ref{fig:improve-resample}, 73\% of traces could benefit from the resampling step with different degrees of improvement. The median improvement on QoE over all traces is 1.5\%. Since the resampling step is adopted for the last mile performance improvement, network operators may choose to skip the step if performance is not a critical issue for them.

\section{Hypergraph Formulations}
\label{sec:models}

We present several other formulations of different application scenarios presented in Table~\ref{tab:hypergraph} (\S\ref{sec:app}).

\subsection{Network Function Placement System}
\label{sec:nfv}

\begin{figure}
	\centering
	\subfigure[Original representation.]{
		\includegraphics[height=2.7cm]{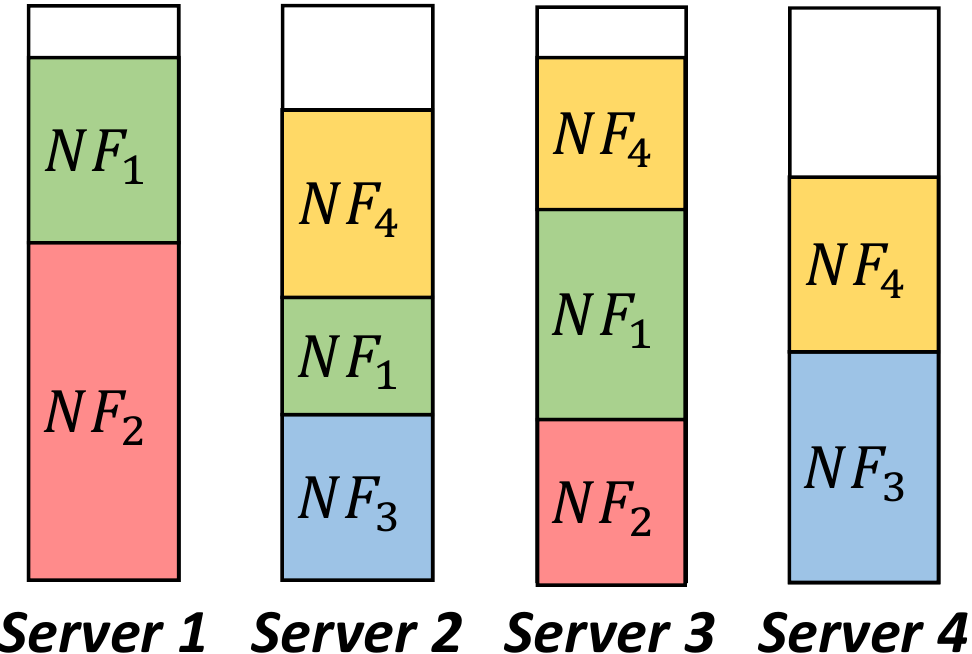}
		\label{fig:nfv-original}
	}
	\subfigure[Hypergraph representation.]{
		\includegraphics[height=2.7cm]{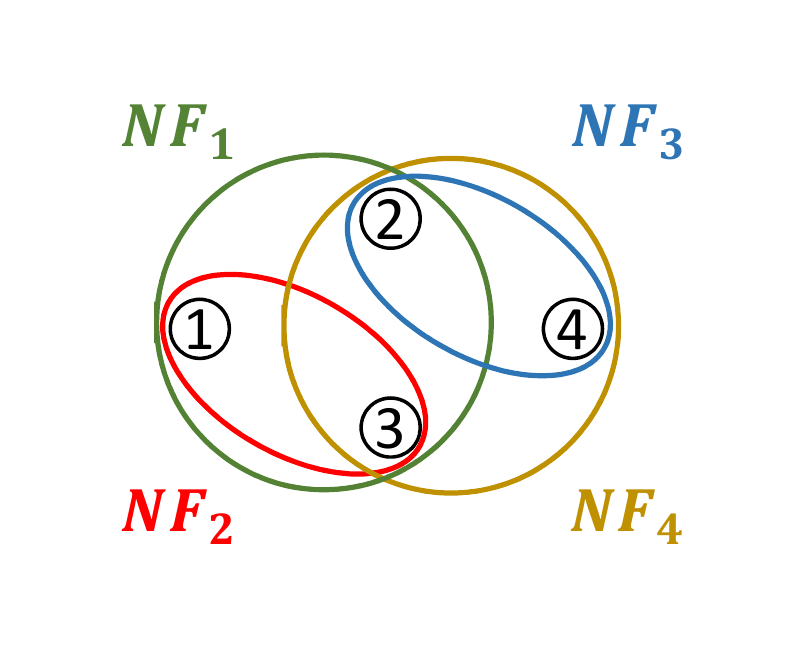}
		\label{fig:nfv-hypergraph}
	}
	\caption{Network function virtualization.}
	\label{fig:nfv}
\end{figure}

Network function virtualization (NFV) is widely adopted to replace dedicated hardware with virtualized network functions (VNFs). Considering the processing ability and fluctuating network demand, network operators can replicate one VNF onto several instances on different servers (devices), and consolidate multiple VNFs onto one server (device). A key problem for network operators is to study where to place their VNF instances~\cite{iwqos2019nfvdeep, icc2018coco}. Traditional methods include different heuristics and integer linear programming (ILP). Our observation is that the consolidation and placement problem in NFV could also be modeled with a hypergraph, with servers as hyperedges and VNF as vertices. An illustration of the hypergraph formulation of NFV placement is presented in Figure~\ref{fig:nfv}. Hyperedge $e$ contains with vertex $v$ indicates that VNF $v$ has an instance placed onto server $e$. Hyperedge features $F_E$ could be the processing capacity of servers, and vertex features $F_V$ could be the processing speed of different types of VNF. \name\ will then interpret the placement results by finding the critical NF instance placement and checking if it is reasonable.

\subsection{Ultra-Dense Cellular Network}
\label{sec:coverage}
\begin{figure}
\centering
\vspace{0pt}
\begin{minipage}[t]{.65\linewidth}
	\centering
	\includegraphics[width=.9\linewidth]{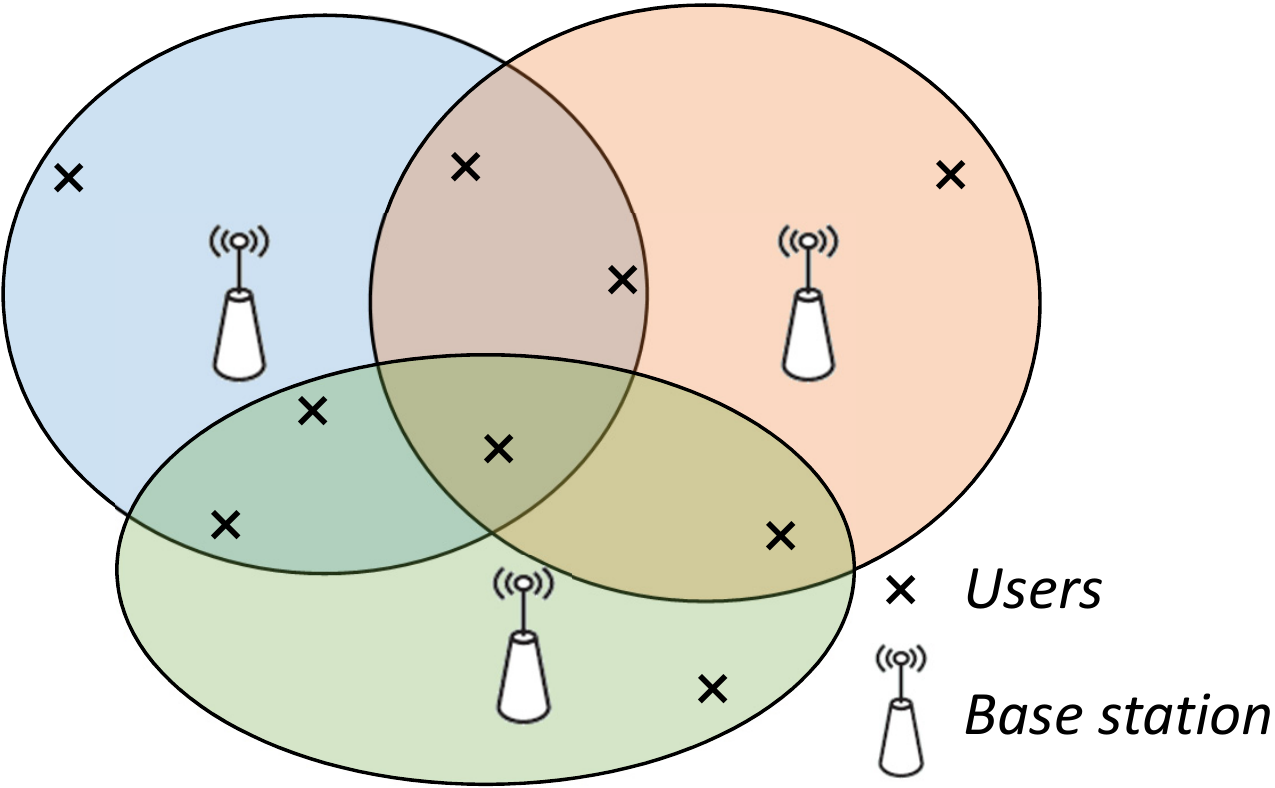}
	\caption{Ultra-dense network.}
	\label{fig:udn}
\end{minipage}
\begin{minipage}[t]{.33\linewidth}
    \includegraphics[width=.9\linewidth]{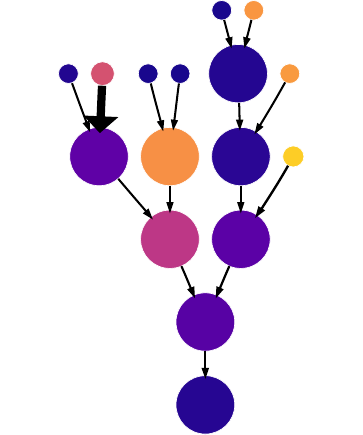}
    \caption{Cluster scheduling jobs.}
    \label{fig:graph-spark}
    \end{minipage}
\end{figure}

In the 5G mobile scenario, one mobile user is usually connected to multiple picocell base stations, which is known as \textit{ultra-dense networking}~\cite{survey2016udn}. Network operators need to decide which base station to connect for each user based on users' traffic demand and base stations' transmission capacity. The scenario could be formulated as a hypergraph~\cite{mcom2017hudn}, with the coverage of the picocell base stations as hyperedges, and mobile users as vertices.

We present an illustration of the hypergraph formulation in Figure~\ref{fig:udn}. The coverage of each base station is shaded with different colors. Hyperedge features $F_E$ could be the capacity of each base station, etc. Vertex features $F_V$ could be the traffic demand of each mobile user. The traffic optimizer will then continuously select the best base station(s) to connect for each mobile user according to the users' locations. \name\ could consequently interpret the system by providing insights on which user-base station connection is critical to the overall performance. For example, connecting a user with high demands to a nearby base station indicates that the system might mimic a nearest-first policy.

\subsection{Cluster Scheduling System}
\label{sec:scheduling}

In cluster scheduling scenarios (e.g., Spark~\cite{sigcomm2019decima}), a job consists of a directed acyclic graph (DAG) whose nodes are the execution stages of the job, as shown in Figure~\ref{fig:graph-spark}. A node's task cannot be executed until the tasks from all its parents have finished. The scheduler needs to decide how to allocate limited resources to different jobs. Since the jobs to schedule are usually represented as dataflow graphs~\cite{sigcomm2019decima}, we can naturally formulate the cluster scheduling scenario with \name. In this case, each vertex represents a set of parallel operations, and each edge represents the dependency between vertices. Vertex features $F_V$ are the work of nodes, and hyperedge features $F_E$ are the data transmission between nodes~\cite{sigcomm2019decima}. \name\ can interpret the scheduling system by finding out which scheduling decisions (allocating a specific job node to a certain number of executors) are significant to the performance.

\section{Implementation Details}
\label{sec:app-param}

\begin{table}
	\centering
	\small
	\begin{tabular}{lll}
		\hline
		Pensieve & $M$ & 200\\
		\hline
		AuTO & $M$ (lRLA) & 2000\\
		 & $M$ (sRLA) & 2000\\
		\hline
		RouteNet* & $\lambda_1$ & 0.25\\
		 & $\lambda_2$ & 1\\
		\hline
	\end{tabular}
	\caption{\name\ hyperparameters.}
	\label{tab:param}
\end{table}

We introduce the parameter settings of \name\ and three DL-based networking systems, together with the details of our testbed, in this section.

\parahead{Parameter settings.} We present the hyperparameter settings of \name\ in Table~\ref{tab:param}. For the DNN in Pensieve, we set the number of leaf nodes ($M$) to 200. Our experiments on the sensitivity of $M$ in Appendix~\ref{sec:app-scs-sensitivity} shows that a wide range of $M$ perform well. For two DNNs in AuTO (lRLA and sRLA), we set the number of leaf nodes to 2000. This is because the state spaces of lRLA (143 states) and sRLA (700 states) are much larger than that of Pensieve (25 states).

For the hypergraph-based interpretation method, network operators can set the hyperparameters $\lambda_1$ and $\lambda_2$ based on their ability to understand the interpreted structure and application scenarios, as discussed in \S\ref{sec:action}. For example, with the settings in Table~\ref{tab:param} results, only 10\% of connections of RouteNet* have mask values greater than 0.8. Further improving $\lambda_1$ will increase the ratio of connections with high mask values and expose more critical connections to network operators. We present the details of sensitivity analysis of $\lambda_1$ and $\lambda_2$ in Appendix~\ref{sec:app-gcs-sensitivity}. 

Note that the five hyperparameters in Table~\ref{tab:param} are the hyperparameters for three systems in total. In practice, network operators only need to set one or two to employ \name\ on their own DL-based networking system.

\parahead{Testbed details.} We train the decision tree with \texttt{sklearn}~\cite{scikit-learn} and modify it to support the CCP. The server for Pensieve and RouteNet* is equipped with an Intel Core i7-8700 CPU (6 physical cores) and an Nvidia Titan Xp GPU. The switchs used in AuTO are two H3C-S6300 48-port switches. 




\section{Pensieve Debugging Deep Dive}
\label{sec:missing}

We also provide more details on the experiments of two links with bandwidth fixed to 3000kbps and 1300kbps in \S\ref{sec:case-troubleshoot}.

\begin{figure}
	\centering
	\includegraphics[width=7cm]{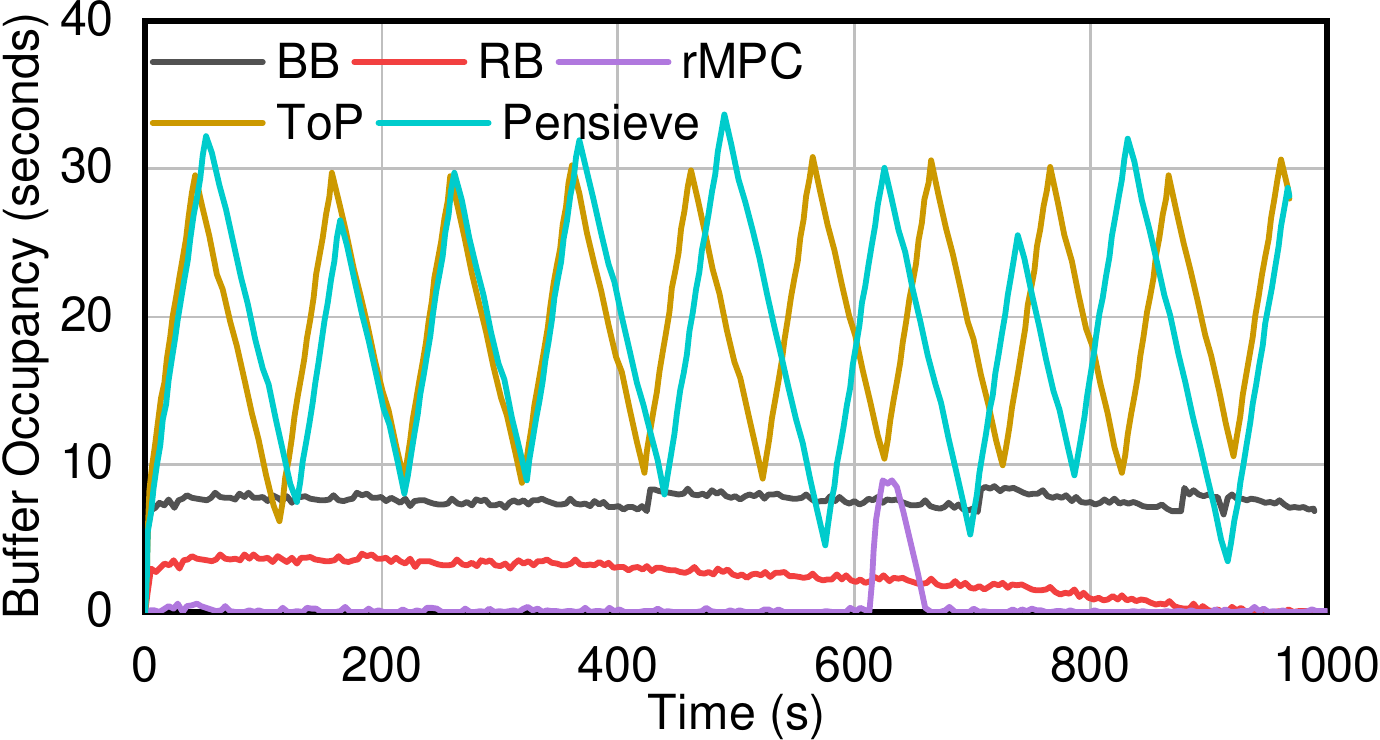}
	\caption{Buffer Occupancy at 3000kbps Link.}
	\label{fig:buffer}
\end{figure}

\begin{figure}
	\centering
	\includegraphics[width=7cm]{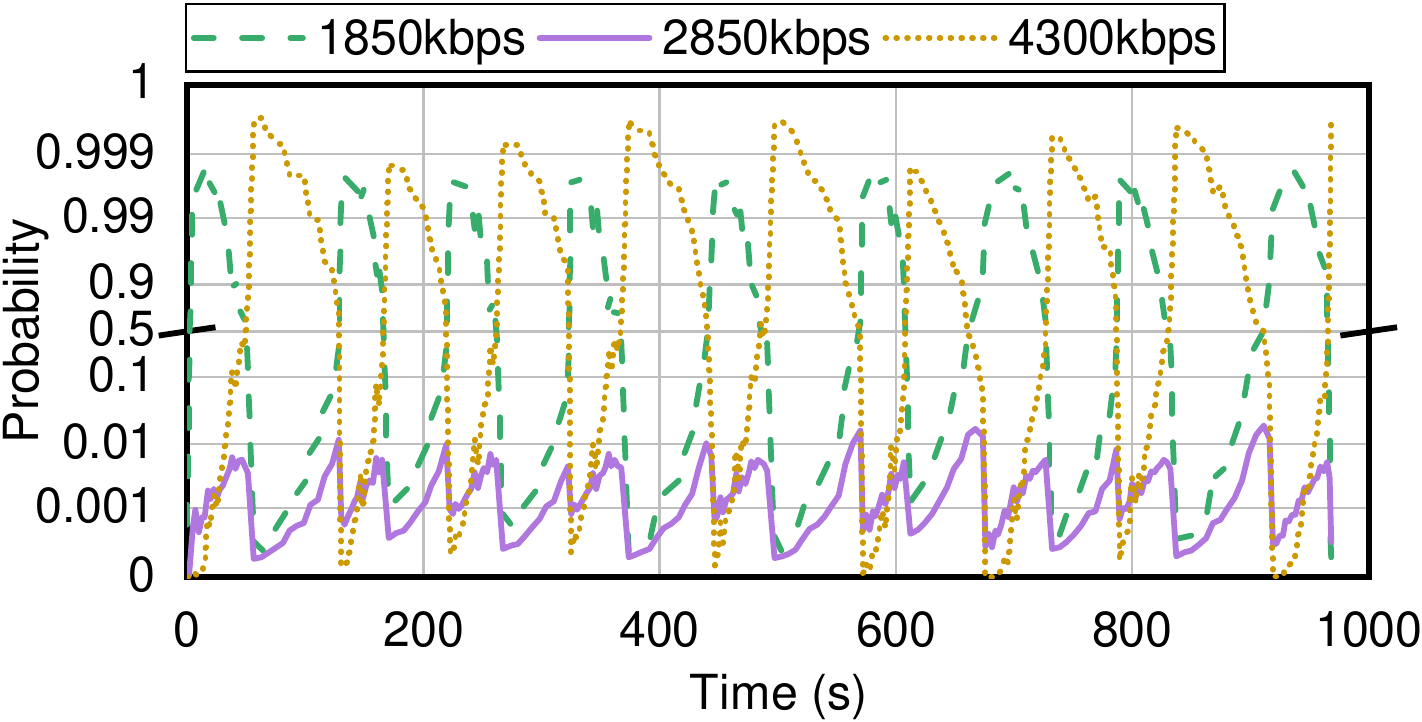}
	\caption{Probabilities of selecting 1850kbps, 2850kbps, 4300kbps qualities. The probability of selecting other three qualities is less than $10^{-4}$ thus not presented.}
	\label{fig:confidence}
\end{figure}

\begin{figure}
	\centering
	\subfigure[Bitrate]{
		\includegraphics[width=7cm]{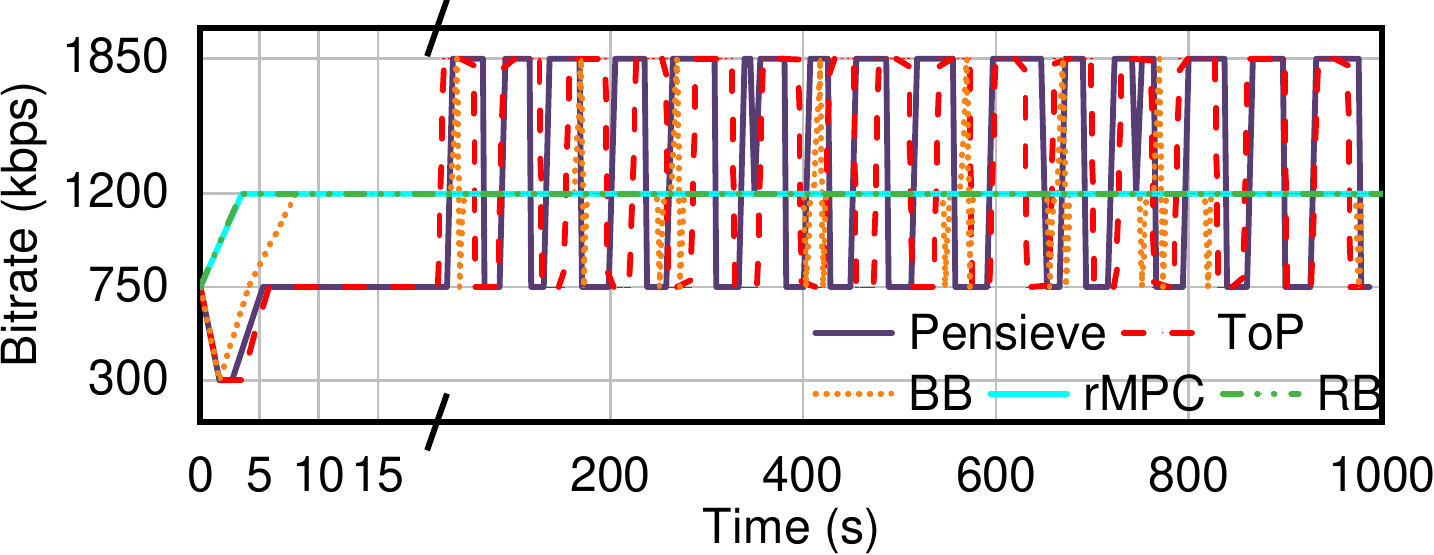}
		\label{fig:bw-variation-1333kbps}
	}
	\subfigure[Buffer Occupancy]{
		\includegraphics[width=7cm]{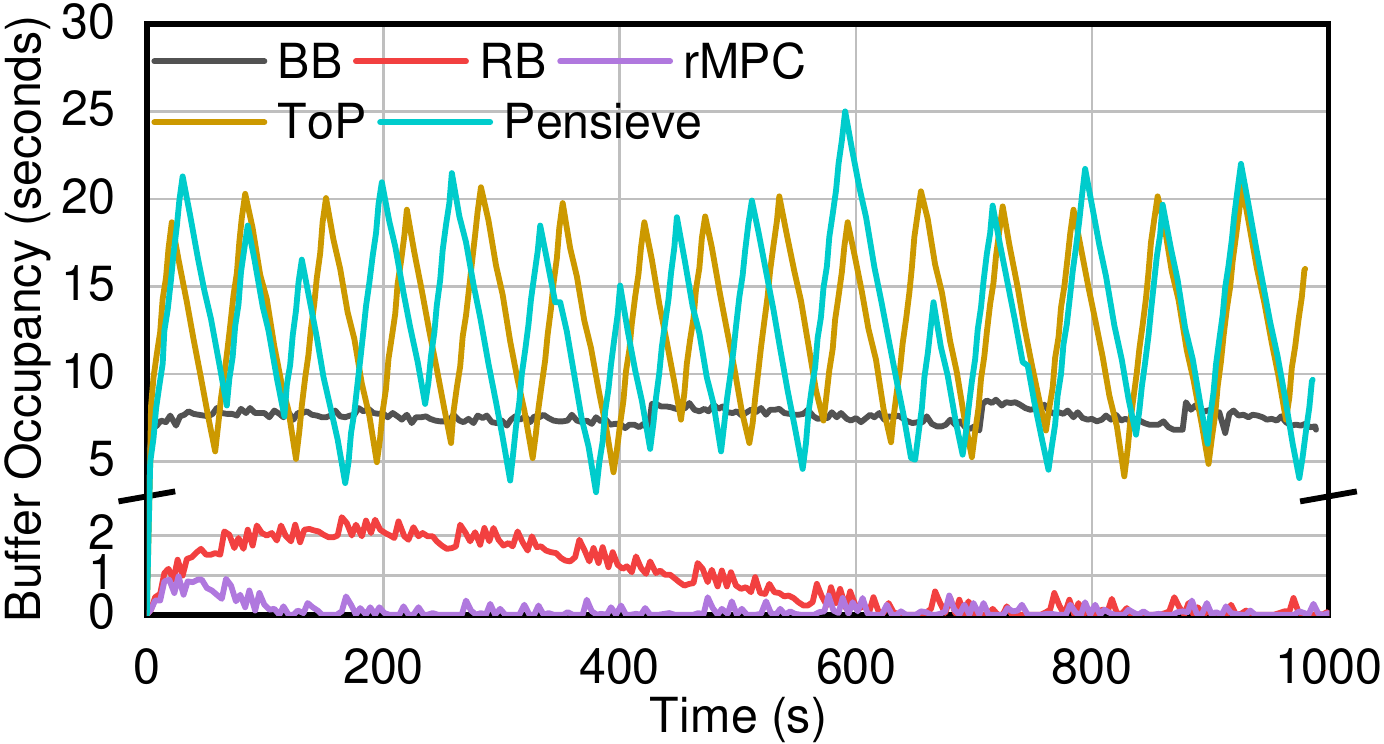}
		\label{fig:buffer-1333kbps}
	}
	\caption{Results on a 1300kbps link. Better viewed in color.}
	\label{fig:1333kbps}
\end{figure}

\begin{figure*}
	\centering
	\subfigure[Accruracy (Pensieve).]{
		\includegraphics[height=3.1cm]{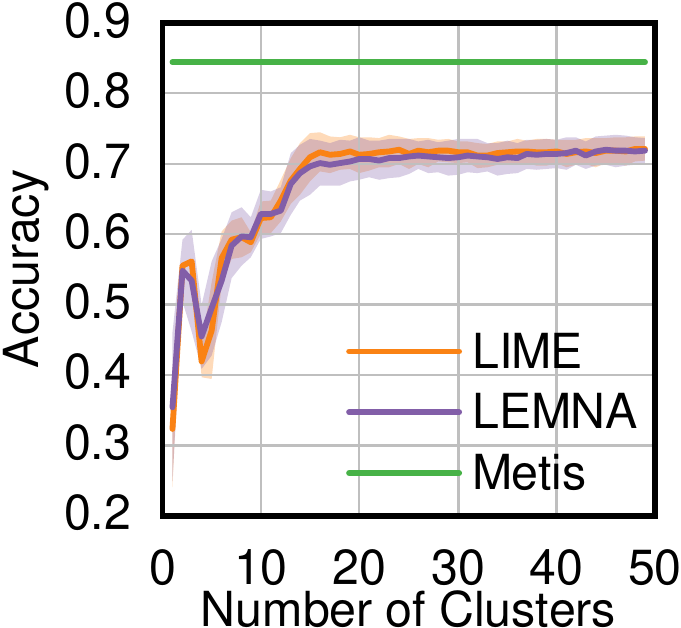}
		\label{fig:accuracy-pensieve}
	}
	\subfigure[RMSE (Pensieve).]{
		\includegraphics[height=3.1cm]{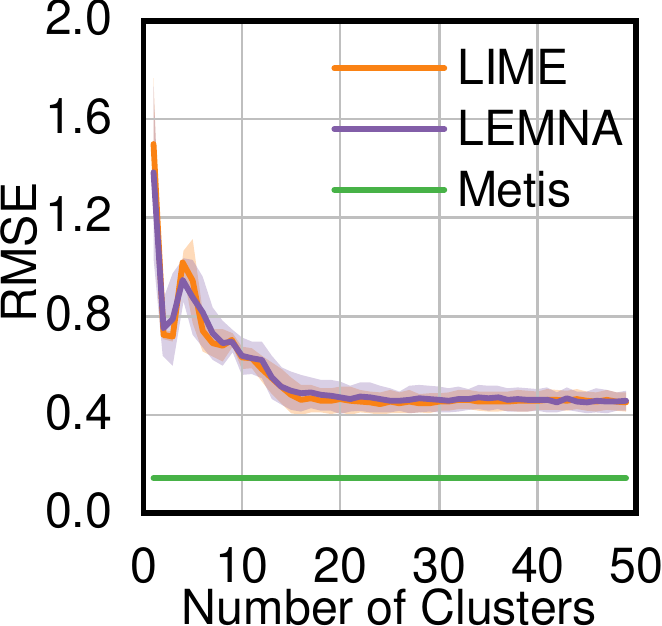}
		\label{fig:rmse-pensieve}
	}
	\subfigure[Accruracy (AuTO-lRLA).]{
		\includegraphics[height=3.1cm]{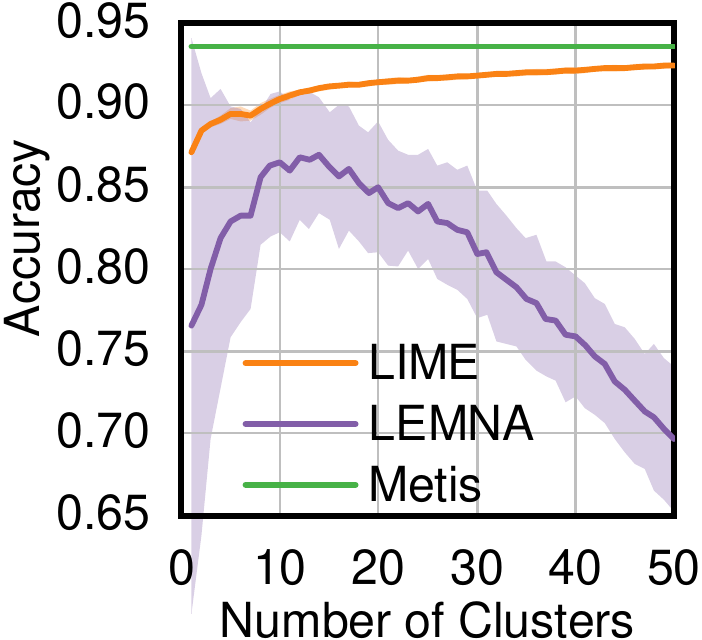}
		\label{fig:accuracy-toa-lrla}
	}
	\subfigure[RMSE (AuTO-lRLA).]{
		\includegraphics[height=3.1cm]{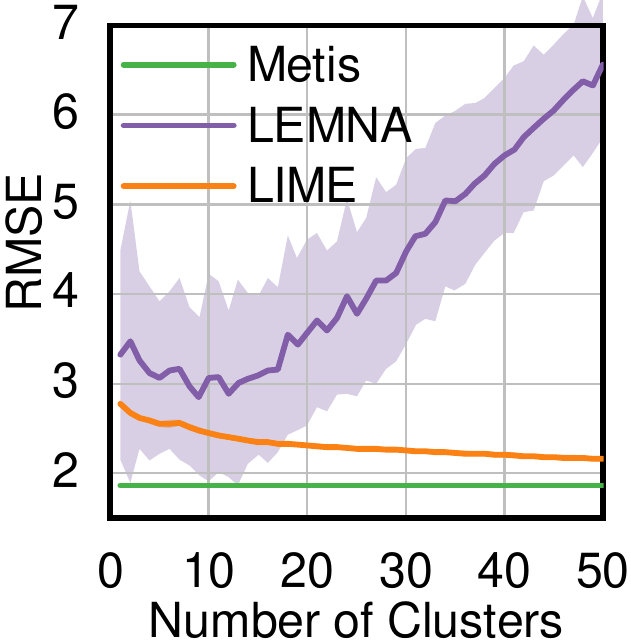}
		\label{fig:rmse-toa-lrla}
	}
	\subfigure[RMSE (AuTO-sRLA).]{
		\includegraphics[height=3.1cm]{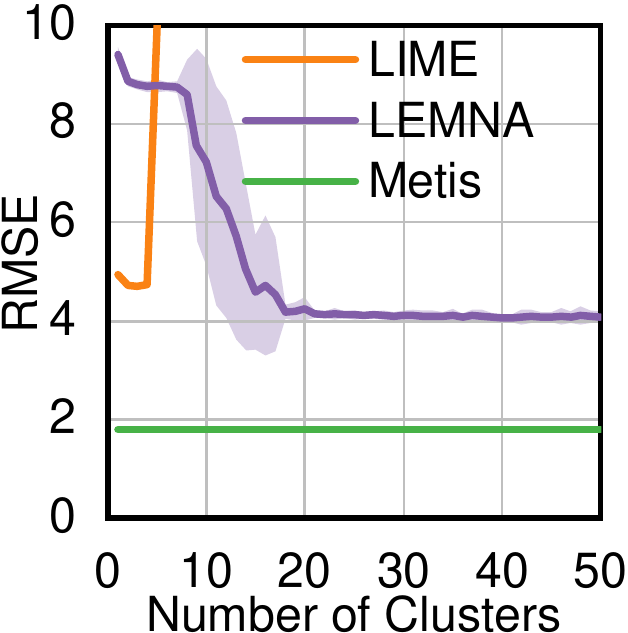}
		\label{fig:rmse-toa-srla}
	}
	\caption{Faithfulness of \name. Shaded area spans $\pm$ std. AuTO-sRLA predicts real values thus does not have accuracy. Results of LIME over sRLA diverges with $\geqslant$5 clusters. Higher accuracy and lower RMSE indicate a better performance. Better viewed in color.}
	\label{fig:faithfulness-pensieve}
\end{figure*}

\parahead{3000kbps link.} Except for the experiments in \S\ref{sec:case-troubleshoot}, we also investigate the runtime buffer occupancy over the 3000kbps link. As shown in Figure~\ref{fig:buffer}, the buffer occupancy of Pensieve fluctuates: buffer increases when 1850kbps is selected and decreases when 4300kbps is selected, which is also faithfully mimicked by \name+Pensieve. The oscillation leads to a drastic smoothness penalty. Meanwhile, the buffer occupancy can also interpret the poor performance of rMPC in Figure~\ref{fig:bw-variation}: rMPC converges at the beginning. Thus, there is no enough buffer against the fluctuation of chunk size since the size of each video chunk is not the same. Thus a substantial rebuffer penalty is imposed on rMPC. The buffer of BB and RB decreases slightly during the total 1000 seconds experiment as the goodput is not exactly 2850kbps (the average bitrate of sample video).

As the raw outputs of the DNNs in Pensieve are the normalized probabilities of selecting each action, we further investigate those probabilities of Pensieve on the 3000kbps link and present the results in Figure~\ref{fig:confidence}. A higher probability close to 1 indicates higher confidence in the decision. We can see that Pensieve does not have enough confidence in the decision it made, which suggests that Pensieve might not experience similar conditions in training; thus, it does not know how to make a decision.

\begin{table}[h]
	\centering
	\small
	\begin{tabular}{ccccc}
		\hline
		BB    & RB    & rMPC  & \name+Pensieve   & Pensieve  \\
		\hline
		1.050 & 0.904 & 0.803 & 0.986 & 0.983     \\
		\hline 
	\end{tabular}
	\caption{QoE on the 1300kbps link.}
	\label{tab:qoe-1333}
\end{table}

\parahead{1300kbps link.} We also provide the details about the experiments in Figure~\ref{fig:freq-bw} on a 1300kbps link and present the results in Figure~\ref{fig:1333kbps} and Table~\ref{tab:qoe-1333}. The results are similar to the 3000kbps experiment, except that the performance of RB is worse since it converges faster.

\section{Interpretation Baseline Comparison}
\label{sec:app-faith}

We further want to know the reason for the performance maintenance of \name. We measure the accuracy and root-mean-square error (RMSE) of the decisions made by \name\ compared to the original decisions made by DNNs. As baselines, we compare the faithfulness of \name\ over three DNNs (\name+Pensieve, \name+AuTO-lRLA, \name+AuTO-sRLA) with two recent interpretation methods:
\begin{itemize}
	\item LIME~\cite{kdd2016lime} is one of the most widely used blackbox interpretation method in the machine learning community
	. LIME interprets the blackbox model with the linear regression of the inputs and outputs.
	\item LEMNA~\cite{ccs2018lemna} is an interpretation method proposed in 2018 and designed to interpret DL models based on time-series inputs (e.g., RNN)
	. LEMNA employs a mixture regression to handle the dependency between inputs. We employ LEMNA as a baseline since some networking systems also handle time-series inputs.
\end{itemize}
As both methods are designed based on regressions around a certain sample, to make a fair comparison, we run the baselines in the following way: At the training stage, we first use $k$-means clustering~\cite{macqueen1967kmeans} to cluster the input-output samples of the DL-based networking system into $k$ groups. We then interpret the results inside each group with LIME and LEMNA. We vary $k$ from 1 to 50 and repeat the experiments for 100 times to eliminate the randomness during training. Results are shown in Figure~\ref{fig:faithfulness-pensieve}. Since the decision tree interpretations of \name\ do not rely on a particular sample, they do not need to be clustered and are constant lines. 

\begin{figure*}
	\begin{minipage}{.33\linewidth}
		\centering
		\subfigure[Normalized Accuracy.]{
			\includegraphics[height=3.6cm]{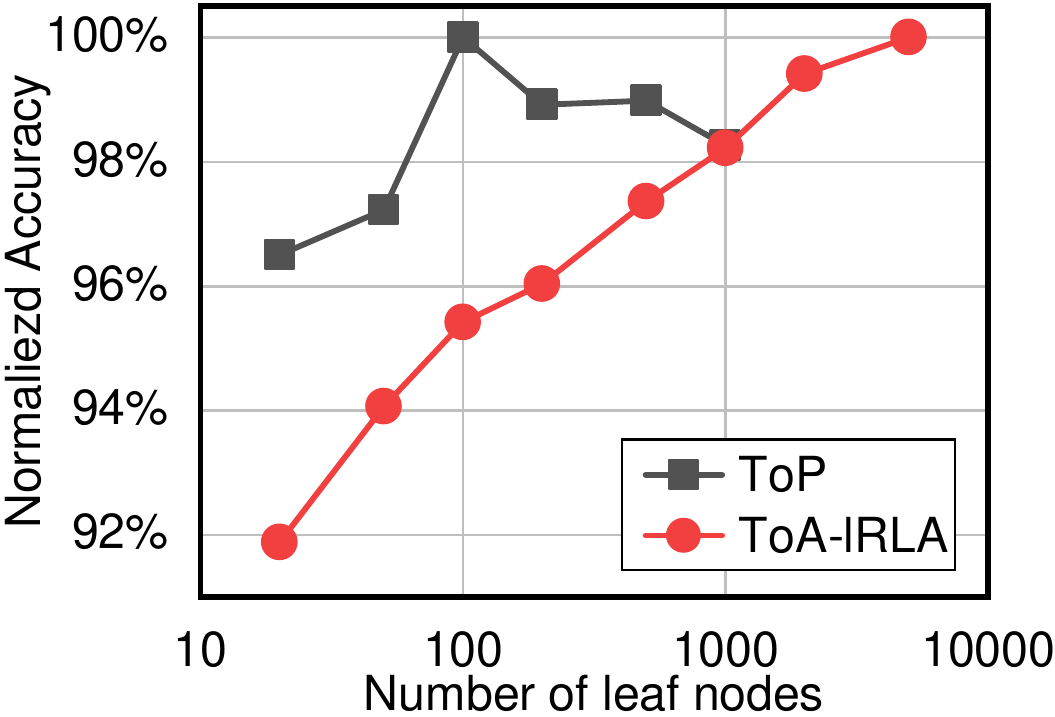}
			\label{fig:robust-accuracy}
		}
		\subfigure[Normalized RMSE.]{
			\includegraphics[height=3.6cm]{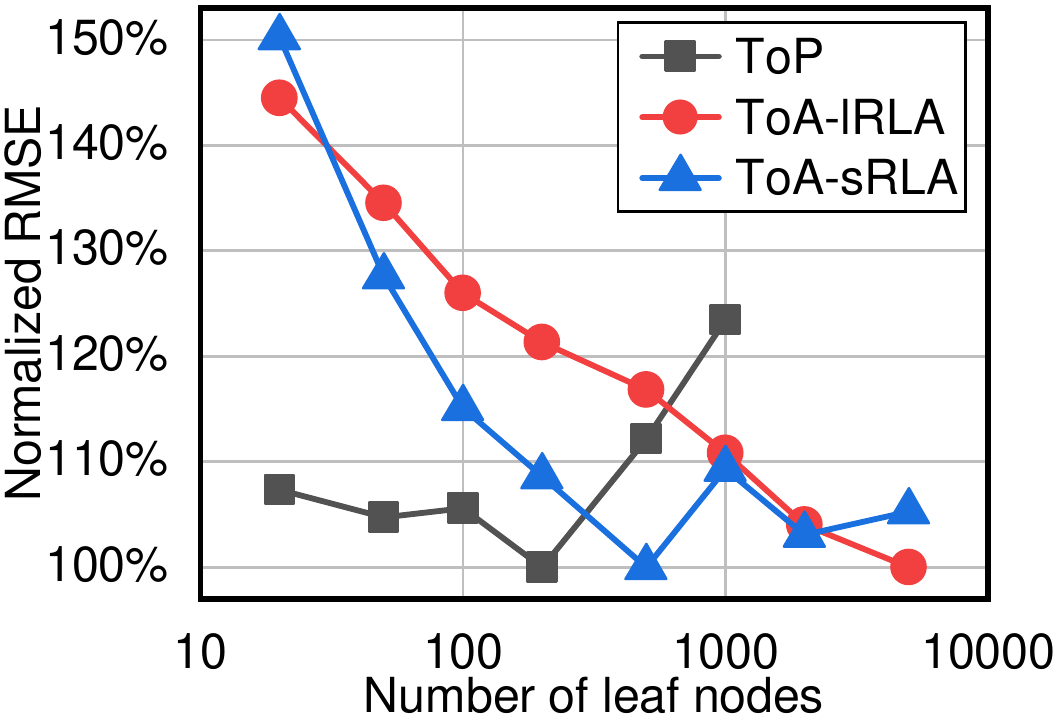}
			\label{fig:robust-rmse}
		}
		\caption{Sensitivity of leaf nodes on prediction accuracy and RMSE. Results are normalized by the best value on each curve.}
		\label{fig:robust}
	\end{minipage}
	\begin{minipage}{.33\linewidth}
		\centering
		\subfigure[Mask distribution when varying $\lambda_1$.]{
			\includegraphics[height=3.6cm]{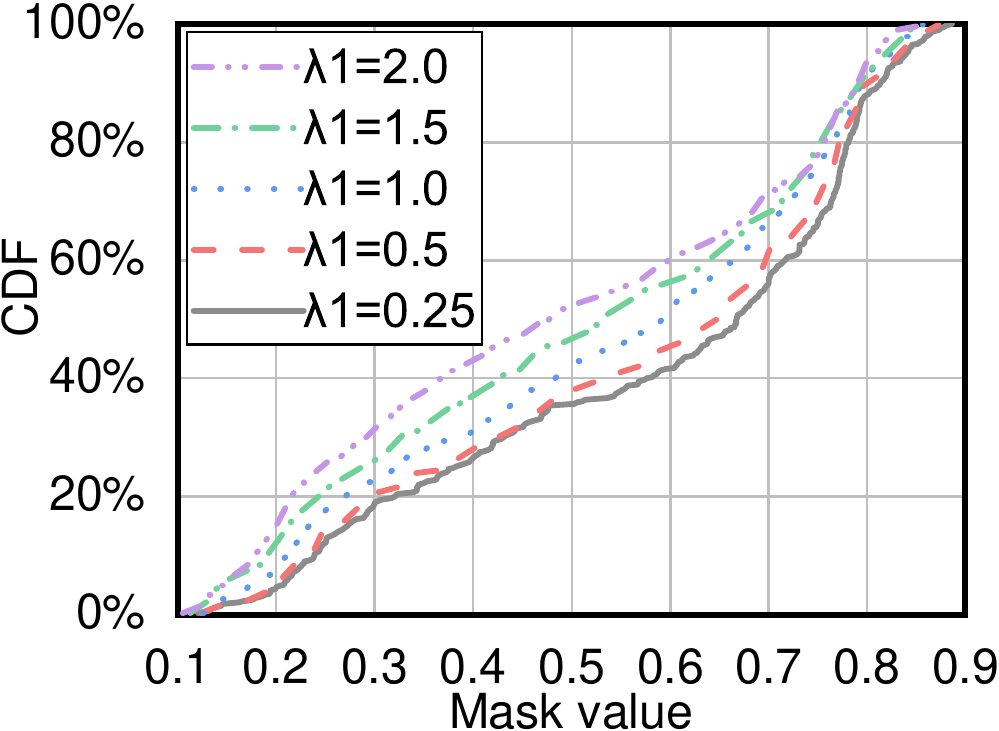}
			\label{fig:mask-lambda1}
		}
		\subfigure[Mask distribution when varying $\lambda_2$.]{
			\includegraphics[height=3.6cm]{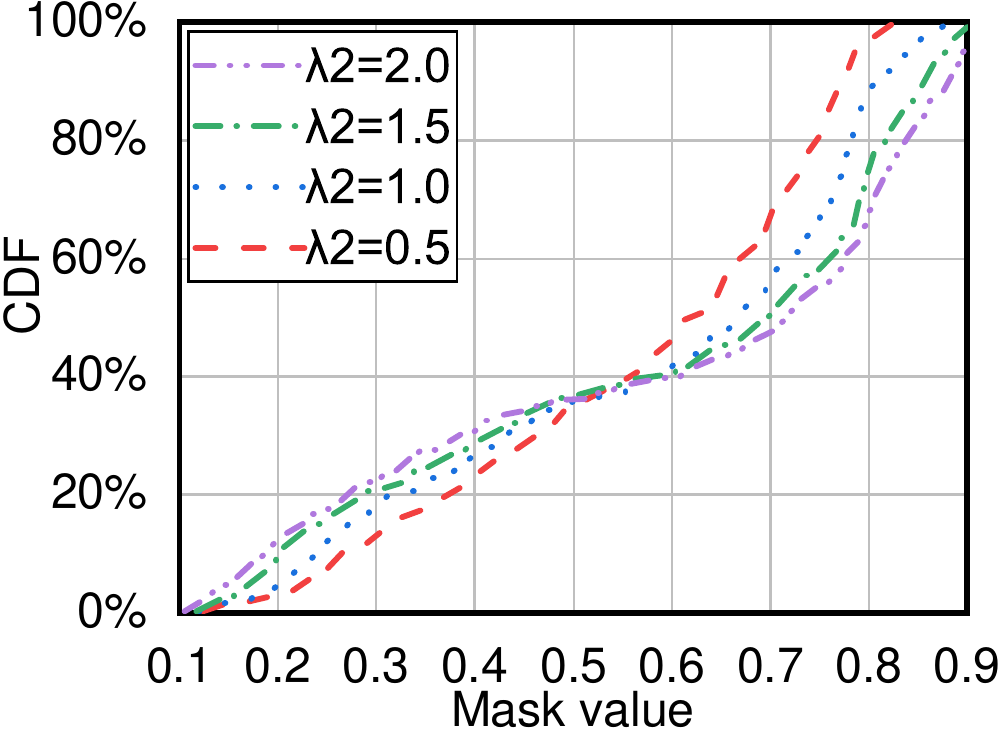}
			\label{fig:mask-lambda2}
		}
		\caption{The masks optimized by \name\ could effectively respond to the variation of hyperparameters $\lambda_1$ and $\lambda_2$.}
		\label{fig:gcs-sensitivity}
	\end{minipage}
\begin{minipage}{.33\linewidth}
		\centering
		\includegraphics[height=3.6cm]{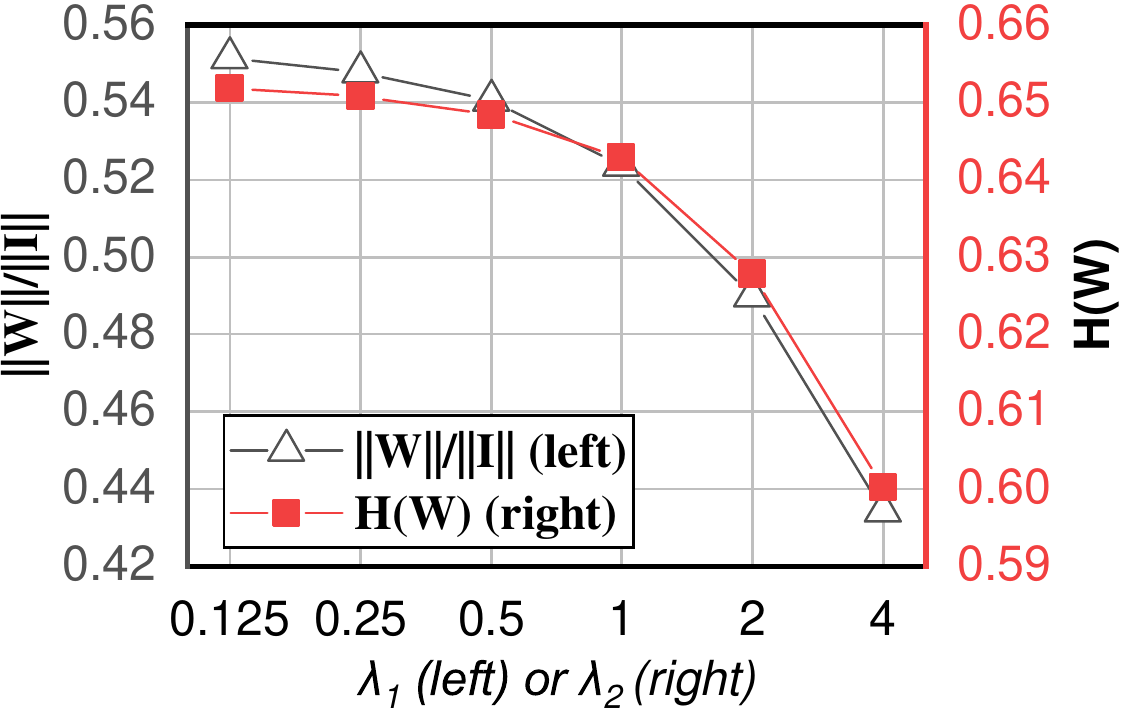}
		\caption{The value of different terms in Equation~\ref{eq:objective} reacts to the change of $\lambda_1$ and $\lambda_2$.}
		\label{fig:mask-goal}

		\includegraphics[height=3.6cm]{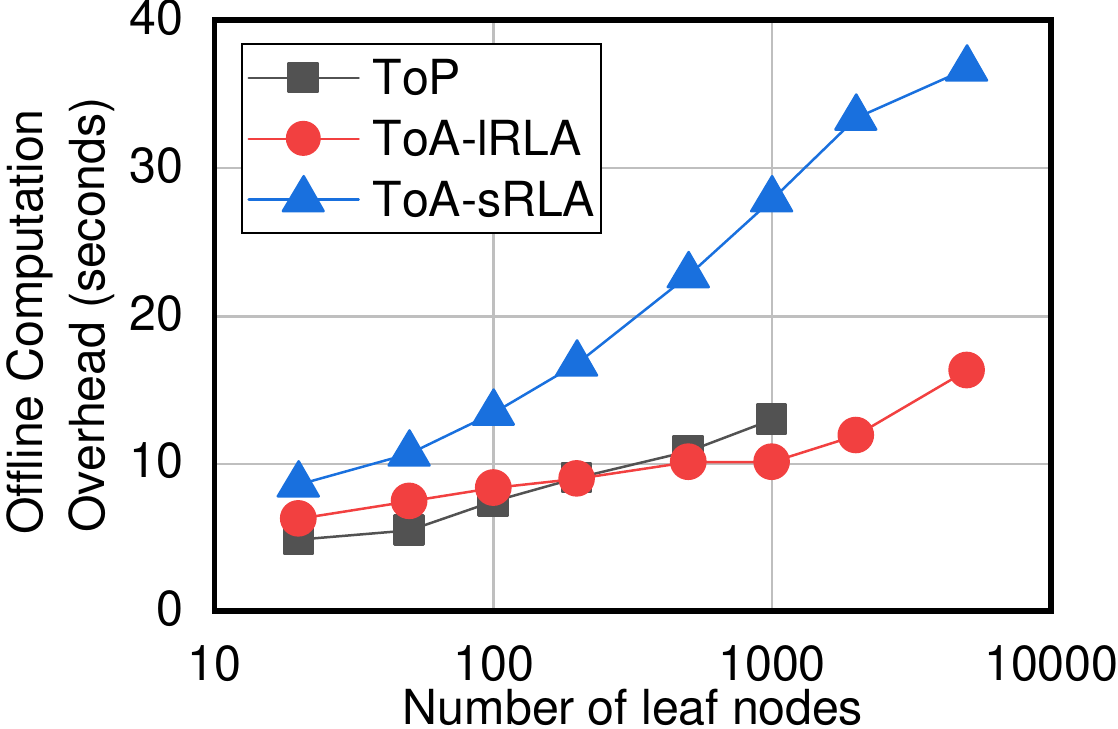}
		\caption{Offline Computation Overhead of \name\ with different number of leaf nodes.}
		\label{fig:overhead}
\end{minipage}
\end{figure*}

From Figures~\ref{fig:accuracy-pensieve} and \ref{fig:accuracy-toa-lrla}, \name+Pensieve and \name+AuTO-lRLA respectively achieve high accuracy of 84.3\% and 93.6\% compared to original DNNs. As the underlying decision logics of state-of-the-art algorithms in flow scheduling~\cite{sigcomm2013pfabric, nsdi2015pias} are much simpler than those of video bitrate adaption (e.g., stochastic optimization~\cite{sigcomm2015robustmpc}, Lyapunov optimization~\cite{infocom2016bola}), the accuracy of \name+AuTO-lRLA is a little higher than that of \name+Pensieve. The low decision errors in Figures~\ref{fig:rmse-pensieve}, \ref{fig:rmse-toa-lrla}, and \ref{fig:rmse-toa-srla} indicate that even for those decision tree decisions that are different from DNNs, the error made by \name\ is acceptable, which will not lead to drastic performance degradation. The accurate imitation of original DNNs with decision trees results in the negligible application-level performance loss in \S\ref{sec:case-lightweight}. Meanwhile, the accuracy and RMSE of \name\ are much better than those of LIME and LEMNA. Our design choice in \S\ref{sec:xai} is thus verified: decision trees can provide richer expressiveness and are suitable for networking systems. The performance of LEMNA is unstable for two agents of AuTO since the states of AuTO is highly centralized at several places from our experiments, which degrades the performance of expectation-maximization iterations in LEMNA~\cite{ccs2018lemna}.


\section{Sensitivity Analysis}
\label{sec:app-sensitivity}

In this section, we present the sensitivity analysis results on the hyperparameters of \name\ when applied to three DL-based networking systems.

\subsection{Pensieve and AuTO}
\label{sec:app-scs-sensitivity}


To test the robustness of \name\ against the number of leaf nodes, we vary the number of leaf nodes from 20 to 5000 and measure the accuracy and RMSE for the three agents evaluated in Appendix~\ref{sec:app-faith} (Pensieve, AuTO-sRLA, AuTO-lRLA). The results are presented in Figure~\ref{fig:robust}. The accuracy and RMSE of \name+Pensieve with the number of leaf nodes varying from 20 to 5000 are better than the best results of LIME and LEMNA in Figure~\ref{fig:faithfulness-pensieve} in Appendix~\ref{sec:app-faith}. \name+AuTO-lRLA and \name+AuTO-sRLA outperform the best value of LIME and LEMNA in a wide range from 200 to 5000 leaf nodes. The robustness indicates that network operators do not need to spend a lot of time in finetuning the hyper-parameter: a wide range of settings all perform well.

\subsection{RouteNet*}
\label{sec:app-gcs-sensitivity}

We measure the sensitivity of two hyperparameters $\lambda_1$ and $\lambda_2$ when interpreting the hypergraph-based global systems with \name\ as introduced in \S\ref{sec:action}. As presented in Figure~\ref{fig:mask-lambda1}, when network operators increase $\lambda_1$, $||W||$ will therefore be penalized. The generated mask values will also be reduced, shifting the cumulative distribution curve up. Those essentially critical connections will be revealed to network operators. Experiments of varying $\lambda_2$ demonstrate similar results, as presented in Figure~\ref{fig:mask-lambda2}. A higher $\lambda_2$ will make mask values concentrated at 0 or 1, resulting in a steeper cumulative distribution curve.

We further measure how will the specific values in Equation~\ref{eq:loss} change when network operators vary the hyperparameters. We measure the $\frac{||W||}{||I||}$ (scale of $W$) after training when varying $\lambda_1$ and keeping $\lambda_2$ unchanged in different experiments and present the results as the black line in Figure~\ref{fig:mask-goal}. We also measure the $H(W)$ (entropy of $W$) by varying $\lambda_2$ and keeping $\lambda_1$ unchanged and present the results in red in Figure~\ref{fig:mask-goal}. From the results, we can see that different terms in the optimization goal all actively respond to the changes in respective hyperparameters.

\section{Computation Overhead}
\label{sec:app-overhead}

We further examine the computation overhead of \name\ in decision tree extraction. We measure the decision tree computation time of Pensieve, AuTO-lRLA, and AuTO-sRLA at different numbers of leaf nodes on our testbed. As the action space of Pensieve (6 actions) is much smaller than those of AuTO-lRLA (108 actions) and AuTO-sRLA (real values), the decision tree of \name+Pensieve has completely been separated with around 1000 leaf nodes.  Thus we cannot generate decision trees for \name+Pensieve with more leaf nodes without enlarging the training set. As shown in Figure~\ref{fig:overhead}, even when we set the number of leaf nodes to 5000, the computation time is still less than one minute. Since decision tree extraction is executed offline after DNN training, the additional time is negligible compared to the training time of DNN models (e.g., at least 4 hours in Pensieve with 16 parallel agents~\cite{sigcomm2017pensieve} and 8 hours in AuTO~\cite{sigcomm2018auto}). \name\ can convert the DNNs into decision trees with negligible computation overhead. 

For RouteNet*, we also measure the computation time of the optimization of the mask matrix $W$. For hypergraph interpretations with 50 different traffic demand samples, the computation time of mask matrices is 80 seconds on average. For offline interpretations of RouteNet*, the computation time is negligible compared to the training time of DNNs. Even for online inspections, the interpretation time is acceptable for most cases since the routing information in a well-configured network rarely changes every tens of seconds~\cite{ton2007fir}. In the future, we will also investigate the incremental computation of the mask values to further accelerate the interpretation.

\end{document}